\documentclass[10pt]{article}
\usepackage{cite}
\usepackage{epsfig}
\usepackage[dvips,usenames]{color}
\usepackage{color}
\usepackage{rotating}
\usepackage{amsmath,amssymb,amsfonts}
\usepackage{latexsym}
\begin{document}
\setcounter{section}{0}
\setcounter{equation}{0}
\setcounter{figure}{0}
\setcounter{table}{0}
\setcounter{footnote}{0}
\begin{center}
{\bf\Large A Pedestrian Introduction to}

\vspace{5pt}

{\bf\Large the Mathematical Concepts of Quantum Physics}\footnote{Contribution to the Proceedings of the Fifth International
Workshop on Contemporary Problems in Mathematical Physics, Cotonou, Republic of Benin, October 27--November 2, 2007,
eds. Jan Govaerts and M. Norbert Hounkonnou (International Chair in Mathematical Physics and Applications,
ICMPA-UNESCO, Cotonou, Republic of Benin, 2008), pp.~36--114.}
\end{center}
\vspace{10pt}
\begin{center}
Jan GOVAERTS\\
\vspace{5pt}
{\sl Center for Particle Physics and Phenomenology (CP3),\\
Institut de Physique Nucl\'eaire, Universit\'e catholique de Louvain (U.C.L.),\\
2, Chemin du Cyclotron, B-1348 Louvain-la-Neuve, Belgium}\\
{\it E-Mail: Jan.Govaerts@uclouvain.be}\\
\vspace{7pt}
{\sl Fellow, Stellenbosch Institute for Advanced Study (STIAS),\\
7600 Stellenbosch, Republic of South Africa}\\
\vspace{7pt}
{\sl International Chair in Mathematical Physics and Applications (ICMPA-UNESCO Chair),\\
University of Abomey--Calavi, 072 B. P. 50, Cotonou, Republic of Benin}
\end{center}

\vspace{15pt}

\begin{quote}
These notes offer a basic introduction to the primary mathematical concepts of quantum physics, and their physical significance,
from the operator and Hilbert space point of view, highlighting more what are essentially the abstract algebraic aspects of
quantisation in contrast to more standard treatments of such issues, while also bridging towards the path
integral formulation of quantisation. A discussion of the (first) Noether theorem and Lie symmetries is also
included to complement the presentation. Emphasis is put throughout, as illustrative examples threading the presentation,
on the quantum harmonic oscillator and the dynamics of a charged particle coupled to the electromagnetic field, with the ambition
to bring the reader onto the threshold of relativistic quantum field theories with their local gauge invariances as
a natural framework for describing relativistic quantum particles in interaction and carrying specific conserved charges.
\end{quote}

\vspace{10pt}

\section{Introduction}
\label{Gov1.Sec1}

\subsection{Motivations}

With the Summer 2008, the world community of high energy physicists is eagerly awaiting to see the first proton beams
circulate in the Large Hadron Collider (LHC), located at the ``Centre Europ\'een, Organisation\
europ\'eenne pour la Recherche Nucl\'eaire" (CERN, Geneva, Switzerland; see {\tt http://www.cern.ch}).
Using fully electronics equipped detectors filling up huge caverns one hundred meters under ground,
at different intersection points of two oppositely moving proton beams in a 27 kilometers long circular collider,
these beams being held in their tracks by thousands of over ten meters long superconducting magnets kept cold at 1.9 K by a true liquid helium
factory---in itself an engineering feat heretofore never witnessed on the face of this Earth---,
teams of thousands of physicists and engineers will start taking data at rates and at energies never before
contemplated, to be analysed in search for the few tell-tale signs of discoveries to be made, those expected like those unexpected.
Among the primary motivations for this truly encompassing international enterprise bridging peoples from all continents,
the hopes are to finally settle some of the crucial questions that our present day theories for the fundamental interactions
and the Universe have unearthed and on whose answers so much hinges for any progress forward, and to catch the first glimpses
of what must be lying beyond, beyond the known energy and distance and time scales, and thereby reach towards those that prevailed
at the origins of time and of our physical Universe.

\begin{figure}[t!]
\begin{center}
\includegraphics[height=2.5cm,width=4cm]{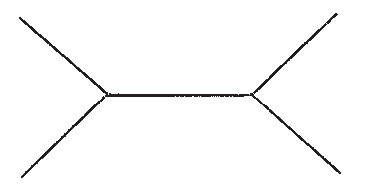}
\caption[]{A Feynman diagram: the interaction between two particles through the exchange of the energy and the momentum carried
by another particle, the carrier of the interaction.}
\label{Govaerts1.Fig1}
\end{center}
\end{figure}

One such question is that of the origin of mass, specifically inertial mass of all the known and unknown fundamental
constituents of matter and interactions, the elementary particles such as the electron or the quarks
(see Tables \ref{Table1.Govaerts1} and \ref{Table2.Govaerts1}). Because of its simplicity, and inspired by the dynamics
underlying the phenomenon of superconductivity in condensed matter physics with its condensation of electron Cooper pairs,
a large consensus favours the so-called Higgs mechanism,
in which particles acquire mass, namely inertia, through their interaction energy with a condensate of particles of some other
type, the so-called higgs boson. If the higgs boson does indeed exist by Mother Nature's own choice, and not by
the theorists' fancies however imaginative and elegant their intellectual inventions may be, it will be discovered at the LHC.
When wanting to extend to matter particles the elegance displayed through the so-called gauge symmetries of the fundamental interactions
with their specific carriers (see Figure~\ref{Govaerts1.Fig1}), expectations are high for the discovery of a new world prevalent
at the higher energies in which each of the known particles of spin 1/2 or 1 is accompanied by a supersymmetric partner
of spin 0 or 1/2, respectively.
Such new structures lie beyond the limits of the present day accepted model, the so-called Standard Model, having so far survived beyond
anyone's best bets the most stringent experimental tests imaginable. More generally, one expects that some form of new physics
beyond that Model will start to unravel at the LHC, such as supersymmetry but possibly something we actually never thought
of since Mother Nature has this supreme knack for always outwitting us for She seems to never be using the same trick twice.
More speculative or fascinating phenomena, as one may like to call
them, are also foreseen by some, though based on less solid and more tantalising ideas and conjectures, such as
the possibility of ``large" extra spatial dimensions having remained hidden to our senses until now, or the
production of light black holes if gravity acts at some energy scale in such extra spatial dimensions lower than that
at which it appears to be acting in our four dimensional spacetime world. Many more such fascinating features are potential
possibilities, all having in common that they would leave behind in LHC data subtle tell-tale signs to be brought
to the fore through painstaking analysis, especially through precision studies of the electroweak flavour interactions.

\begin{table}
\begin{center}
\begin{tabular}{|c|c|c|c|c|}
\hline
 & \multicolumn{3}{c|}{Families} & $Q$ \\
\hline
 & $\nu_e$ & $\nu_\mu$ & $\nu_\tau$ & $0$ \\
Leptons & & & & \\
 & $e^-$ & $\mu^-$ & $\tau^-$ & $-1$ \\
\hline
 & & & & \\
 & $u$ (up) & $c$ (charm) & $t$ (top) & $+\frac{2}{3}$ \\
Quarks & & & & \\
 & $d$ (down) & $s$ (strange) & $b$ (bottom) & $-\frac{1}{3}$ \\
 & & & & \\
\hline
\end{tabular}
\caption[]{The structure of matter: the periodic table of the elementary particles (all of spin $\frac{1}{2}$).}
\label{Table1.Govaerts1}
\end{center}
\end{table}

\begin{table}
\begin{center}
\begin{tabular}{|l|c|c|c|}
\hline
Interaction & Carrier & Spin & Local gauge symmetry \\
\hline\hline
Strong & $8$ gluons & $1$ & SU(3)$_c$ \\
\hline
Electromagnetic & $\gamma$ (photon) & $1$ & Electroweak interaction \\
Weak & $W^\pm$, $Z_0$ & $1$ & SU(2)$_L\times$U(1)$_Y$ \\
\hline
Gravity & graviton (?) & $2$ & Spacetime reparametrisations \\
 & (spacetime curvature) & & (local Poincar\'e group ISO(3,1)) \\
\hline
\end{tabular}
\caption[]{The structure of the fundamental interactions: from the strongest to the weakest.}
\label{Table2.Govaerts1}
\end{center}
\end{table}

But on which paper music are the scales of the Standard Model and its description of all known
fundamental particles and interactions written? By which rules have the harmonies performed by this huge
and richly diversified and dynamic symphonic orchestra of the microscopic universe been organised and directed
ever since the creation of the physical Universe? Building on the legacy of XIX$^{\rm th}$ century physics,
through the marriage of these two fundamental conceptual revolutions which have been quantum physics, involving
as fundamental physical constant $\hbar=h/2\pi$---the reduced Planck constant---, and special relativity,
involving as fundamental physical constant $c$---the speed of light in vacuum as the limiting speed for any actual
exchange of energy and momentum, namely any interaction---, physics of the XX$^{\rm th}$ century has come
to realise that relativistic quantum field theory (QFT), though a conceptual revolution still to be completed
in the XXI$^{\rm st}$ century \cite{Gov1.Gov1,Gov1.Gov2}, provides the right harmonies by which to scale up
the fundamental dynamics and matter content of our world (see Table~\ref{Table3.Govaerts1}). A dynamical field
possesses both at the same time an oscillatory character and a wavelike spacetime propagation behaviour inclusive of interference properties.
As is true for any oscillatory system, when quantised a dynamical field possesses quantum states of quantised energy,
the number of these quanta being modified through exchanges of energy, namely interactions through the exchanges of particles.
Yet its spacetime dynamics remains wavelike with all its ensuing correlation and interference properties. A quantum field
thus achieves the description of a system which at the same time is constituted of particles---when one measures
its energy content and its quantum states---and of waves---when one considers its spacetime dynamics and correlations,
thereby by-passing any possible issue of conceptual inconsistencies of these dual perceptions of the physical reality.

On account of Noether's theorem which associates a conserved dynamical quantity to each generator of a symmetry of a dynamics,
relativistic fields evolving within a Minkowski spacetime with its invariance properties under spacetime translations and rotations
also possess conserved energy, momentum and angular-momentum. The quantum states of a relativistic
quantum field theory thus carry all the hallmarks defining a relativistic quantum particle, namely its energy,
momentum---hence relativistic invariant mass $m$ through the relation $E^2-(\vec{p}c)^2=(mc^2)^2$, $E$ and $\vec{p}$
being the particle's energy and momentum, respectively---, and also spin. Yet in their spacetime dynamics these particles
possess all the properties of waves as well. Furthermore, other conserved quantities that particles may carry, such as the
electric charge, are then also associated to some other, non spacetime symmetries, so-called internal symmetries.
When these symmetries are realised in such a manner that the effected transformations may be different at each
point in spacetime, even though in a continuous fashion throughout spacetime---namely, gauge symmetry transformations---,
the result is necessarily the existence of an interaction---a gauge interaction---with its own field of which the quanta are the carrier
of that interaction acting between those particles carrying the charge associated to that symmetry. All known
fundamental interactions are understood within that framework, given some symmetry group (see Table~\ref{Table3.Govaerts1}).

\begin{table}
\begin{center}
\begin{tabular}{|c|c|c|c|c|}
\hline
Quantum & & Special Relativity & & Gravity \\
 & & & & \\
$\hbar$ & & $c$ & & $G_N$ \\
$\searrow$ & & $\swarrow$ $\searrow$ & & $\swarrow$ \\
\hline
 & Relativistic & & General Relativity & \\
 & Quantum Field Theory & & (not quantum) & \\
\hline
Local & & & & \\
gauge & SU(3)$_c\times$SU(2)$_L\times$U(1)$_Y$ & & Local spacetime & \\
symmetries & & & diffeomorphisms & \\
 & $\downarrow$ & & & \\
 & ? Grand Unification & & & \\
 & $\searrow$ & & $\swarrow$ & \\
\hline
 & & & & \\
 & & {\bf ? Quantum Gravity ?} & & \\
\hline
\end{tabular}
\caption[]{General conceptual framework: road map to the final unification.}
\label{Table3.Govaerts1}
\end{center}
\end{table}

The conceptual framework of relativistic quantum field theory---the music paper for the harmonies of the
microscopic universe---is thus a most encompassing one producing, through the marriage of $\hbar$ and $c$,
quite a unification of concepts accounting for observational facts of the physical Universe. For instance,
note that rather than having to speak and consider all possible particles in the Universe of a given species,
say all electrons of the Universe, it suffices to consider the single field filling all of spacetime
of which the quanta are that specific particle but existing in different quantum states, hence with
different energy-momentum and spin values. All electrons of the Universe are indistinguishable because
they are harmonic quantum excitations of a same and single underlying basic physical reality, the electron field.
This totally new view point on the reality of the physical world is but one illustration of the
unification of concepts unravelled through the construction of quantum field theories as an appropriate
framework for the understanding of the fundamental particles and their interactions. Quite a towering
achievement of XX$^{\rm th}$ century physics, its fourth but yet unfinished conceptual revolution of the
harmonies of the physical Universe!

Indeed, this achievement remains partially unfulfilled since, as is well known, when computing within
these theories corrections to quantum processes generated by higher order exchanges of particules, one always
encounters divergences generated by fluctuating harmonic modes of the fields at the highest energy
or shortest distances scales. However, for all the above gauge theories describing all the fundamental interactions
except for gravity, these ultra-violet (UV) divergences may be renormalised away, when expressing
all observables in terms of the physical parameters of particles rather than those un-normalised parameters
defining the theory before considering its quantised version. Thus even though on a practical level one may
live comfortably with such a situation, the fact remains somewhat unsettling, and may well beg for a final
theory within which relativistic quantum field theory would find its rightful place in a certain limit,
somewhat like Newton's nonrelativistic dynamics is a limiting subcase of Einstein's special relativity dynamics
for velocities small compared to $c$. However, gravity remains an exception in that respect. Even though
the third conceptual revolution of XX$^{\rm th}$ century physics, namely General Relativity (GR) resulting from the
marriage of $c$ and Newton's constant $G_N$ for the gravitational interaction, has achieved a great unification
of concepts in its own right with regards to the Universe at its largest scales where gravity reigns supreme, all
attempts so far at a final marriage of all three fundamental constants $\hbar$, $c$ and $G_N$ within the
field theory context have failed. Even though GR itself is also a gauge dynamics, though of a different type since
the symmetries involved are those of spacetime rather than internal symmetries as is the case for all other three
fundamental interactions, the renormalisation programme has not resolved the issue of the UV divergences
within a field theory approach to quantum gravity including all of matter and interactions (see Table~\ref{Table2.Govaerts1}).
Thus indeed, the final unification of gravity with the other three fundamental interactions and all forms of
matter particles within a single quantum framework remains one of the formidable challenges offered to the young and creative
physicists of the XXI$^{\rm st}$ century. XXI$^{\rm st}$ century physics is in search of a next Einstein \cite{Gov1.Gov1,Gov1.Gov2}.
Let us all dream and work hard that she will come from Africa!

The purpose of these notes is mainly to provide an introduction, put may be in more mathematical terms and concepts than what is usual,
to the basics of quantum physics, whatever the type of dynamical system being considered provided it possesses
a classical variational principle description based on an action (see for instance Ref.~\cite{Gov1.GovBook}).
This framework encompasses both nonrelativistic
as well as relativistic systems, as well as systems with finite or infinite numbers of degrees of freedom, field
theories being of course of the latter class. In order to also reach out more to our more mathematically
inclined readers, emphasis is laid on the algebraic and geometric aspects of quantisation rather than more traditional
trodden paths through the Schr\"odinger wave equation for instance. Yet, physics illustrations and applications
are never far behind, while at the same time aiming towards an understanding of the basics of relativistic quantum field theory,
even though the latter as such are not addressed in the present pages. Previous lectures in these Series
have already discussed such aspects, with notes available in Refs.\cite{Gov1.Gov1,Gov1.Gov2}.
An excellent treatise is Ref.\cite{Gov1.Peskin}. No attempt shall be made to provide here
any further references, and even less an exhaustive list of reviews on these topics. Such material is
readily available through references given in those above. In particular, the basics of special relativity are
not included either, since many sources for thorough, detailed and clear presentations are easily found as well
(for a web site of interest, see {\tt www.phys.unsw.edu.au/einsteinlight/}).

\subsection{An outline}

\vspace{10pt}

\begin{center}
\fbox{
\begin{tabular}{ccc}
Lagrangian & $\longleftrightarrow$ & Hamiltonian \\
formulation & (Legendre transform) & formulation\\
 & $\nwarrow \qquad\qquad\qquad\qquad\qquad\qquad \nearrow$ & \\
$\hbar$ $\updownarrow$ & \fbox{\bf ACTION (Symmetries)} & $\updownarrow$ $\hbar$\\
 &  $\swarrow \qquad\qquad\qquad\qquad\qquad\qquad \searrow$ & \\
Path/Functional integral &  & Canonical/Operator \\
quantisation & $\longleftrightarrow$ & quantisation
\end{tabular}
}
\end{center}

\vspace{10pt}

As illustrated in the above diagram, the quantity central to our entire discussion is the action of the system.
Its dynamical equations of motion follow from it through the variational principle which requires that
classical trajectories be extremal points of the action in configuration space. The action also embodies
most elegantly the existence of symmetries by being invariant, possibly up to a surface term, under the
corresponding transformations of the degrees of freedom of the system. As a consequence, through Noether's theorem,
one readily identifies conserved quantities. Because of the necessary requirement of (space)time locality,
the action is given by the (space)time integral of a local function of configuration space, namely of
the degrees of freedom of the system and their (space)time derivatives. This function is known as the Lagrange
function or Lagrangian of the system, or Lagrangian density in the case of fields for which a space and time
integration is effected to obtain the action.

One possible path towards quantisation is to directly move from the classical dynamics in its Lagrangian form
to the path integral, or functional integral representation of the quantum dynamics (the l.h.s. of the above
diagram). Here rather we shall take the ``canonical path", namely move from the Lagrangian formulation to the
Hamiltonian one for a given system, to which the rules of canonical operator quantisation are then applied
to result in a purely abstract algebraic construction defining the quantised system (the r.h.s. of the above
diagram). However, we shall show how both realisations of a quantum system are actually equivalent, as much from
the mathematical as from the physical point of view. Simply, depending on the problem at hand, one representation
may be more convenient than the other to address some issues or computations, while they also speak differently
even though in complementary ways to one's physical intuition. The path integral approach has provided much fruitful
insight into the nonperturbative dynamics of nonlinear quantum field theories, for instance. Here the operator approach
is rather emphasized, first, because it allows to quickly develop a language which mathematicians and physicists alike
can easily share, and second, because it is the most straightforward path to grasping why it is that relativistic quantum field theories
are in fact theories of relativistic quantum particles, and {\it vice versa\/}, simply by relying on the quantum
physics of the ordinary one dimensional harmonic oscillator, certainly the simplest of all nontrivial dynamical systems,
and in a certain manner thus the ``mother" of all quantum field theories, at least in as far as perturbative quantisation
and renormalisation of field theories are concerned. This is no small wonder! Indeed, when applied for instance
to the electromagnetic interactions of electrons and photons, namely the theory of quantum electrodynamics (QED),
such techniques have produced theoretical results precise to eleven decimal places in perfect agreement
with experimental results known to the same amazing degree of accuracy. When the art of quantum physics is pushed to
such extremes of excellence, any shadow of a discrepancy could be the harbinger for new physics beyond the Standard Model.

It is here that these notes find another of their motivations. Namely, to bring the reader onto the threshold
of quantum field theory, a study that he/she could then hopefully embark on afterwards on his/her own, with the help
possibly of references such as those in Refs.\cite{Gov1.Gov1,Gov1.Gov2,Gov1.Peskin}. One of the ambitions of these
notes is to demystify, if necessary, what relativistic quantum field theory actually is in its basic essentials
and physical representations, simply as a natural framework for theories of relativistic quantum particles in interaction and characterised
by conserved charges in direct correspondence with important symmetries whether of spacetime or
internal space geometries.

The outline of the notes is as follows. In the next Section, the Lagrangian formulation of dynamical systems
is briefly reviewed. In Section~\ref{Gov1.Sec3}, through a Legendre transform, one moves over to the Hamiltonian
formulation of the same dynamics, leading to the phase space representation of its degrees of freedom and
the ensuing geometrical structures. Section~\ref{Gov1.Sec4} then addresses the canonical quantisation of the
Hamiltonian formulation, emphasizing mostly the purely abstract aspects to that programme, as well as the
different representations possible for those abstract structures. It is also at that point that the junction with
the path integral quantisation of the classial Lagrangian dynamics is made. Finally, Section~\ref{Gov1.Sec5}
discusses the content and consequences of the Noether theorem, a most important result in relation to the
existence of continuous (Lie) symmetry groups of transformations leaving the equations of motion of the
system invariant and with as further consequence the existence of conservation laws.
Some concluding comments are offered in Section~\ref{Gov1.Sec6}.

\noindent
\section{Lagrangian Dynamics}
\label{Gov1.Sec2}

\subsection{The Action Principle}

Let us consider some dynamical system. Its configuration space (the space of all its possible configurations)
may be characterised in terms of some manifold $M_N$ of dimension $N$. For example, in the case of a single point particle
moving along an infinite straight line, the possible configurations of the system are any of the positions along
that line; its configuration space is thus the real line, $M_N=\mathbb{R}$, a one dimensional space, $N=1$,
for this one degree of freedom system. By extension, the configuration space of a single point
particle moving in an $N$ dimensional Euclidean space is $M_N=\mathbb{R}^N$. Likewise, a point particle
constrained to be moving on a circle has that circle as configuration space. More generally, constrained to
be moving on a sphere of dimension $N$, its configuration space is the $N$-sphere, $M_N=S^N$. Such a
point particle could also be moving on a torus of dimension $N$, the cartesian product of $N$ circles,
corresponding to a configuration space which is the $N$ dimensional torus, $M_N=T^N$.

The configuration space manifold $M_N$ is also taken to be connected, for if it were to have multiple components,
each component could be seen to correspond to a different system, each of these systems being decoupled from
one another. Generally, configuration space also comes equipped with some geometry, defined in terms of some
metric structure, usually a positive definite one.

Being a manifold, a local system of coordinates may be defined over configuration space,
which is also required in order to perform actual calculations and for representing trajectories of the system
throughout its configuration space. Generally, such local coordinates are denoted $q^n\in\mathbb{R}$ with $n=1,2,\ldots,N$,
corresponding to generalised coordinates in configuration space. For example, if configuration space is
Euclidean, $M_N=\mathbb{R}^N$, these coordinates could be taken to be cartesian coordinates with respect to some
scalar product, or positive definite metric structure defined over configuration space. In the case of
the $N$-sphere, $M_N=S^N$, or the $N$-torus, $M_N=T^N$, natural coordinates would correspond to some angular
parametrisation of the configuration space manifold.

For a realistic system, the number, $N$, of such degrees of freedom $q^n$ $(n=1,2,\ldots,N)$ may be extremely large indeed.
Imagine a system of $N$ particles moving in three dimensional Euclidean space representing physical space. The associated configuration
space is $\mathbb{R}^{3N}$. Choosing to work in terms of cartesian coordinates, these $3N$ degrees of freedom
may be expressed as time dependent functions $x^i_\alpha(t)$, with $i=1,2,3$ and $\alpha=1,2,\ldots,N$.
But in practice for a cubic centimeter of ordinary matter, $N$ is given essentially by Avogadro's number,
$N_A\simeq 6\times 10^{23}$. Note also that the cartesian choice of coordinates may not be the most convenient
one for the general $N$-body system, even already for the $N=2$ two-body problem for which relative and center-of-mass
coordinates are more relevant.

Any time history of the system is associated to some trajectory in configuration space, representing all
the successive configurations of the system as it evolves in time. Having specified a local coordinate system
over configuration space, such a trajectory is then given by a set of time parametrised functions $q^n(t)$
$(n=1,2,\ldots,N)$, assumed to be sufficiently smooth for all practical purposes. These functions
thus correspond to the degrees of freedom of the system, since they represent the freedom the system has
in moving throughout its space of possible configurations as time evolves. Note that the parameter $t$
need not necessarily be the physical time as measured on a clock, even though quite generally $t$ is
chosen to be such a physical observable. Indeed $t$ serves the purpose of parametrising the dynamical
evolution of the system along its trajectories in configuration space, and any such parametrisation is
acceptable. In particular for systems invariant under arbitrary transformations of spacetime coordinates,
such as General Relativity, string and M-theory, or even the single parametrised relativistic particle \cite{Gov1.GovBook},
the time evolution parameter $t$ is such an arbitrary choice, of which the physical time itself is then a particular function.

Note that the number $N$ of degrees of freedom need not be finite. The above framework remains applicable
for systems with an infinite number of degrees of freedom, including field theories. This number could
be infinie countable or even non countable. Consider for instance a real scalar field, $\phi(t,\vec{x}\,)\in\mathbb{R}$,
defined over spacetime, $\vec{x}$ parametrising the space dependence of the field. As a matter of fact,
that space variable may be seen to correspond to an index, albeit a continuous one, labelling
the different degrees of freedom of the system,
\begin{equation}
\phi(t,\vec{x}\,)=\phi^{\vec{x}}(t)\leftrightarrow q^{\vec{x}}(t)\leftrightarrow q^n(t).
\end{equation}
Even if the label $\vec{x}$ takes values in a continuous set, the number of degrees of freedom
is not necessarily infinite non countable. In the case for instance of a torus topology for the spatial
directions, through Fourier mode analysis in $\vec{x}$ the number of degrees of freedom is infinite discrete,
thus infinite countable with $N\to\infty$. Consequently, at least formally the entire discussion to be developed in these
notes extends also to field theories, keeping aware of possible difficulties arising because of an infinite
number of degrees of freedom.

Given such a parametrisation of configuration space, how does one determine the dynamics of the system
within its configuration space? Through which equations of motion governing its time evolution? As mentioned
previously, such dynamics follows from the variational principle applied to a specific quantity known
as the action of the system, $S[q^n]$. This action is a {\bf functional} of configuration space,
namely a number which is a function of functions, constructed out of the time dependent functions $q^n(t)$
characterising any trajectory of the system in its configuration space. As already indicated, the
action is a most encompassing and powerful concept, and plays a central r\^ole in accounting for the dynamics
of the system and all the properties thereof, since,
\begin{itemize}
\item[1.] the dynamical equations of motion follow from the action, $S[q^n]$, through the variational principle, and,
\item[2.] as will be discussed later on, through Noether's first theorem, invariance of the action
under symmetry transformations embodies the existence of conservation laws, namely the existence
of conserved quantities for the classical system, corresponding to conserved quantum numbers
(such as the electric charge, the spin and angular-momentum, etc.) for the quantised system.
\end{itemize}
Requiring locality in time, the action must of the form
\begin{equation}
S[q^n]=\int_{t_i}^{t_f}dt\,L(q^n(t),\dot{q}^n(t)),
\end{equation}
where $L(q^n,\dot{q}^n)$ is some function of the variables $q^n$ and $\dot{q}^n$---viewed at this stage as
independent variables with respect to which separate partial derivatives of the function $L(q^n,\dot{q}^n)$
may be taken---, known as the {\bf Lagrange function} or {\bf Lagrangian} of the system.
When substituted in the above integral, the Lagrange
function is composed with the time dependence $q^n(t)$ representing the flow of the system
in time along any specific trajectory in configuration space represented by the functions $q^n(t)$ $(n=1,2,\ldots,N)$.
In that case, $\dot{q}^n(t)$ then stands for the ordinary derivative of the function $q^n(t)$ with respect to time $t$,
a notation customary in mechanics,
\begin{equation}
\dot{q}^n(t)\equiv \frac{dq^n(t)}{dt}\qquad [{\rm generalised\ velocities}].
\end{equation}
The value for the action, $S[q^n]$, is thus associated to a specific trajectory in configuration space
corresponding to the time interval $[t_i,t_f]$ for which one has the initial and final configurations
$q^n(t_i)=q^n_i$ and $q^n(t_f)=q^n_f$. This trajectory may be pictured as some curved line in configuration space
connecting these two end points.

\vspace{10pt}

\noindent
\underline{\bf Remarks}

\vspace{10pt}

\noindent
1. In the case of a field theory, locality in both time and space implies that
the Lagrange function should itself be given as the space integral of a {\bf Lagrangian density},
${\cal L}(\phi,\partial_t\phi,\partial_{\vec{x}}\phi)$, the latter being a local function of the
field and its time and spatial derivatives, to be then composed with the time and space dependence
of the field, $\phi(t,\vec{x}\,)$, when evaluating the corresponding action,
\clearpage
\begin{equation}
S[\phi]=\int_{t_i}^{t_f}dt\int_{\rm space}\,d\vec{x}\,{\cal L}(\phi(t,\vec{x}\,),
\partial_t\phi(t,\vec{x}\,),\partial_{\vec{x}}\phi(t,\vec{x}\,)),\ \
L=\int_{\rm space}\,d\vec{x}\,{\cal L}(\phi(t,\vec{x}\,),
\partial_t\phi(t,\vec{x}\,),\partial_{\vec{x}}\phi(t,\vec{x}\,)).
\end{equation}
Essentially, the Lagrange function of the system remains given as a sum over the degrees of freedom
of the system distinguished by the vector index $\vec{x}$.

\vspace{10pt}

\noindent
2. The Lagrange function may, in general, also have an explicit time dependence,
$L(q^n,\dot{q}^n;t)$. However in such a situation, if $t$ stands for the physical time,
the energy of the system is not conserved, a situation which does not apply at a fundamental physical
level. It is for this reason that at the outset we take the Lagrange function not to have any explicit
time dependence, $L(q^n,\dot{q}^n)$. As will be established later on, the Hamiltonian (which
coincides with the energy when the time evolution parameter $t$ coincides with the physical time)
is then indeed a constant of motion, namely a conserved quantity.

\vspace{10pt}

\noindent
3. One could consider systems for which the Lagrange function may depend on derivatives of the
degrees of freedom of order exceeding one, $L(q^n,\dot{q}^n,\ddot{q}^n,\cdots)$. However, through
the introduction of an appropriate choice of auxiliary variables of which the equations of motion
are such that these extraneous degrees of freedom coincide with the successive time derivatives of the
original ones, $q^n$, it is always possible to bring the description of the system into the
above general form, in which the Lagrange function depends on the complete set of degrees of
freedom, inclusive of the auxiliary ones, and their first order time derivatives only. Hence
no loss of generality is incurred through the above choice of parametrisation for the Lagrange function.

\vspace{10pt}

Given the specification of the action functional for the system, its dynamics then follows from the
local variational principle, which states that,
\begin{quote}
{\bf Variational Principle}:
Classical trajectories of the system correspond to local minima (which is possible in the best of cases only,
otherwise more generally, they correspond to local extrema or even just stationary points) of the action,
$S[q^n]$, of the system.
\end{quote}

As shall become clear hereafter, this principle implies differential equations of order two in time,
one for each independent degree of freedom $q^n(t)$ of the system, namely specific equations of motion of which
the solutions represent the dynamics or time evolution of the classical configurations of the system. 
Being differential equations of second order in time means also that one must specify for each of these
equations of motion two boundary conditions, or integration constants, in order to determine in a unique
fashion a specific solution.

The practical implementation of the variational principle proceeds as follows. Imagine that a given
(still unknown) classical trajectory $q^n(t)$ determines such a minimum (or stationary point, in general) of the
action, and consider then an arbitrary but ``infinitesimal" variation $\delta q^n(t)$ in the
neighbourhood of $q^n(t)$. The value for the action will thus change accordingly. Expanding the latter
to first order in the variation $\delta q^n(t)$, the requirement is that to first order the variation
of the action should vanish identically. As we shall see, this latter requirement is to be considered
up to surface terms in time, stemming from a contribution to the variation which is a total time derivative.
Thus more explicitly and precisely, one has the following representation of the variational principle,
\begin{eqnarray}
q^n(t)&\longrightarrow& q^n(t)+\delta q^n(t),\quad \delta q^n(t):\ \mbox{infinitesimal variation}, \nonumber\\ 
S[q^n]&\longrightarrow& S[q^n+\delta q^n]=S[q^n]+\delta S[q^n]+(\mbox{higher order terms}),
\end{eqnarray}
with
\begin{equation}
\delta S[q^n]=0,\qquad \mbox{up to surface terms, or total derivatives}.
\end{equation}
As mentioned above, the latter requirement translates into a set of differential equations
of motion for the functions $q^n(t)$, the Euler--Lagrange equations of motion.

\subsection{\bf The Euler--Lagrange equations of motion}

Let us proceed with the explicit evaluation of $\delta S[q^n]$ to first order in $\delta q^n(t)$,
\begin{eqnarray}
\delta S[q^n] &=& S[q^n+\delta q^n]-S[q^n]\qquad\qquad\qquad\qquad\qquad\qquad
 [{\rm only\ to\ first\ order\ in\ } \delta q^n] \nonumber\\
 &=& \int_{t_i}^{t_f}dt\left\{L(q^n+\delta q^n,\dot{q}^n+\delta\dot{q}^n)-
L(q^n,\dot{q}^n)\right\}\qquad\qquad \left[\delta\dot{q}^n(t)=\frac{d}{dt}\delta q^n(t)\right] \nonumber \\
 &=& \int_{t_i}^{t_f}dt\left\{\delta q^n\frac{\partial L}{\partial q^n}
+\frac{d}{dt}\delta q^n\frac{\partial L}{\partial\dot{q}^n}\right\} \nonumber \\
 &=& \int_{t_i}^{t_f} dt\,\delta q^n
\left[\frac{\partial L}{\partial q^n}-\frac{d}{dt}\frac{\partial L}{\partial\dot{q}^n}\right]\,+\,
\int_{t_i}^{t_f}dt\frac{d}{dt}\left[\delta q^n\frac{\partial L}{\partial\dot{q}^n}\right],
\label{J1.eq:deltaS}
\end{eqnarray}
where in the last line an integration by parts was effected in order to isolate all terms in
$\delta q^n$ as factorised contributions. In (\ref{J1.eq:deltaS}), the first term is given as an integral
over the entire ``volume" of the time interval $[t_i,t_f]$ (in the case of field theory, it is given
as an integral also over the volume of space), hence that term is the ``volume" term contribution to
$\delta S[q^n]$. On the other hand, the second contribution in (\ref{J1.eq:deltaS}) is a surface
term in time, being given by the time integration of the total time derivative of the specific
combination $\delta q^n\partial L/\partial\dot{q}^n$, of which the value thus depends only on the
values of $q^n(t)$, $\dot{q}^n(t)$ and $\delta q^n(t)$ at $t_i$ and $t_f$, namely the boundary
or ``surface" of the time interval $[t_i,t_f]$ (again in the case of field theory, one then gets
genuine surface terms of space and time). Hence that second contribution is the ``surface" term contribution
to $\delta S[q^n]$.

\noindent
Another remark is worth to be made explicit here. In the above expressions, the so-called Einstein
convention for summation is used. Namely whenever an identical index $n$ appears in two quantities
that are multiplied with one another, a summation over $n=1,2,\ldots,N$ is implicit. For instance
\begin{equation}
\delta q^n\frac{\partial L}{\partial q^n}\equiv
\sum_{n=1}^N\delta q^n\frac{\partial L}{\partial q^n}.
\end{equation}
This notation is very common and most widely used in the physics literature.
The same practice is thus followed
throughout these notes, unless otherwise specified.

\noindent
Since according to the variational principle the condition $\delta S[q^n]=0$ (up to the surface term
contributions) is to be met whatever the variations $\delta q^n(t)$, a vanishing ``volume" contribution
is guaranteed only provided the following equations are obeyed for each of the degrees of freedom $q^n$,
\begin{equation}
\frac{d}{dt}\frac{\partial L(q^n(t),\dot{q}^n(t))}{\partial\dot{q}^n}\,-\,
\frac{\partial L(q^n(t),\dot{q}^n(t))}{\partial q^n}=0,\qquad \mbox{for all}\ n=1,2,\ldots,N.
\label{J1.eq:EL}
\end{equation}
Even though discussed explicitly hereafter, it should already be clear that these equations are
in general second order differential equations in time for the functions $q^n(t)$, of which the solutions
thus determine the possible classical trajectories in configuration space, depending on a specific choice
of boundary conditions. These equations are the {\bf Euler--Lagrange equations of motion} of the system.

\noindent
The specification of boundary conditions may or may not be done according to whether one wishes
also the ``surface" term contribution in (\ref{J1.eq:deltaS}) to vanish identically or not. For instance,
keeping the end values $q^n_i=q^n(t_i)$ and $q^n_f=q^n(t_f)$ fixed for the time interval $[t_i,t_f]$
amounts to considering arbitrary variations $\delta q^n(t)$ which are necessarily such that
\begin{equation}
\delta q^n(t_i)=0,\quad \delta q^n(t_f)=0.
\end{equation}
In other words, requiring the variational principle in a ``strong" sense, meaning that $\delta S[q^n]$
ought to vanish including the ``surface" contribution, imposes as boundary conditions in order to solve the
Euler--Lagrange equations of motion the following integration constants
\begin{equation}
q^n(t_i)=q^n_i,\quad q^n(t_f)=q^n_f.
\end{equation}
However quite often such a choice is not convenient, or does not correspond to the actual physical
situation being considered. This is the case for instance when both the initial configuration and
generalised velocity of the system are specified, $q^n(t_i)=q^n_i$ and $\dot{q}^n(t_i)=\dot{q}^n_i$,
and under such a circumstance the ``surface" contribution to $\delta S[q^n]$ generally does not
vanish. Consequently, it is usually preferable to impose the variational principle in a ``weak" sense,
meaning that only the ``volume" contribution to $\delta S[q^n]$ is required to vanish, as was discussed
above. Imposing the variational principle in a strong sense by also requiring the ``surface" term
to vanish is very often too restrictive.

\vspace{10pt}

\noindent
\underline{\bf Comments}

\vspace{10pt}

\noindent
1. It may readily be established that two Lagrange functions for a same configuration space $q^n$
which differ by a total time derivative of an arbitrary function $F(q^n)$ of the configuration space
coordinates in fact lead to identical Euler--Lagrange equations of motion, hence describe identical
dynamics in that configuration space,
\begin{equation}
L'(q^n,\dot{q}^n)=L(q^n,\dot{q}^n)+\frac{dF(q^n)}{dt}=
L(q^n,\dot{q}^n)+\dot{q}^n\frac{\partial F(q^n)}{\partial q^n}.
\end{equation}
This result follows by considering the difference of the Euler--Lagrange equations
associated to the two Lagrange functions, and establishing that this difference vanishes identically
irrespective of the choice for $F(q^n)$ (in the context of the exterior differential calculus
on the configuration space manifold, the calculation is equivalent to showing that the squared
exterior derivative vanishes identically). But a more immediate proof notices that the
additional term $dF(q^n)/dt$ simply induces an additional surface term contribution to the action $S[q^n]$.
Since the Euler--Lagrange equations follow from a volume contribution only, such a surface term contribution
simply cannot affect these equations of motion. This result also shows that when the dynamics of a
system follows from the variational principle, the associated action is at best defined up to such
total derivative contributions.

The fact that the Euler--Lagrange equations of motion are left invariant under such a change in
action implies only that the {\bf classical} dynamics is independent of such redefinitions of the
action. However, this is not necessarily the case at the quantum level. Indeed, when configuration
space possesses nontrivial topology, in particular non contractible cycles (in other words, when
the first homotopy group or fundamental group of the configuration space is nontrivial), this
arbitrariness in the choice of Lagrange function carries some physical and in principle observable
consequences, often leading to extra parameters of a purely quantum character of which the value must be quantised.

The issue of the inverse variational problem, namely the determination of a Lagrange function
given a set of equations of motion, is also an interesting one, but is not discussed here.
Let us only say that generically, the choice is unique modulo the arbitrariness discussed above,
even though there exist large classes of exceptions for which even an infinite number of different actions all
lead to identical equations of motion.

\vspace{10pt}

\noindent
2. Let us now make more explicit the expression for the Euler--Lagrange equations of motion (\ref{J1.eq:EL}),
\begin{equation}
\frac{\partial^2 L}{\partial\dot{q}^{n_1}\partial\dot{q}^{n_2}}\,\ddot{q}^{n_2}\,+\,
\frac{\partial^2 L}{\partial\dot{q}^{n_1}\partial q^{n_2}}\,\dot{q}^{n_2}\,-\,
\frac{\partial L}{\partial q^{n_1}}=0.
\end{equation}
In this collection of equations, one for each $n=1,2,\ldots,N$, the terms multiplying the
generalised accelerations $\ddot{q}^{n_2}$ define, at each point in the velocity phase space
$(q^n,\dot{q}^n)$, a $N\times N$ square matrix known as the Hessian of the Lagrange function,
\begin{equation}
H_{n_1n_2}(q^n,\dot{q}^n)=
\frac{\partial^2 L(q^n,\dot{q}^n)}{\partial\dot{q}^{n_1}\partial\dot{q}^{n_2}}.
\end{equation}
Thus depending on whether this matrix is regular or not, these equations may or may not be used
to express all accelerations in terms of the generalised positions and velocities, in the form
\begin{equation}
\ddot{q}^n(t)=g^n(q^n(t),\dot{q}^n(t)),
\end{equation}
for some functions $g^n(q^n,\dot{q}^n)$. Clearly in such a situation, one recovers equations
of motion of the Newton type, $m\ddot{\vec{r}}(t)=\vec{F}(\vec{r}(t),\dot{\vec{r}}(t))$,
which are second order in time derivatives for each of the degrees of freedom. Thus a system for which
the Lagrange function has a regular Hessian,
\begin{equation}
{\rm det}\,\frac{\partial^2 L}{\partial\dot{q}^{n_1}\partial\dot{q}^{n_2}}\ne 0,
\end{equation}
is said to be a {\bf regular} system (strictly speaking, it is the Lagrangian used to
described the dynamics of that system which is regular).

In contradistinction, a {\bf singular} system is one for which the Hessian is singular,
\begin{equation}
{\rm det}\,\frac{\partial^2 L}{\partial\dot{q}^{n_1}\partial\dot{q}^{n_2}}=0.
\end{equation}
Consequently, the Hessian then possesses at least one eigenvector of vanishing eigenvalue.
Denoting such zero eigenvectors by $V^n_\alpha(q^n,\dot{q}^n)$ with the label $\alpha$
distinguishing all such independent zero eigenvectors, and projecting the Euler--Lagrange equations
onto any one of these eigenvectors, one obtains a series of constraints for the coordinates $q^n$
and their velocities $\dot{q}^n$,
\begin{equation}
V^{n_1}_\alpha(q^n,\dot{q}^n)
\left[\frac{\partial^2 L(q^n,\dot{q}^n)}{\partial\dot{q}^{n_1}\partial q^{n_2}}\dot{q}^{n_2}
-\frac{\partial L(q^n,\dot{q}^n)}{\partial q^{n_1}}\right]=0.
\end{equation}
Hence singular systems are constrained systems. An important class of constrained systems
is that of gauge invariant systems, of which General Relativity, Yang-Mills theories or string
theories are famous and most relevant and interesting examples.

\vspace{10pt}

\noindent
3. At the classical level, the absolute (numerical) and physical (physical dimension) normalisation
of the action or Lagrange function are totally irrelevant. Rescaling these quantities by a dimensional
or dimensionless factor does not affect the Euler--Lagrange equations of motion which are homogeneous
(of weight one) in the Lagrange function. However when it comes to quantum mechanics, this is no
longer irrelevant and as a matter of fact it is essential that the physical dimension of the action
be that of Planck's (reduced) constant $\hbar=h/2\pi$. Likewise, rescaling the absolute normalisation
of the action by a dimensionless factor also has physical consequences or significance. Furthermore, in order
that the quantum dynamics be unitary and thus preserves quantum probabilities, it is a sufficient condition
that the action be a real quantity under complex conjugation, even in the presence of complex valued degrees of
freedom.

\subsection{Illustrative examples}

\subsubsection{Newton's mechanics of conservative systems}

Consider a system of $N$ nonrelativistic massive particles of masses $m_\alpha$, $\alpha=1,2,\ldots,N$,
and of position vectors $\vec{r}_\alpha(t)$ with respect to some choice of inertial frame (these
position vectors may be decomposed in terms of their cartesian coordinates, since space is taken
to be Euclidean in Newton's mechanics). These particles are subjected to a collection of forces which
are all conservative (which means that those forces that may not necessarily be conservative
develop anyway an identically vanishing power or work, hence do not contribute to the energy
balance of the system; a typical example is that of a force perpendicular at all times to
the velocity $\dot{\vec{r}}(t)$). Consequently, these forces may be characterised by their total
potential energy $V(\vec{r}_\alpha)$.

Furthermore, each of the particles possesses a kinetic energy associated to its velocity, leading
to the total kinetic energy of the system,
\begin{equation}
T(\dot{\vec{r}}_\alpha)=\sum_{\alpha=1}^N\frac{1}{2}m_\alpha\,\dot{\vec{r}}_\alpha\,^2.
\end{equation}

In order to reproduce Newton's equations of motion for this system as Euler--Lagrange equations
of motion for some choice of Lagrange function, let us consider the following combination of $T$ and $V$,
\begin{equation}
L(\vec{r}_\alpha,\dot{\vec{r}}_\alpha)=T(\dot{\vec{r}}_\alpha)\,-\,V(\vec{r}_\alpha)=
\sum_{\alpha=1}^N\frac{1}{2}m_\alpha\dot{\vec{r}}_\alpha\,^2\,-\,V(\vec{r}_\alpha).
\end{equation}
To establish the Euler--Lagrange equations of motion for this choice of Lagrange function,
one needs to consider the partial derivatives of $L$ separately with respect to each of the cartesian
coordinates of either $\vec{r}_\alpha$ or $\dot{\vec{r}}_\alpha$. For a given particle, namely value
of $\alpha$, these quantities combine into a vector quantity again, of which the cartesian components
are these partial derivatives. Hence the compact notation used hereafter for such partial derivatives,
in which a partial derivative with respect to a vector stands for the vector of which the components
are the successive partial derivatives with respect to the components of the vector with respect to which
the vector partial derivative is taken. Therefore, one obtains
\begin{equation}
\frac{\partial L}{\partial\vec{r}_\alpha}=-\frac{\partial V}{\partial\vec{r}_\alpha},\quad
\frac{\partial L}{\partial\dot{\vec{r}}_\alpha}=m_\alpha\dot{\vec{r}}_\alpha,
\end{equation}
leading to the Euler--Lagrange equations of motion
\begin{equation}
m_\alpha\ddot{\vec{r}}_\alpha(t)=-\frac{\partial V(\vec{r}_\alpha(t))}{\partial\vec{r}_\alpha}=
\vec{F}_\alpha(\vec{r}_\alpha(t)).
\end{equation}
These are indeed precisely Newton's equations of motion for the system.

Note that by taking the scalar product of the equation of motion for the particle $\alpha$
with its velocity $\dot{\vec{r}}_\alpha$, and then summing over all particles, any solution $\vec{r}_\alpha(t)$
to the equations of motion obeys the following identity,
\begin{equation}
\sum_{\alpha=1}^N\dot{\vec{r}}_\alpha\cdot\left(m_\alpha\ddot{\vec{r}}_\alpha\right)=
\sum_{\alpha=1}^N\dot{\vec{r}}_\alpha\cdot\vec{F}_\alpha=
-\sum_{\alpha=1}^N\dot{\vec{r}}_\alpha\cdot\frac{\partial V}{\partial\vec{r}_\alpha}.
\end{equation}
However since on both sides of this identity one recognises total time derivatives, it may also
be expressed as,
\begin{equation}
\frac{d}{dt}\left[\sum_{\alpha=1}^N\frac{1}{2}m_\alpha\dot{\vec{r}}_\alpha\,^2\,+\,V(\vec{r}_\alpha)\right]=0,
\end{equation}
namely
\begin{equation}
\frac{d}{dt}\left[T+V\right]=0.
\end{equation}
In other words, the quantity
\begin{equation}
E=T+V,
\end{equation}
which defines the total mechanical energy of the system is a constant of motion. This means that the value
it takes for a given solution to the equations of motion is time independent, a constant in time, even though
the specific value that is obtained varies from one solution to another since it depends for example on the initial
values for both the positions, $\vec{r}_\alpha$, and velocities, $\dot{\vec{r}}_\alpha$. Later on, we shall understand
on the basis of Noether's (first) theorem that the existence of a such a conserved energy is consequence of
a symmetry of the system, namely its invariance under arbitrary constant translations in time.

\subsubsection{The free nonrelativistic particle}

According to the previous general discussion, the Lagrange function for a single nonrelativistic
massive particle free of the action of any forces is simply
\begin{equation}
L=\frac{1}{2}m\dot{\vec{r}}\,^2.
\end{equation}
It thus follows that the equation of motion is
\begin{equation}
m\ddot{\vec{r}}(t)=\vec{0}.
\end{equation}
Specifying as boundary conditions for these second order differential equations
the initial values
\begin{equation}
\vec{r}_0=\vec{r}(t_0),\qquad
\vec{v}_0=\dot{\vec{r}}(t_0),
\end{equation}
the solution reads
\begin{equation}
\vec{r}(t)=\vec{r}_0+\vec{v}_0\,(t-t_0),\qquad
\dot{\vec{r}}(t)=\vec{v}_0,
\end{equation}
representing a straight line trajectory at a constant velocity.
Hence not only is the total energy of the particle a constant of motion,
\begin{equation}
E=T=\frac{1}{2}m\dot{\vec{r}}\,^2=\frac{1}{2}m\vec{v}_0\,^2,
\end{equation}
but so is its linear or velocity {\bf momentum},
\begin{equation}
\vec{p}(t)=m\dot{\vec{r}}(t)=m\vec{v}_0.
\end{equation}
Furthermore its {\bf angular-momentum},
\begin{equation}
\vec{L}(t)=\vec{r}(t)\times\vec{p}(t)=m\vec{r}(t)\times\dot{\vec{r}}(t)=
m\vec{r}_0\times\vec{v}_0,
\end{equation}
is then also a constant of motion. In the same manner as for the energy, we shall understand
later on how conservation of momentum is related to invariance of the system under constant translations
in space, and conservation of angular-momentum to invariance under constant rotations in space.

\subsubsection{The nonrelativistic one dimensional harmonic oscillator}

Consider a spring of spring constant $k$ allowed to be deformed along a single cartesian
direction, with a mass $m$ attached at one of its ends, the other being kept fixed.
Denoting the elongation of the spring by $x$ ($x=0$ being the value for $x$ when the spring in its
natural undeformed state, while $x<0$ corresponds to a contracted state of the spring), the force developed by the spring is
\begin{equation}
F(x)=-kx,\qquad F(x)=-kx=-\frac{dV(x)}{dx},
\end{equation}
with potential energy
\begin{equation}
V(x)=\frac{1}{2}kx^2.
\end{equation}
Consequently the Lagrange function for this system is simply
\begin{equation}
L(x,\dot{x})=\frac{1}{2}m\dot{x}^2-\frac{1}{2}kx^2,
\end{equation}
with as equation of motion
\begin{equation}
m\ddot{x}(t)=-kx(t).
\end{equation}
Defining the quantity
\begin{equation}
\omega=\sqrt{\frac{k}{m}}>0,
\end{equation}
one thus obtains the simple linear harmonic equation
\begin{equation}
\ddot{x}(t)\,+\,\omega^2\,x(t)=0.
\end{equation}

Being a linear differential equation of order two, its general solution is the superposition
of any two linearly independent elements in the set of its solutions. Taking for the latter $\cos\omega(t-t_0)$ and
$\sin\omega(t-t_0)$, where $t_0$ is some time at which the following initial values are
specified as boundary conditions,
\begin{equation}
x(t_0)=x_0,\quad
\dot{x}(t_0)=v_0,
\end{equation}
it readily follows that the solution is given as
\begin{equation}
x(t)=x_0\cos\omega(t-t_0)+\frac{v_0}{\omega}\sin\omega(t-t_0),\quad
\dot{x}(t)=v_0\cos\omega(t-t_0)-\omega x_0\sin\omega(t-t_0),
\end{equation}
or equivalently
\begin{equation}
x(t)=C\cos\left[\omega(t-t_0)-\varphi_0\right],\quad
\dot{x}(t)=-C\omega\sin\left[\omega(t-t_0)-\varphi_0\right],
\end{equation}
with
\begin{equation}
C=\sqrt{x^2_0+\frac{v^2_0}{\omega^2}},\qquad
\cos\varphi_0=\frac{x_0}{C},\qquad
\sin\varphi_0=\frac{v_0/\omega}{C},\qquad
\tan\varphi_0=\frac{v_0/\omega}{x_0}.
\end{equation}
The latter form makes explicit that the solution is periodic with period $T=2\pi/\omega=2\pi\sqrt{m/k}$,
and purely harmonic since only the mode with the frequency $\nu=\omega/2\pi$ is involved, $\omega$ thus
being its angular frequency.

Because of the acting force of the spring the momentum of the particle is not conserved. However its energy is
and takes the value
\begin{equation}
E=\frac{1}{2}m\dot{x}^2(t)+\frac{1}{2}kx^2(t)=\frac{1}{2}m v^2_0+\frac{1}{2}kx^2_0.
\end{equation}

Alternative representations of the general solution are of course possible. One which will become of
relevance when quantising the system is obtained by using as generating basis of the solutions the
pure imaginary exponentials, namely
\begin{equation}
x(t)=\frac{1}{\sqrt{2m\omega}}\left[\,
\alpha_0\,e^{-i\omega(t-t_0)}\,+\,\alpha^*_0\,e^{i\omega(t-t_0)}\,\right].
\end{equation}
Here, the normalisation is chosen for later convenience, while $\alpha_0$ stands for a complex valued
integration constant. That the second contribution in this sum involves the complex conjugate coefficient
$\alpha^*_0$ follows from the requirement that the solution $x(t)$ be real under complex conjugation.
Hence the two real boundary conditions necessary to uniquely specify a solution to the equation of
motion are traded for a single complex valued boundary condition. Of course, the value for $\alpha_0$ may
be expressed in terms of a different choice of integration constants, such as for instance the one
made above in terms of $x_0$ and $v_0$. One finds in that case,
\begin{equation}
\alpha_0=\sqrt{\frac{m\omega}{2}}\left[x_0+\frac{i}{\omega}v_0\right],\qquad
\alpha^*_0=\sqrt{\frac{m\omega}{2}}\left[x_0-\frac{i}{\omega}v_0\right].
\end{equation}
More generally, defining
\begin{equation}
\alpha(t)=\alpha_0\,e^{-i\omega(t-t_0)},\qquad
\alpha^*(t)=\alpha^*_0\,e^{i\omega(t-t_0)},
\end{equation}
one has
\begin{equation}
x(t)=\frac{1}{\sqrt{2m\omega}}\left[\,\alpha(t)\,+\,\alpha^*(t)\,\right],\qquad
p(t)=m\dot{x}(t)=-\frac{im\omega}{\sqrt{2m\omega}}
\left[\,\alpha(t)\,-\,\alpha^*(t)\,\right],
\end{equation}
and conversely
\begin{equation}
\alpha(t)=\sqrt{\frac{m\omega}{2}}\left[x(t)+\frac{i}{m\omega}p(t)\,\right],\qquad
\alpha^*(t)=\sqrt{\frac{m\omega}{2}}\left[x(t)-\frac{i}{m\omega}p(t)\,\right].
\end{equation}
These expressions will become of relevance when considering the Hamiltonian formulation
of this system, and subsequently its canonical quantisation.

\subsubsection{The simple pendulum}

As an example of a coordinate which is not cartesian, let us turn now to the simple
pendulum. Namely a string of inextensible length $\ell$ fixed at one end and with a point mass $m$
attached at the other, free to oscillate in a fixed vertical plane and being kept straight because
of its inner tension. The motion of the mass $m$ is thus
circular and of radius $\ell$, requiring a single degree of freedom to specify its configuration at
any instant in time, for which we shall take the angular position $\theta(t)$ of the mass $m$ measured
with respect to the downward vertical direction. Ignoring any possible friction, the mass $m$ is subjected to
only two forces. One of these is the force of gravity or weight, $m\vec{g}$, of the mass $m$,
a conservative force of potential energy $V(\theta)=mg\ell(1-\cos\theta)$. The other is the tension $T(t)$
in the string, but being perpendicular at all times to the velocity of the particle, it does not develop
any work nor power and is thus irrelevant as far as the balance of energy of the system is concerned.

Knowing, on basis of the kinematics of the system, that the norm of the velocity of the mass is
$\ell|\dot{\theta}(t)|$, it follows from our previous discussion that the Lagrange function
for this system is simply
\begin{equation}
L(\theta,\dot{\theta})=\frac{1}{2}m\ell^2\dot{\theta}^2-mg\ell(1-\cos\theta).
\end{equation}
Consequently
\begin{equation}
\frac{\partial L}{\partial\theta}=-mg\ell\sin\theta,\qquad
\frac{\partial L}{\partial\dot{\theta}}=m\ell^2\dot{\theta},
\end{equation}
so that the Euler--Lagrange equation of motion of the pendulum is
\begin{equation}
\ddot{\theta}(t)\,+\,\frac{g}{\ell}\,\sin\theta(t)=0,
\end{equation}
a nonlinear differential equation of order two at the basis of the construction of Jacobi's
elliptic functions. This is indeed also the equation which follows from Newton's equation
of motion for this system. From the latter equation, one may also establish the value
for the tension $T(t)$ in the string,
\begin{equation}
T(t)=mg\left[\frac{\ell}{g}\dot{\theta}^2(t)+\cos\theta(t)\right].
\end{equation}

In the limit of small oscillations, such that $\theta\ll 1$ radian, by linearisation with
$\sin\theta\simeq\theta+\cdots$ and $\cos\theta\simeq 1-\theta^2/2+\cdots$, the Lagrange
function and equation of motion become
\begin{equation}
L\simeq\frac{1}{2}m\left(\ell\dot{\theta}\right)^2-\frac{1}{2}m\left(\frac{g}{\ell}\right)
\left(\ell\theta\right)^2,
\end{equation}
\begin{equation}
\ddot{\theta}+\frac{g}{\ell}\theta=0.
\end{equation}
These expressions are recognised to be equivalent to those for a harmonic oscillator of degree of freedom
$(\ell\theta)$ and angular frequency
\begin{equation}
\omega=\sqrt{\frac{g}{\ell}}.
\end{equation}
Hence the period of a simple pendulum in the limit of small oscillations is
\begin{equation}
T=2\pi\sqrt{\frac{\ell}{g}}.
\end{equation}
It is by observing and then measuring such a period of a chandelier in a church in Pisa where
he attended Mass, that Galilei Galileo started thinking about mechanics... The rest is history.

\subsubsection{The charged particle in a background electromagnetic field}

Let us consider a nonrelativistic particle of mass $m$ and position vector $\vec{r}(t)$ (with
respect to some inertial frame), subjected to conservative forces of which the total potential
energy is denoted $V(\vec{r}\,)$. In addition, the particle possesses a charge $q$, and is
subjected to a background electromagnetic field, of electric field $\vec{E}(t,\vec{r}\,)$
and magnetic field $\vec{B}(t,\vec{r}\,)$. Associated to these fields one has 
the scalar and vector electromagnetic potentials $\Phi(t,\vec{r}\,)$
and $\vec{A}(t,\vec{r}\,)$, respectively. One of the purposes of the present illustration
is to recover the relation existing between these potentials and the
electric and magnetic fields. This will be done by deriving the equations of
motion given a Lagrange function, and identifying these equations with the Lorentz force
developed by the electric and magnetic fields.

In that spirit, let us consider the following Lagrange function,
\begin{equation}
L(\vec{r},\dot{\vec{r}};t)=\frac{1}{2}m\dot{\vec{r}}\,^2-q\Phi(t,\vec{r}\,)
+q\dot{\vec{r}}\cdot\vec{A}(t,\vec{r}\,)-V(\vec{r}\,).
\label{J1.eq:LEB}
\end{equation}
Except perhaps for the term involving the vector potential $\vec{A}(t,\vec{r}\,)$,
this Lagrange function is recognised once again to be of the form $T-V$, with $T$ the
nonrelativistic kinetic energy of the particle, and $V$ standing for the total potential energy
comprised, in the present case, of the potential energy $V(\vec{r}\,)$ as well as the term
$q\Phi(t,\vec{r}\,)$. Indeed, in the static case, it is well known that the scalar electromagnetic
potential $\Phi(\vec{r}\,)$ is related to the electric field by $\vec{E}(\vec{r}\,)=-\vec{\nabla}\Phi(\vec{r}\,)$,
while the potential energy of a charge $q$ in such a field is $q\Phi(\vec{r}\,)$. Even for
a time dependent electromagnetic scalar potential, we have kept this contribution in the Lagrange function.
Note also that we have here an example of a Lagrange function which carries an explicit time dependence
when the background fields $\Phi(t,\vec{r}\,)$ and $\vec{A}(t,\vec{r}\,)$ 
vary in time, which is certainly the case when the electromagnetic fields $\vec{E}(t,\vec{r}\,)$
and $\vec{B}(t,\vec{r}\,)$ vary in time. An example is that of a passing electromagnetic wave
acting on the charged particle.

Since we are to take partial variations with respect to the components of the position and
velocity vectors, $\vec{r}$ and $\dot{\vec{r}}$, let us now make these contributions explicit
in the expression of the Lagrange function, and denote the associated cartesian components
as $x^i$ and $\dot{x}^i$, with $i=1,2,3$, while the index $i$ may be freely raised or lowered
(since the metric is Euclidean, given by $\delta_{ij}$). Then
\begin{equation}
L(x^i,\dot{x}^i;t)=\frac{1}{2}m(\dot{x}^i)^2-q\Phi(t,x^i)+q\dot{x}^i\,A_i(t,x^i)-V(x^i).
\end{equation}
Once again, here it is understood that the implicit summations over repeated indices in a
product are to be summed over all their values, $i=1,2,3$ (hence this also applies to the
term in $(\dot{x}^i)^2$).

One then readily has
\begin{equation}
\frac{\partial L(x^i,\dot{x}^i;t)}{\partial x^i}=-q\frac{\partial\Phi(t,x^i)}{\partial x^i}
+q\dot{x}^j\frac{\partial A_j(t,x^i)}{\partial x^i}-\frac{\partial V(x^i)}{\partial x^i},
\end{equation}
\begin{equation}
\frac{\partial L(x^i,\dot{x}^i;t)}{\partial\dot{x}^i}=m\dot{x}_i+qA_i(t,x^i),
\end{equation}
hence
\begin{equation}
\frac{d}{dt}\frac{\partial L(x^i,\dot{x}^i;t)}{\partial\dot{x}^i}=m\ddot{x}_i
+q\dot{x}^j\frac{\partial A_i(t,x^i)}{\partial x^j}+q\frac{\partial A_i(t,x^i)}{\partial t}.
\end{equation}
Consequently, the Euler--Lagrange equations of motion become,
\begin{equation}
m\ddot{x}_i(t)=-q\frac{\partial\Phi(t,x^i(t))}{\partial x^i}-q\frac{\partial A_i(t,x^i(t))}{\partial t}
+q\dot{x}^j(t)\left[\frac{\partial A_j(t,x^i(t))}{\partial x^i}\,-\,
\frac{\partial A_i(t,x^i(t))}{\partial x^j}\right]\,-\,\frac{\partial V(x^i(t))}{\partial x^i}.
\end{equation}
The term in the gradient of the potential energy $V(x^i)$ is of course identified with the (sum of
the) mechanical force(s) to which the particle is subjected,
\begin{equation}
F_i(x^i)=-\frac{\partial V(x^i)}{\partial x^i},\qquad
\vec{F}(\vec{r}\,)=-\vec{\nabla}V(\vec{r}\,).
\end{equation}
The remaining contributions in the r.h.s. of the above equations of motion should thus be identifiable with
the Lorentz force,
\begin{equation}
\vec{F}_{\rm Lorentz}(\vec{r},\dot{\vec{r}};t)=q\vec{E}(t,\vec{r}\,)\,+\,
q\dot{\vec{r}}\times\vec{B}(t,\vec{r}\,).
\end{equation}
Consequently, the electric field is to be defined according to
\begin{equation}
E_i(t,x^i)=-\frac{\partial\Phi(t,x^i)}{\partial x^i}-\frac{\partial A_i(t,x^i)}{\partial t},\quad
\vec{E}(t,\vec{r}\,)=-\vec{\nabla}\Phi(t,\vec{r}\,)-\frac{\partial\vec{A}(t,\vec{r}\,)}{\partial t}.
\end{equation}
For what concerns the magnetic field, let us set
\begin{equation}
\vec{B}(t,\vec{r}\,)=\vec{\nabla}\times\vec{A}(t,\vec{r}\,),\quad
B_i(t,x^i)=\epsilon^{ijk}\frac{\partial A_k(t,x^i)}{\partial x^j},
\end{equation}
$\epsilon^{ijk}$ being the totally antisymmetry invariant tensor in three dimensional Euclidean
space, with the value $\epsilon^{123}=+1$. Conversely, one then has
\begin{equation}
\frac{\partial A_j(t,x^i)}{\partial x^i}\,-\,
\frac{\partial A_i(t,x^i)}{\partial x^j}=\epsilon^{ijk}\,B_k.
\end{equation}
Consequently, the contributions in $\dot{x}^i$ to the equations of motion are identified as
\begin{equation}
\dot{x}^j\left[\frac{\partial A_j(t,x^i)}{\partial x^i}-\frac{\partial A_i(t,x^i)}{\partial x^j}\right]=
\dot{x}^j\epsilon^{ijk}B_k(t,x^i)=\left(\dot{\vec{r}}\times\vec{B}(t,\vec{r}\,)\right)^i.
\end{equation}

In conclusion, combining all the contributions and writing the equations in vector form again, we have finally
obtained for the vector equation of motion of the particle
\begin{equation}
m\ddot{\vec{r}}(t)=q\vec{E}(t,\vec{r}(t))+q\dot{\vec{r}}(t)\times\vec{B}(t,\vec{r}(t))+\vec{F}(\vec{r}(t)),
\end{equation}
where the electric and magnetic fields making up the electromagnetic field are related to the scalar and
vector electromagnetic potentials by,
\begin{equation}
\vec{E}(t,\vec{r}\,)=-\vec{\nabla}\Phi(t,\vec{r}\,)\,-\,\frac{\partial\vec{A}(t,\vec{r}\,)}{\partial t},\qquad
\vec{B}(t,\vec{r}\,)=\vec{\nabla}\times\vec{A}(t,\vec{r}\,).
\label{J1.eq:EBPhiA}
\end{equation}
This analysis has thus established that the Lagrange function (\ref{J1.eq:LEB}) indeed describes the
dynamics of a nonrelativistic charged massive particle subjected to a background electromagnetic field,
as well as some mechanical conservative forces. But additionally, we have identified the relations
between the electromagnetic field, namely its electric and magnetic components $\vec{E}$ and $\vec{B}$,
with the associated scalar and vector components, $\Phi$ and $\vec{A}$, of the electromagnetic potential.
Note that the Lagrange function is given in terms of the latter and not the electric and magnetic fields.
The reasons for this feature are far reaching and physically most significant, but are not discussed here.

Let us also take the opportunity to discuss here different aspects related to these electromagnetic potential
components. Given the fundamental identities of vector analysis,
\begin{equation}
\vec{\nabla}\cdot(\vec{\nabla}\times\vec{V}(\vec{r}\,))=0,\qquad
\vec{\nabla}\times\vec{\nabla}S(\vec{r}\,)=\vec{0},
\label{J1.eq:vectoriden}
\end{equation}
valid for any vector, $\vec{V}(\vec{r}\,)$, and scalar, $S(\vec{r}\,)$, fields, it follows that given the relations
in (\ref{J1.eq:EBPhiA}) one has for the electric and magnetic fields associated to the scalar and vector potentials
$\Phi$ and $\vec{A}$
\begin{equation}
\vec{\nabla}\times\vec{E}+\frac{\partial\vec{B}}{\partial t}=\vec{0},\qquad
\vec{\nabla}\cdot\vec{B}=0,
\label{J1.eq:MaxHom}
\end{equation}
precisely the two homogeneous Maxwell equations of electromagnetism. In other words, the general
solution to the homogeneous Maxwell equations is given in the form (\ref{J1.eq:EBPhiA}) in terms of
the scalar and vector electromagnetic potentials $\Phi$ and $\vec{A}$ (nevertheless, this still
leaves to solve the two inhomogeneous Maxwell equations in which the source terms contribute, namely
the scalar charge and vector current densities). This then raises the issue of the uniqueness of
these electromagnetic potentials.

Once again because of the second identity in (\ref{J1.eq:vectoriden}), in fact given any two fields $\vec{E}$ and $\vec{B}$
obeying the homogeneous Maxwell equations (\ref{J1.eq:MaxHom}) there exist an infinity of electromagnetic
potentials reproducing these fields through the relations (\ref{J1.eq:EBPhiA}). Indeed, it is readily
checked that the following redefinition of the electromagnetic potentials
\begin{equation}
\Phi'(t,\vec{r}\,)=\Phi(t,\vec{r}\,)-\frac{\partial\chi(t,\vec{r}\,)}{\partial t},\qquad
\vec{A}'(t,\vec{r}\,)=\vec{A}(t,\vec{r}\,)+\vec{\nabla}\chi(t,\vec{r}\,),
\label{J1.eq:EMgauge}
\end{equation}
where $\chi(t,\vec{r}\,)$ is an arbitrary function of time and space (possibly subjected to boundary
conditions at infinity), leads back to the same electric and magnetic fields as do the potentials
$\Phi$ and $\vec{A}$. This symmetry transformation represents a huge freedom in the choice of
electromagnetic potentials, known as a {\bf local gauge symmetry}. The idea of a local or gauge symmetry
entails the idea that the parameters of the symmetry transformation may be not only constants (as in
ordinary symmetry transformations; think of a rotation in space of fixed rotation angle and direction),
but may be in general arbitrary functions of space and time, thus corresponding to symmetry transformations
which differ, though in a continuous fashion, from one point to the next in space or in time. This is
certainly an extreme realisation of the concept of symmetry. And in fact this concept of local or gauged
symmetries ({\it i.e.\/}, made local in time and space) has proved to be central to all the fundamental
interactions. As it turns out, the electromagnetic interaction is in fact the physics of the
electromagnetic potentials $\Phi$ and $\vec{A}$ viewed as scalar and vector fields (in a relativistic
context, they are indeed the components of a 4-vector, $A^\mu=(\Phi/c,\vec{A})$, $\mu=0,1,2,3$,
of which the time component, $\mu=0$, is the scalar potential, and the space components, $\mu=i=1,2,3$,
the vector potential; here $c$ is the velocity of light in vacuum) rather than that of the electric and magnetic
fields $\vec{E}$ and $\vec{B}$ which are ``only" derived quantities but not the fundamental
quantum fields of the electromagnetic interaction (the photon is the quantum of the electromagnetic
potential fields, and not as such that of the electric or magnetic fields). The dynamics of the
electromagnetic potential fields $\Phi$ and $\vec{A}$ must be formulated
in such a manner that it is invariant under the gauge symmetry transformations (\ref{J1.eq:EMgauge}).
As a passing remark, let us also mention that it is precisely this gauge symmetry which is the reason
why the photon, the quantum of the electromagnetic potential fields, must be exactly massless.

Having identified the gauge symmetry of the scalar and vector potentials, let us also consider how
the Lagrange function (\ref{J1.eq:LEB}) transforms under a gauge transformation (\ref{J1.eq:EMgauge}).
Denoting by $L'$ the Lagrange function in (\ref{J1.eq:LEB}) associated to the transformed potentials
$\Phi'$ and $\vec{A}'$, and by $L$ that associated to the non transformed potentials $\Phi$ and $\vec{A}$,
one simply has,
\begin{equation}
L'-L=-q\left(-\frac{\partial\chi}{\partial t}\right)+q\dot{\vec{r}}\cdot\vec{\nabla}\chi=
\frac{d}{dt}\left(q\chi\right).
\end{equation}
Hence the two Lagrange functions differ only by a total time derivative. We know that in such
a case they share identical equations of motion, and indeed these equations of motion are expressed
solely in terms of the electric and magnetic fields which are gauge invariant in the first place.
The gauge symmetry of the electromagnetic interaction is thus an example of a symmetry transformation
under which the Lagrange function is invariant up to a surface term.

It is also of interest to consider the evolution in time of the energy of the particle for the
present system. Knowing the forces to which it is subjected, we may easily identify those forces
which develop a nonvanishing power or work, by projecting the forces onto, say, the velocity
$\dot{\vec{r}}(t)$ (in the case of an evaluation of the power of the force). Since the
magnetic component of the Lorentz force, $q\dot{\vec{r}}\times\vec{B}$, is always perpendicular
to the velocity, it is clear that the magnetic field, or this magnetic force never develops
any power nor work, and thus does not contribute to the energy balance in the system. Thus
only the (total) potential energy $V(\vec{r}\,)$ of the mechanical force(s) as well as the electromagnetic
potential energy $q\Phi(t,\vec{r}\,)$ associated to the scalar potential should be considered
in combination with the kinetic energy to characterise the total mechanical energy of the particle,
\begin{equation}
E=\frac{1}{2}m\dot{\vec{r}}\,^2+q\Phi(t,\vec{r}\,)+V(\vec{r}\,).
\end{equation}
However, in the case that at least one field among the electric or the magnetic fields is not static, namely
carries a time dependence, and thus so do also the scalar and vector potentials, one ought to expect
that this total mechanical energy of the particle is not conserved (imagine a passing electromagnetic
wave, thus setting into motion, or at least a different motion, the particle with otherwise a
conserved energy). Indeed, a simple calculation of the rate of change of the total mechanical energy finds
\begin{equation}
\frac{dE}{dt}=q\frac{\partial\Phi}{\partial t}-q\dot{\vec{r}}\cdot\frac{\partial\vec{A}}{\partial t}.
\end{equation}
Hence indeed, it is only when both the scalar and vector potentials are static, and thus so are also
the electric and magnetic fields, namely are time independent, that the total mechanical energy of the
particle is a constant of motion.

\section{Hamiltonian Dynamics}
\label{Gov1.Sec3}

As a motivation for the first of the definitions given hereafter and laying the
basis for the Hamiltonian formulation of dynamical systems, let us just say here that
one of the purposes of this formalism is to turn the second order in time differential
Euler--Lagrange equations of motion into first order ones, which is {\it a priori\/}
an advantage when it comes to constructing explicit solutions. However, this comes
with the necessity to consider twice as many functions of time to solve for in comparison
with the original set of generalised coordinate functions $q^n(t)$ (but on the other hand after all, the Lagrange
function already depends on the variables $q^n$ and $\dot{q}^n$), hence the next definition.
Let us also mention here in passing that all mathematical studies of chaotic and dynamical systems
are best considered within the Hamiltonian formalism, since it allows for powerful mathematical
concepts and tools to be brought to bear on difficult issues, leading to the beautiful and still largely
to be explored and understood field of symplectic geometry in differential geometry, which, in conjunction
with concepts and techniques developed within quantum physics and quantum field theories has seen in
recent years important progress and some profound results.

\subsection{Conjugate momenta and phase space}

Given the system's configuration space parametrised by the coordinates $q^n$, and its dynamics
determined from the Lagrange function $L(q^n,\dot{q}^n)$, by definition the {\bf conjugate momenta},
$p_n$, of the system, namely a set of variables each of which is conjugate to one of the degrees of freedom, $q^n$,
are defined by the relations,
\begin{equation}
p_n(q^n,\dot{q}^n)=\frac{\partial L(q^n,\dot{q}^n)}{\partial\dot{q}^n},\qquad n=1,2,\ldots,N.
\end{equation}
Rather than the so-called velocity phase space spanned by the pairs of variables $(q^n,\dot{q}^n)$ $(n=1,2,\ldots,N$),
one then considers the (momentum) {\bf phase space}, the $2N$ dimensional manifold
spanned by the pairs of variables $(q^n,p_n)$ $(n=1,2,\ldots,N)$, namely the configuration space coordinates and
their conjugate momenta. In the general case, phase space is the cotangent bundle of the configuration
space manifold $M_N$. However, there are dynamical systems of great interest, both to mathematics and
to physics, for which this is not the case, for instance systems of which phase space is compact on account
of some nontrivial symmetries.

This definition and concept of conjugate momentum calls for a series of comments.

\vspace{5pt}

\noindent
\underline{\bf Comments}

\vspace{5pt}

\noindent
1. Based on the definition of the conjugate momenta, one readily notices that the
Euler--Lagrange equations of motion may also be written as
\begin{equation}
\dot{p}_n=\frac{\partial L}{\partial q^n}.
\end{equation}

\vspace{5pt}

\noindent
2. {\it A priori\/}, the conjugate momenta are functions of the velocity phase space $(q^n,\dot{q}^n)$,
$p_n(q^n,\dot{q}^n)$. But since one would like to trade the velocity phase space for the momentum phase
space spanned by the local variables $(q^n,p_n)$, one needs to consider the conditions under which it
is possible to express the generalised velocities $\dot{q}^n$ in terms of the phase space variables $(q^n,p_n)$.
Thus locally in velocity phase space this requires to consider the relations between the $p_n$'s and
the $\dot{q}^n$'s, which in the neighbourhood of any point $q^n$ of configuration space requires to consider the following
$N\times N$ square matrix,
\begin{equation}
\frac{\partial p_{n_1}}{\partial\dot{q}^{n_2}}=
\frac{\partial^2 L}{\partial\dot{q}^{n_1}\partial\dot{q}^{n_2}},
\end{equation}
which thus coincides with the Hessian of the Lagrange function of the system. Consequently in the case of
a regular system all such relations may be inverted and all generalised velocities $\dot{q}^n$ expressed
as functions of the phase space coordinate variables $(q^n,p_n)$,
\begin{equation}
p_n(q^n,\dot{q}^n)\,\longleftrightarrow\,\dot{q}^n(q^n,p_n).
\end{equation}
As an illustration of this argument in a most simple case, imagine that the dependence of the conjugate momenta
on the velocities is purely linear, of the form
\begin{equation}
p_n(q^n,\dot{q}^n)=H_{nn'}(q^n)\dot{q}^{n'}.
\end{equation}
In such a case it is clear that these relations are invertible provided the $N\times N$ square matrix
of coefficients $H_{nn'}(q^n)$ is regular. But this matrix coincides precisely with the quantities
$\partial p_n/\partial\dot{q}^{n'}$ considered in the general discussion. In this example, the
relations between the $p_n$'s and the $\dot{q}^n$'s are linear, whereas in the general case they are not.
But the argument still applies since the problem of inversion is to be considered at each point locally
in velocity phase space, and by working in the tangent space to that point precisely similar linear relations
are involved of which the coefficients are the entries of the Hessian matrix of $L(q^n,\dot{q}^n)$.

On the other hand, if the Hessian is singular, it then follows that the relations between the $p_n$'s and
the $\dot{q}^n$'s may not all be inverted. As we have seen previously, such a situation is characteristic of
constrained systems, and indeed this lack of independence of the conjugate momenta $p_n$ then translates
into a series of constraints on phase space of the form $\phi(q^n,p_n)=0$ which must properly be dealt with
when considering the Hamiltonian formulation of such systems, which include gauge invariant dynamics, and
their subsequent quantisation.

\vspace{10pt}

\noindent
3. Even though two Lagrange functions for a same configuration space which differ by a total time
derivative of an arbitrary function of $q^n$ lead to identical equations of motion,
\begin{equation}
L'(q^n,\dot{q}^n)=L(q^n,\dot{q}^n)+\frac{dF(q^n)}{dt}=L(q^n,\dot{q}^n)+\dot{q}^n\frac{\partial F(q^n)}{\partial q^n},
\end{equation}
the conjugate momenta associated to each do not coincide,
\begin{equation}
p'_n=p_n+\frac{\partial F(q^n)}{\partial q^n}.
\end{equation}
Once again, even though this fact does not lead to physical consequences at the classical level,
when configuration space has a nontrivial topology such redefinitions of the action lead to
observable effects at the quantum level.

\subsection{Canonical Hamiltonian and Hamiltonian equations of motion}

As a motivation for the next definition, let us consider the differential of the Lagrange function,
and aim to make the r\^ole of the conjugate momenta explicit in that quantity,
\begin{eqnarray}
dL(q^n,\dot{q}^n)&=&dq^n\frac{\partial L}{\partial q^n}+d\dot{q}^n\frac{\partial L}{\partial\dot{q}^n}\nonumber\\
 &=&dq^n\frac{\partial L}{\partial q^n}+d\left[\dot{q}^n\frac{\partial L}{\partial\dot{q}^n}\right]
-\dot{q}^nd\frac{\partial L}{\partial\dot{q}^n},
\end{eqnarray}
hence
\begin{equation}
d\left[\dot{q}^n\frac{\partial L}{\partial\dot{q}^n}-L\right]=
\dot{q}^nd\frac{\partial L}{\partial\dot{q}^n}-dq^n\frac{\partial L}{\partial q^n}.
\end{equation}
Since in each of these terms except for the very last one a contribution of $p_n=\partial L/\partial\dot{q}^n$
is now made explicit, let us use the Euler--Lagrange equation to complete the last term as follows,
\begin{eqnarray}
d\left[\dot{q}^np_n-L\right]&=&\dot{q}^ndp_n-dq^n\frac{d}{dt}\frac{\partial L}{\partial\dot{q}^n}
+dq^n\left[\frac{d}{dt}\frac{\partial L}{\partial\dot{q}^n}-\frac{\partial L}{\partial q^n}\right]\nonumber\\
 &=&\dot{q}^ndp_n-dq^n\dot{p}_n+dq^n\left[\dot{p}_n-\frac{\partial L}{\partial q^n}\right],
\label{J1.eq:diffL}
\end{eqnarray}
hence the following definition.

The {\bf canonical Hamiltonian} of the system is the quantity defined hereafter over phase space,
constructed through the Legendre transform of the Lagrange function with respect to the
conjugate momenta~$p_n$,
\begin{equation}
H_0(q^n,p_n)=\dot{q}^np_n-L(q^n,\dot{q}^n)\qquad[{\rm Legendre\ transform\ of\ } L],
\label{J1.eq:H0}
\end{equation}
\begin{equation}
dH_0=\dot{q}^ndp_n-dq^n\dot{p}_n\,\left[+dq^n\left(\dot{p}_n-\frac{\partial L}{\partial q^n}\right)\right].
\end{equation}

Given the above discussion the following point should be emphasized. It may appear odd that in the definition
of the canonical Hamiltonian its dependence is explicitly given to be in terms of the phase space coordinates,
$(q^n,p_n)$, rather than the velocity phase space ones, $(q^n,\dot{q}^n)$, since indeed the r.h.s. of the definition
in (\ref{J1.eq:H0}) is a combination of quantities which in the general case are functions of the latter variables
and not the phase space ones. The truth of the matter is that the calculations leading to (\ref{J1.eq:diffL})
show that the quantity defining $H_0$ is indeed a function of phase space, since its differential is
expressible solely in terms of the differentials in $dq^n$ and $dp_n$ only. This means that any dependence
of $H_0$ on $\dot{q}^n$, even in the case of a singular system for which the relations between the
$p_n$'s and the $\dot{q}^n$'s may not all be inverted, is through the dependence of $H_0$ on $p_n$ only and
the latter's dependence on the $\dot{q}^n$'s. Irrespective of whether the system is regular or singular,
the canonical Hamiltonian always reduces to a function defined over phase space. As we shall see hereafter,
the Hamiltonian generates time evolution of the dynamics in phase space. In the case of a singular systems
one has to extend the canonical Hamiltonian in order to induce a time evolution consistent with the
constraints (Dirac's analysis of constraints \cite{Gov1.Gov1,Gov1.GovBook}). In the case of a regular system,
the canonical Hamiltonian suffices. Only the latter case will thus explicitly be discussed hereafter.

\vspace{10pt}

\noindent
\underline{\bf Consequences}

\vspace{10pt}

\noindent
1. When one considers classical trajectories which thus obey the Euler--Lagrange equations
of motion, the differential of the quantity which defines the canonical Hamiltonian reads
\begin{equation}
dH_0=\dot{q}^n\,dp_n\,-\,\dot{p}_n\,dq^n.
\end{equation}
Consequently, by considering separate variations of the canonical Hamiltonian $H_0(q^n,p_n)$
in either one of the $q^n$'s or one of the $p_n$'s,
one identifies from this relation the {\bf Hamiltonian equations of motion} of the system for each
of its degrees of freedom labelled by $n=1,2,\ldots,N$,
\begin{equation}
\dot{q}^n(t)=\frac{\partial H_0(q^n(t),p_n(t))}{\partial p_n},\qquad
\dot{p}_n(t)=-\frac{\partial H_0(q^n(t),p_n(t))}{\partial q^n},\qquad n=1,2,\ldots,N.
\end{equation}
Note that these are indeed first order in time differential equations. For each of the
degrees of freedom $n=1,2,\ldots,N$, the second order Euler--Lagrange differential equations
have been transformed into twice as many first order Hamiltonian differential equations.
These equations have to be supplemented with a choice of boundary conditions. The number
of these boundary conditions thus remains the same in both cases, namely two boundary conditions
per degree of freedom.

\vspace{10pt}

\noindent
2. {\bf The inverse Legendre transformation or Hamiltonian reduction of phase space}.
In the case of regular systems, the definition
of the conjugate momenta, $p_n(q^n,\dot{q}^n)=\partial L(q^n,\dot{q}^n)/\partial\dot{q}^n$,
may be inverted to express the generalised velocities in terms of the phase space
coordinates, $\dot{q}^n(q^n,p_n)$. On the other hand, we now also have, among the
Hamiltonian equations of motion, those for the degrees of freedom $q^n(t)$ given in terms of
equations for $\dot{q}^n(t)$ in which the r.h.s. is a function of the phase space variables again.
Therefore, one may conversely use the first ensemble of Hamiltonian equations of motion
to solve for the conjugate momenta in terms of the variables $(q^n,\dot{q}^n)$.
Doing so, one is bound to recover the dependence $p_n(q^n,\dot{q}^n)$ obtained
from the definition of the conjugate momenta,
\begin{equation}
\dot{q}^n(q^n,p_n)=\frac{\partial H_0(q^n,p_n)}{\partial p_n}\ \longleftrightarrow\
p_n(q^n,\dot{q}^n)=\frac{\partial L(q^n,\dot{q})}{\partial\dot{q}^n}.
\end{equation}
One may then substitute this expression for the conjugate momenta $p_n(q^n,\dot{q}^n)$
back into the second ensemble of Hamiltonian equations of motion, $\dot{p}_n=-\partial H_0/\partial q_n$,
to obtain again the Euler--Lagrange equations of motion of the Lagrangian formulation of the
dynamics based on the original Lagrange function $L(q^n,\dot{q}^n)$. Establishing this fact is
straightforward and is left to the reader.

\vspace{10pt}

\noindent
3. From the Hamiltonian point of view, phase space defines the {\bf space of states} of the system.
Indeed, given initial values for both $q^n(t)$ and $p_n(t)$, the corresponding solution to the
Hamiltonian equations of motion defines in a unique manner a specific trajectory in phase space
along which the system is evolving in time. Any point on that trajectory then describes the state
in which the system is to be found at that time. By extension, phase space is the ensemble of all
possible states in which the system may be found. In the Lagrangian formulation, configuration space
as such is not sufficient to completely characterise the states of the system, since information
either for the velocities or the configurations at different times are also required because of the
second order nature of the equations of motion. This identification of phase space with the space
of states of the system will extend later to the quantum context in terms of a space of quantum states.

\clearpage

\subsection{Phase space dynamics and Poisson brackets}

\subsubsection{Poisson brackets}

Let us now consider an arbitrary observable defined over phase space, $f(q^n,p_n;t)$, which may
even possess some explicit time dependence. Most observables are constructed as composite quantities
out of the ``elementary", ``fundamental" or ``basic" phase space coordinates $(q^n,p_n)$. As an example think
of the total mechanical energy of some ensemble of particles. Clearly, it could be that such an
observable includes an explicit time dependence (the previous discussion of the particle in a time
dependent background field provides an illustration with as observable the energy), which is the reason
why the possibility is allowed in the discussion hereafter.

The rate of change in time in the value of the observable, in other words its equation of motion, is readily
established,
\begin{equation}
\frac{df}{dt} = \frac{\partial f}{\partial t}\,+\,
\frac{\partial f}{\partial q^n}\dot{q}^n\,+\,
\frac{\partial f}{\partial p_n}\dot{p}_n
 = \frac{\partial f}{\partial t}\,+\,
\frac{\partial f}{\partial q^n}\frac{\partial H_0}{\partial p_n}\,-\,
\frac{\partial f}{\partial p_n}\frac{\partial H_0}{\partial q^n}.
\end{equation}
This results thus justifies the following definition.

The {\bf Poisson bracket} of two phase space observables $f$ and $g$ is, by definition, the
quantity,
\begin{equation}
\left\{f,g\right\}=\frac{\partial f}{\partial q^n}\,\frac{\partial g}{\partial p_n}\,-\,
\frac{\partial f}{\partial p_n}\,\frac{\partial g}{\partial q^n}.
\end{equation}\
This definition calls for a series of comments.
\vspace{10pt}

\noindent
\underline{\bf Comments}

\vspace{10pt}

\noindent
1. As always in these notes unless otherwise specified, whenever indices are repeated
in a product, it is implicitly understood they are summed over their whole range of values.
Thus in the above definition, the indices $n$ appearing in the partial derivatives are summed over the range $n=1,2,\ldots,N$.

\vspace{10pt}

\noindent
2. One should also keep in mind that these Poisson brackets are defined purely in terms of the
dependence of the observables on the phase space variables $q^n$ and $p_n$, irrespective of their dynamics or
whatever their time dependence. Thus in fact these Poisson brackets are defined ``at equal time",
meaning that the arguments $q^n$ and $p_n$ of the observables should be considered at an
identical time $t$ for both observables. Even though the notation does not make that explicit,
this point has to be kept in mind. In particular, at the quantum level Poisson brackets are
put into correspondence with commutation relations of operators, and in the same manner, these
commutations relations are then defined ``at equal time".

\vspace{10pt}

\noindent
3. In terms of the definition of Poisson brackets, it is clear that the equation of motion
for any observable $f(q^n,p_n;t)$ may be expressed more compactly as,
\begin{equation}
\frac{df}{dt}=\frac{\partial f}{\partial t}\,+\,
\left\{f,H_0\right\}.
\end{equation}
Besides their obvious notational advantage, Poisson brackets embody essential features of dynamics
and the geometry of phase space for dynamical systems, not to mention their central r\^ole in the
programme of canonical quantisation through the correspondence principle.

As examples of the above general discussion, let us reconsider the Hamiltonian equations of motion
for the ``elementary" phase space degrees of freedom. By a direct evaluation of the Poisson brackets
using their definition, one readily finds
\begin{equation}
\dot{q}^n=\left\{q^n,H_0\right\}=\frac{\partial H_0}{\partial p_n},\qquad
\dot{p}_n=\left\{p_n,H_0\right\}=-\frac{\partial H_0}{\partial q^n}.
\end{equation}
These are indeed the correct expressions. Furthermore, let us consider as observable the
(canonical) Hamiltonian itself,
\begin{equation}
\frac{dH_0}{dt}=\left\{H_0,H_0\right\}=0.
\end{equation}
Hence as mentioned previously, having chosen the Lagrange function not to possess any explicit
time dependence, it follows that the Hamiltonian of the system is always a constant of motion and
conserved. When the evolution parameter $t$ coincides with the physical time, the Hamiltonian
coincides with the energy of the system, which is then conserved. The fact that the Hamiltonian
coincides with the energy in such circumstances will be illustrated hereafter.

\subsubsection{Algebraic properties of Poisson brackets}

A direct evaluation of the Poisson brackets for the ``elementary" phase space coordinates finds
\begin{eqnarray}
\left\{q^{n_1},q^{n_2}\right\}=0,\qquad\qquad & &
\left\{q^{n_1},p_{n_2}\right\}=\delta^{n_1}_{n_2},\nonumber\\
\left\{p_{n_1},q^{n_2}\right\}=-\delta^{n_2}_{n_1},\,\qquad & & 
\left\{p_{n_1},p_{n_2}\right\}=0.
\label{J1.eq:canonicalbrackets}
\end{eqnarray}
These brackets are known as {\bf canonical brackets}, while phase space coordinates
obeying such brackets are known as {\bf canonical coordinates}.

Given phase space observables $f$, $g$ and $h$, and constants $c$, $c_1$ and $c_2$, it may be
shown that Poisson brackets obey the following properties which are purely of an algebraic character,
\begin{itemize}
\item[a)] Antisymmetry: $\left\{f,g\right\}=-\left\{g,f\right\}$.
\item[b)] Neutral element: $\left\{f,c\right\}=0$.
\item[c)] Linearity: $\left\{c_1f+c_2g,h\right\}=c_1\left\{f,h\right\}+c_2\left\{g,h\right\}$.
\item[d)] Leibnitz rule: $\left\{fg,h\right\}=\left\{f,h\right\}g+f\left\{g,h\right\}$.
\item[e)] Jacobi identity:
$\left\{\left\{f,g\right\},h\right\}+\left\{\left\{g,h\right\},f\right\}+
\left\{\left\{h,f\right\},g\right\}=0$.
\end{itemize}
Given these properties and the values (\ref{J1.eq:canonicalbrackets}) of the Poisson brackets for the ``elementary"
phase space coordinates $q^n$ and $p_n$, the evaluation of Poisson brackets of any two composite
observables becomes a purely algebraic problem, with which one quickly becomes familiar through
some solid practice.

\vspace{10pt}

\noindent
\underline{\bf Remarks}

\vspace{10pt}

\noindent
1. Even though this point will not at all be discussed here, let us only mention that the structure
of Poisson brackets with which phase space comes equipped is in fact directly related to the
existence of a symplectic geometry on phase space. This fundamental property enables a purely geometric and
coordinate free approach to dynamical systems, which has provided profound and important
insights into the dynamics of complicated nonlinear systems, and is an essential tool in
the mathematical studies of chaotic dynamical systems.

\vspace{10pt}

\noindent
2. One may abstract from the above specific context the algebraic properties of Poisson brackets.
There exist other mathematical contexts where identical algebraic properties of a ``bracket"
arise. In particular, note that the algebra of commutators of matrices, and more generally
of linear operators on a vector space, share precisely the same properties as those listed
above for Poisson brackets. This remark is at the basis of the correspondence principle
between classical and quantum physics, as laid out by P. A. M. Dirac in 1931 in his famous book
on quantum mechanics, {\sl The Principles of Quantum Mechanics} (Oxford University Press, 1931),
which has known many reprintings. Reading its first few chapters is a must, and a jewel
of clarity very much characteristic of most of Dirac's writings. This very point will also be our starting point
when addressing the canonical quantisation of a system of which the dynamics is defined through the action principle.

\vspace{10pt}

\noindent
3. There exists a famous theorem due to Darboux, which states that whenever one has a phase
space which thus comes equipped with such a bracket structure, namely a symplectic geometry,
one may always find locally at each point of phase space a system of canonical coordinates,
namely a system of coordinates for which the brackets take the values in (\ref{J1.eq:canonicalbrackets}).
In the above discussion starting from the Lagrange function and introducing conjugate momenta,
the pairs of phase space coordinates $(q^n,p_n)$ for each $n=1,2,\ldots,N$ are always canonical.

\subsubsection{The Hamiltonian formulation of dynamical systems}

In conclusion, the above discussion has established that the Hamiltonian formulation
of any dynamical system is characterised by three essential data, which will be 
put in direct correspondence with analogous data when considering the quantum dynamics
of such a system. These data are:
\begin{center}
\begin{tabular}{l l l}
$\cdot$ space of states: & phase space: & $(q^n,p_n)$ \\
 & & \\
$\cdot$ algebraic structure: & Poisson brackets: & Canonical brackets\\
\ \ (symplectic geometry) & & $\left\{q^{n_1},p_{n_2}\right\}=\delta^{n_1}_{n_2}$ \\
 & & \\
$\cdot$ time evolution: & Hamiltonian $H$: & 
$\frac{df}{dt}=\frac{\partial f}{\partial t}+\left\{f,H\right\}$
\end{tabular}
\end{center}

\vspace{10pt}

\noindent
In fact, all this information may finally be ``encoded" into an action principle but this
time defined on phase space rather than configuration space only, in terms of a first-order
action (first-order because it depends only linearly on the first order time derivatives
of the phase space coordinates). Namely, it is straightforward to check that the
Hamiltonian equations of motion for both $\dot{q}^n$ and $\dot{p}_n$ follow from the
variational principle (in a weak sense again) applied to the following first-order phase space
action,
\begin{equation}
S[q^n,p_n]=\int_{t_i}^{t_f}dt\,
\left[\dot{q}^np_n\,-\,H(q^n,p_n)\,+\,\frac{dF(q^n,p_n)}{dt}\right].
\end{equation}
What is remarkable about this action is that all three data listed above play a r\^ole in its expression.
First, there is the space of states through the
coordinates $(q^n,p_n)$ parametrising that space. Next, the Poisson bracket structure
is directly related\footnote{Proof of this statement is not provided here, but may be found
in Ref.[3]. In any case, this fact may be established without too much difficulty.}
to the terms linear in the time derivatives of $q^n$ or $p_n$
(in the above expression, this is the term in $\dot{q}^np_n$). And finally the
generator of time evolution through Poisson brackets, namely the Hamiltonian is explicitly
the opposite of the sum of all those terms in the action which do not involve any time derivatives
of either $q^n$ or $p_n$.

In the above action, the function $F(q^n,p_n)$ is arbitrary, and is introduced
once again because actions differing by total time derivatives possess identical equations of
motion. By adjusting the choice of that total time derivative, alternative and sometimes more
convenient forms of the action may be considered. In the case of nontrivial topology in
configuration (and phase) space, such redefinitions have physical consequences at the quantum
level. Note also that in contradistinction to the Lagrangian formulation, the function $F(q^n,p_n)$
may now be a function of both the $q^n$'s and the $p_n$'s, which allows to specify through the
variational principle in a strong sense larger classes of boundary conditions than is possible
with the Lagrangian action. Finally, let us just mention that the Hamiltonian formulation of
a dynamical system is in a certain sense more ``fundamental" than its Lagrangian formulation,
especially when singular systems are being considered.

As an example of a redefinition by a total time derivative, consider the function
$F(q^n,p_n)=-\frac{1}{2}q^np_n$, in which case the above action becomes
\begin{equation}
S_2[q^n,p_n]=\int_{t_i}^{t_f}dt\left[
\frac{1}{2}\left(\dot{q}^np_n-q^n\dot{p}_n\right)-H(q^n,p_n)\right].
\end{equation}
It is interesting to put this expression in relation to that of the action for a charged
nonrelativistic particle confined to a plane and subjected to a static and homogeneous magnetic field
$\vec{B}$ perpendicular to that plane, namely the system of the ``pure Landau problem",
\begin{equation}
S[x,y]=\int_{t_i}^{t_f}dt\left[\frac{1}{2}m\left(\dot{x}^2+\dot{y}^2\right)
\,-\,\frac{1}{2}qB\left(\dot{x}y-x\dot{y}\right)\right],
\label{J1.eq:Landau}
\end{equation}
where $(x,y)$ denote cartesian coordinates in the plane. In this expression the choice of gauge
for the vector potential is such that $\vec{A}(\vec{r}\,)=\vec{B}\times\vec{r}/2$, in a three dimensional notation,
which is known as the circular or symmetric gauge, since it is covariant under rotations in the plane
perpendicular to the magnetic field. Other choices of gauge, in the form
$\vec{A}(t,\vec{r}\,)=\vec{B}\times\vec{r}/2+\vec{\nabla}\chi(t,\vec{r}\,)$,
$\chi(t,\vec{r}\,)$ being an arbitrary function, are also possible, and simply correspond to redefining the Lagrange
function by a total time derivative term with $F(t,\vec{r}\,)=q\chi(t,\vec{r}\,)$. In the limit of a vanishing mass $m$
(or a ratio $B/m$ growing infinite), this action reduces to
\begin{equation}
\lim_{m\to 0}S_2[x,y]=\int_{t_i}^{t_f}dt\left[-\frac{1}{2}qB\left(\dot{x}y-x\dot{y}\right)\right].
\end{equation}
Compared to the above first-order Hamiltonian action, we see that the system is already in
Hamiltonian form, with a two dimensional phase space which coincides with the original
configuration space of the plane, with canonically conjugate coordinates that may be taken
to be, for instance, $x$ and $(-qB y)$, and with an identically vanishing Hamiltonian
(had a potential energy term $V(x,y)$ been introduced in the action (\ref{J1.eq:Landau}), this
potential energy would now play the r\^ole of the Hamiltonian). Consequently, no classical
dynamics survives in this limit (the particle remains pinned to a position, or in actual fact onto the
equipotentials of the potential energy $V(x,y)$ were one to be present), while
one has the Poisson bracket $\left\{x,y\right\}=-1/qB$. Since when quantising the
system these brackets become commutation relations and coordinates become operators,
this system provides the simplest illustration of a noncommutative geometry in two
dimensions, since the cartesian coordinates $x$ and $y$ then do no longer commute.

\subsection{Illustrative examples}

\subsubsection{The nonrelativistic particle}

Let us consider the nonrelativistic particle described by the Lagrange function
\begin{equation}
L=\frac{1}{2}m\dot{\vec{r}}\,^2-V(\vec{r}\,).
\end{equation}
The momentum conjugate to the degrees of freedom $\vec{r}$ is
\begin{equation}
\vec{p}=\frac{\partial L}{\partial\dot{\vec{r}}}=m\dot{\vec{r}}.
\end{equation}
This quantity thus coincides, in this specific case, with the ordinary linear or velocity
momentum of a particle. Clearly, since the velocities may be inverted in terms of the
momenta, $\dot{\vec{r}}=\vec{p}/m$, this is also a regular system.

From the general discussion, we know at once that the cartesian components of both the
position vector, $\vec{r}$, namely $x^i$ with $i=1,2,3$, and the conjugate momentum
vector, $\vec{p}$, namely $p_i$, are canonically conjugated coordinates of the phase space
$(\vec{r},\vec{p}\,)$ of this system. Hence we have the Poisson brackets (as is customary,
only the nonvanishing Poisson brackets are displayed),
\begin{equation}
\left\{x^i,p_j\right\}=\delta^i_j .
\end{equation}
Furthermore, a direct calculation finds that the canonical Hamiltonian of the system is,
\begin{equation}
H=\dot{\vec{r}}\cdot\vec{p}-L=\frac{1}{2m}\vec{p}\,^2\,+\,V(\vec{r}\,),
\end{equation}
which is seen to coincide with the total mechanical energy of the particle. This is as it should
be since the time evolution parameter $t$ is in this case also the physical time. And of course,
this energy is a constant of motion for this conservative system.

Having identified the necessary three data for the Hamiltonian formulation of the dynamics, the
Hamiltonian equations of motion readily follow, with
\begin{equation}
\dot{\vec{r}}=\left\{\vec{r},H\right\}=\frac{1}{m}\vec{p},\qquad
\dot{\vec{p}}=\left\{\vec{p},H\right\}=-\vec{\nabla}V(\vec{r}\,).
\end{equation}
The first of these vector equations may indeed be solved for $\vec{p}$ in terms of
$\dot{\vec{r}}$, $\vec{p}=m\dot{\vec{r}}$, a representation which when substituted into
the second Hamiltonian equation of motion recovers the Newton equation for this system,
\begin{equation}
m\ddot{\vec{r}}=-\vec{\nabla}V(\vec{r}\,)=\vec{F}(\vec{r}\,).
\end{equation}
Consequently all the other considerations already developed earlier when this type
of system was discussed follow as well. For what concerns the possible conservation of
the linear momentum or the angular-momentum now in presence of the potential energy $V(\vec{r}\,)$,
one finds
\begin{equation}
\dot{\vec{p}}=-\vec{\nabla}V(\vec{r}\,)=\vec{F}(\vec{r}\,),
\end{equation}
\begin{equation}
\dot{\vec{L}}=\frac{d}{dt}\left[\vec{r}\times\vec{p}\,\right]=
-\vec{r}\times\vec{\nabla}V(\vec{r}\,)=\vec{r}\times\vec{F}(\vec{r}\,).
\end{equation}
Consequently, the linear momentum is conserved only if the particle is free, $\vec{F}=\vec{0}$,
while the angular-momentum may be conserved even in the presence of a nonvanishing force provided only
it is radial, namely always colinear with the position vector $\vec{r}$. In either of these two
situations leading to a conservation law, it is a symmetry law of space which is again at work.

\subsubsection{The one dimensional harmonic oscillator}

Returning to the one dimensional harmonic oscillator, this is but an example of the previous
general discussion particularised to a one degree of freedom system, namely the coordinate $x(t)\in\mathbb{R}$, with
\begin{equation}
L=\frac{1}{2}m\dot{x}^2-\frac{1}{2}m\omega^2 x^2.
\end{equation}
Consequently, it readily follows that
\begin{equation}
p=m\dot{x},\quad \dot{x}=\frac{1}{m}p,\quad
\left\{x,p\right\}=1,\qquad
H=\frac{1}{2m}p^2+\frac{1}{2}m\omega^2x^2,
\end{equation}
leading to the equations of motion,
\begin{equation}
\dot{x}=\left\{x,H\right\}=\frac{1}{m}p,\quad
\dot{p}=\left\{p,H\right\}=-m\omega^2x.
\end{equation}
{}From the previous discussion of this system, the solutions are of the form
\begin{eqnarray}
x(t)&=&\frac{1}{\sqrt{2m\omega}}
\left[\alpha_0e^{-i\omega(t-t_0)}+\alpha^*_0e^{i\omega(t-t_0)}\right],\qquad
p(t)=-i\frac{m\omega}{\sqrt{2m\omega}}
\left[\alpha_0e^{-i\omega(t-t_0)}-\alpha^*_0e^{i\omega(t-t_0)}\right],\nonumber\\
x(t)&=&\frac{1}{\sqrt{2m\omega}}\left[\alpha(t)+\alpha^*(t)\right],\qquad\qquad\qquad\qquad\
p(t)=-i\frac{m\omega}{\sqrt{2m\omega}}\left[\alpha(t)-\alpha^*(t)\right],
\end{eqnarray}
hence
\begin{equation}
\alpha(t)=\sqrt{\frac{m\omega}{2}}\left[x(t)+\frac{i}{m\omega}p(t)\right],\qquad
\alpha^*(t)=\sqrt{\frac{m\omega}{2}}\left[x(t)-\frac{i}{m\omega}p(t)\right].
\end{equation}

Clearly, the complex valued integration constant $\alpha_0$ (and its complex conjugate $\alpha^*_0$)
determines in a unique fashion a trajectory in the phase space of the system. This suggests another
point of view on the space of states and the Poisson bracket structure it comes equipped with.
Specifying classical solutions to the equations of motion is tantamount to choosing the integration
constant $\alpha_0$. Equivalently, one may say that rather than considering phase space as spanned
by the two real coordinates $x$ and $p$, phase space is spanned
by the complex coordinate $\alpha$ which then evolves in time for a specific classical solution
according to the dynamics of the system. Hence let us consider the change of variable determined
by the coordinate $\alpha(t)$ of which the real and imaginary parts are, up to normalisation factors,
the configuration space coordinate $x(t)$ and its conjugate momentum $p(t)$, respectively.

A simple calculation then finds the following Poisson bracket structure,
\begin{equation}
\left\{\alpha,\alpha^*\right\}=-i,
\end{equation}
while the Hamiltonian becomes
\begin{equation}
H=\frac{1}{2}\omega\left[\alpha\,\alpha^*\,+\,\alpha^*\,\alpha\right].
\end{equation}
In obtaining this latter expression, care has been exercised not to commute the variables $\alpha$ and
$\alpha^*$, since at the quantum level they indeed no longer commute, so that we shall be able
to directly take over the above expression for the quantum Hamiltonian of the system.

Given these results, the equation of motion for the complex phase space coordinate is,
\begin{equation}
\dot{\alpha}=\left\{\alpha,H\right\}=-i\omega\,\alpha,\qquad
\dot{\alpha}^*=\left\{\alpha,H\right\}=i\omega\,\alpha^*,
\end{equation}
of which the solution is obviously,
\begin{equation}
\alpha(t)=\alpha_0\,e^{-i\omega(t-t_0)},\qquad
\alpha^*(t)=\alpha^*_0\,e^{i\omega(t-t_0)},
\end{equation}
$\alpha_0$ being again the necessary integration constant.

\subsubsection{The simple pendulum}

{}From a previous discussion, we know the Lagrange function for this system is given as
\begin{equation}
L=\frac{1}{2}m\ell^2\dot{\theta}^2-mg\ell(1-\cos\theta),
\end{equation}
for a pendulum of mass $m$, length $\ell$, $\theta$ being the angular position
of the mass with respect to the downward vertical direction, and also the single degree
of freedom of this system. Consequently, we readily obtain,
\begin{equation}
p_\theta=\frac{\partial L}{\partial\dot{\theta}}=m\ell^2\dot{\theta},\qquad \dot{\theta}=\frac{1}{m\ell^2}p_\theta,\qquad
\left\{\theta,p_\theta\right\}=1,\qquad
H=\frac{1}{2m\ell^2}p^2_\theta+mg\ell(1-\cos\theta).
\end{equation}
The Hamiltonian equations of motion are thus
\begin{equation}
\dot{\theta}=\left\{\theta,H\right\}=\frac{1}{m\ell^2}p_\theta,\qquad
\dot{p}_\theta=\left\{p_\theta,H\right\}=-mg\ell\sin\theta.
\end{equation}
Reducing the first equation reproduces again the relation $p_\theta=m\ell^2\dot{\theta}$,
which upon substitution into the second of these two equations leads back to the
Euler--Lagrange equation of the system, namely
\begin{equation}
\ddot{\theta}+\frac{g}{\ell}\sin\theta=0.
\end{equation}

\subsubsection{The charged nonrelativistic particle in a background electromagnetic field}

Using the same notations as in the previous discussion of this system, its Lagrange function is
\begin{equation}
L=\frac{1}{2}m\dot{\vec{r}}\,^2-q\Phi(t,\vec{r}\,)+q\dot{\vec{r}}\cdot\vec{A}(t,\vec{r}\,)-V(\vec{r}\,).
\end{equation}
The conjugate momentum vector is thus,
\begin{equation}
\vec{p}=\frac{\partial L}{\partial\dot{\vec{r}}}=m\dot{\vec{r}}+q\vec{A}(t,\vec{r}\,),\qquad
\dot{\vec{r}}=\frac{1}{m}\left[\vec{p}-q\vec{A}(t,\vec{r}\,)\right].
\end{equation}
Note that in this system the conjugate, or canonical momentum $\vec{p}$ {\bf does not} coincide
with the ordinary linear or velocity momentum, $m\dot{\vec{r}}$. This is characteristic of systems
in the presence of background fields. Hence the Poisson brackets are,
\begin{equation}
\left\{x^i,p_i\right\}=\delta^i_j,\quad i,j=1,2,3,
\end{equation}
where $x^i$ and $p_i$ stand for the cartesian coordinates of $\vec{r}$ and $\vec{p}$,
respectively.

The determination of the canonical Hamiltonian is straightforward and leads to
\begin{equation}
H=\frac{1}{2m}\left[\vec{p}-q\vec{A}(t,\vec{r}\,)\right]^2\,+\,
q\Phi(t,\vec{r}\,)\,+\,V(\vec{r}\,).
\end{equation}
The Hamiltonian equations of motion are thus
\begin{equation}
\dot{\vec{r}}=\frac{1}{m}\left[\vec{p}-q\vec{A}(t,\vec{r}\,)\right],\qquad
\dot{\vec{p}}=\frac{q}{m}\left[\vec{p}-q\vec{A}(t,\vec{r}\,)\right]^j\cdot
\frac{\partial\vec{A}_j(t,\vec{r}\,)}{\partial \vec{r}}\,-\,q\vec{\nabla}\Phi(t,\vec{r}\,)
\,-\,\vec{\nabla}V(\vec{r}\,).
\end{equation}
Once again, it may shown that by using the first of these equations to reduce for the
conjugate momentum, and by substituting then its expression into the second of these
equations of motion, the Euler--Lagrange equations are recovered in terms of
the Lorentz force associated to the scalar and vector potentials $\Phi(t,\vec{r}\,)$
and $\vec{A}(t,\vec{r}\,)$, as well as the force associated to the potential energy $V(\vec{r}\,)$.

Given the explicit time dependence of the Lagrange function, hence also the Hamiltonian,
let us consider the time evolution of the latter,
\begin{equation}
\frac{dH}{dt}=\frac{\partial H}{\partial t}+\left\{H,H\right\}=
\frac{\partial H}{\partial t}=-\frac{q}{m}\left[\vec{p}-q\vec{A}(t,\vec{r}\,)\right]
\cdot\frac{\partial\vec{A}(t,\vec{r}\,)}{\partial t}+
q\frac{\partial\Phi(t,\vec{r}\,)}{\partial t}=
-q\dot{\vec{r}}\cdot\partial_t\vec{A}(t,\vec{r}\,)+q\partial_t\Phi(t,\vec{r}\,).
\end{equation}
Consequently one has for instance,
\begin{equation}
\frac{d}{dt}\left[\frac{1}{2}m\dot{\vec{r}}\,^2\right]=
q\dot{\vec{r}}\cdot\vec{E}(t,\vec{r}\,)+\dot{\vec{r}}\cdot\vec{F}(\vec{r}\,),
\end{equation}
which is indeed the equation of motion for the kinetic energy of the particle,
the r.h.s. being the sum of the powers developed by the electric field $\vec{E}(t,\vec{r}\,)$,
on the one hand, and the mechanical force $\vec{F}(\vec{r}\,)$, on the other.

\section{Canonical Quantisation}
\label{Gov1.Sec4}

The concept of the action principle is central to the whole discussion and
framework within which all observed properties of the fundamental interactions
and the elementary particles are being described and understood today. Not only does
the action embody in one single expression all the complicated nonlinear equations
of motion associated to these dynamical systems, but in fact it also accounts for
all the conservation laws through the existence of symmetry transformations of the
configurations of these systems which leave the action invariant. Indeed, as follows
from Noether's (first) theorem, to be discussed at a later stage, any continuous
symmetry of the action directly implies conservation laws, which at the quantum level
translate into conserved charges. One famous example is of course the conservation
of the electric charge, in fact related to the invariance of the electromagnetic interaction
under the local gauge transformations considered already previously, and which extend
also to the quantum context and the quantum states of matter degrees of freedom.

\begin{center}
\fbox{
\begin{tabular}{ccc}
Lagrangian & $\longleftrightarrow$ & Hamiltonian \\
formulation & (Legendre transform) & formulation\\
 & $\nwarrow \qquad\qquad\qquad\qquad\qquad\qquad \nearrow$ & \\
$\hbar$ $\updownarrow$ & \fbox{\bf ACTION (Symmetries)} & $\updownarrow$ $\hbar$\\
 &  $\swarrow \qquad\qquad\qquad\qquad\qquad\qquad \searrow$ & \\
Path/Functional integral &  & Canonical/Operator \\
quantisation & $\longleftrightarrow$ & quantisation
\end{tabular}
}
\end{center}

In fact, besides the general framework outlined here, the culmination of all the progress
made throughout the XX$^{\rm th}$ century in fundamental physics may well be considered to
be the concept of local gauge symmetry, a symmetry realised independently at each point of spacetime
though in a continuous fashion. All interactions, whether classical (as still is General
Relativity for the gravitational interaction) or quantum (for all other three fundamental
interactions) have their properties governed by a gauge symmetry principle. Only the
origin of (inertial) mass still escapes that formulation, and is indeed one of the main open
problems today for fundamental physics. The Standard Model of the fundamental interactions
offers an answer through the Higgs mechanism, and predicts the existence of at least one
more scalar particle unobserved so far, known as the higgs. But it remains to be seen
through the experiments to be started at the LHC (Large Hadron Collider, CERN, Geneva)
whether Nature has not outdone us once again with some far more clever trick
than anyone has yet imagined.

Given the general classical frameworks of the Lagrangian and Hamiltonian formulations
of dynamics, we are now ready to discuss how dynamical systems, whether mechanical systems or
field theories, may be quantised. The path we shall follow is that of canonical
quantisation based on the canonical Hamiltonian formalism, which introduces the
fundamental constant of quantum mechanics, namely the reduced Planck constant,
$\hbar=h/2\pi\simeq 1.055\times 10^{-34}$ J$\cdot$s. Nevertheless, it is possible to also set up path or functional
integral representations of quantum matrix elements in which the classical
first-order Hamiltonian or even Lagrangian actions appear again explicitly.
Such path integral representations thus provide an alternative and complementary
approach to quantisation, physically equivalent to the canonical operator approach.
Depending on the type of issue to be addressed, one approach is often far more
convenient than the other, while they each speak differently to our mathematical and
physical intuitions.

The general discussion will be illustrated mostly with the harmonic oscillator, since
a great deal may be learned already from so simple a system. However, in a perturbative
approach, it is in fact also the harmonic oscillator which lies at the basis of the
whole physical interpretation of relativistic quantum field theories as theories
of relativistic quantum point particles as being the quantum states of the fields. This latter
result is, in certain sense, the fourth revolution of XX$^{\rm th}$ century physics,
after those of quantum mechanics, special relativity and General Relativity. Merging
together quantum mechanics and special relativity leads to relativistic quantum field
theory as a dual description of relativistic quantum point particles and relativistic quantum
fields.

\subsection{The canonical quantisation programme}

\vspace{10pt}

\begin{center}
\begin{tabular}{|c|c|c|}
\hline
{\bf Hamiltonian dynamics} & Correspondence & {\bf Quantum dynamics}\\
 & principle (Dirac) & ($\hbar$) \\
\hline
Canonical formalism & & Canonical quantisation \\
\hline
 & & \\
Phase space  & $\longleftarrow$ {\bf Space of states} $\longrightarrow$ & ``{\bf Hilbert}" {\bf space}:\\
$\left(q^n(t),p_n(t)\right)$ & & $|\psi\rangle$, $\langle\psi|\chi\rangle=\langle\chi|\psi\rangle^*$ \\
 & Representation of & \\
 & & \\
Poisson brackets & {\bf Algebraic} & Commutation relations\\
 & & (Equal time, $t=t_0$)\\
$\left\{A,B\right\}=C$ & $\longleftarrow$ {\bf structures} $\longrightarrow$ & $\left[\hat{A},\hat{B}\right]=i\hbar\hat{C}$ \\
 & & \\
Fundamental brackets & & {\bf Heisenberg algebra}: $\hat{q}^{n\dagger}=\hat{q}^n$, $\hat{p}^\dagger_n=\hat{p}_n$ \\
$\left\{q^n,q^{n'}\right\}=0=\left\{p_n,p_{n'}\right\}$ & &
$\left[\hat{q}^{n},\hat{q}^{n'}\right]=0=\left[\hat{p}_n,\hat{p}_{n'}\right]$ \\
$\left\{q^n,p_{n'}\right\}=\delta^n_{n'}$ & &
$\left[\hat{q}^n,\hat{p}_{n'}\right]=i\hbar\delta^n_{n'}$ \\
 & & \\
Hamiltonian equations & $\longleftarrow$ {\bf Dynamics} $\longrightarrow$ & {\bf Schr\"odinger equation} \\
of motion & & \\
 & & $i\hbar\frac{d|\psi,t\rangle}{dt}=\hat{H}_0\,|\psi,t\rangle$ 
[Schr\"odinger picture] \\
 & & \\
$\frac{dA}{dt}=\frac{\partial A}{\partial t}+\left\{A,H_0\right\}$ & &
$i\hbar\frac{d\hat{A}(t)}{dt}=\left[\hat{A}(t),\hat{H}_0\right]$
[Heisenberg picture] \\
 & & \\
 & & $\hat{H}_0$ self-adjoint \\
 & & \\
Composite observables & & Composite operators [Noether charges]: \\
 & Example & ordering ambiguities \\
$qp$ & $\longleftrightarrow$ & $\hat{q}\hat{p}$, $\hat{p}\hat{q}$,
$(\hat{q}\hat{p}+\hat{p}\hat{q})/2$\\
 & & \\
\hline
\end{tabular}
\end{center}

As indicated in the above Table, to each of the three structures inherent to the
Hamiltonian formulation of any classical dynamical system (namely its phase space as the space of states,
its Poisson brackets with their algebraic properties also shared, as abstract properties,
by the algebra of commutators of linear operators or matrices on a vector space, and
finally, its Hamiltonian as the generator of time evolution through the Poisson brackets
for any observable), there correspond, through the {\bf correspondence principle}, three
analogous structures for the quantised system. Quantising a system amounts to constructing
these three data given their classical counterparts. This requires also the introduction of
Planck's (reduced) constant, $\hbar=h/2\pi$. However, there may exist more than a unique
quantum system which, in the classical limit $\hbar\to 0$, reproduces a given classical system.
It is then a matter of experimental investigations to determine which quantum realisation
Nature is ``using" for the system. For instance, a rotationally invariant system in space
may, at the quantum level, be realised in any of an infinite discrete set of spin values.
It is only by measuring the spin value of, say, the electron, that one may determine it
to be 1/2 (in units of $\hbar$). Some people see this multiplicities of quantum realisations as an inconvenience
and wish to identify a more restricted quantisation programme leading, given any dynamical
system, to a single and unique quantum counterpart. The author of these notes rather sees
this issue of multiple quantisations as a virtue, as a riches of opportunities of which
Nature certainly makes good ``use".

In the quantised system, corresponding to the space of quantum states in which the system
may be found and through which it may evolve in time, one now has to consider some ``Hilbert" space
${\cal H}$, namely a vector space over the complex number and equipped with an inner product
which is sesquilinear, hermitian and positive definite. ``Hilbert" is here put in between quotation marks
for the following reason. In mathematics a {\bf Hilbert space} corresponds to a vector
space with all these structures but meeting also a series of further conditions of a more
technical character (the Hilbert space of the harmonic oscillator to be discussed hereafter
is the example ``par excellence" of a genuine Hilbert space). However in physics often it
is not possible to satisfy exactly and specifically all the properties characteristic of
a Hilbert in the sense of the strict mathematical definition of that word and concept,
but physicists proceed nevertheless and achieve nonetheless
most impressive results with which Nature seems to be happy (the anomalous magnetic moment
$(g-2)$ of the electron has been computed based on QED (Quantum Electrodynamics) to a precision
in eleven decimal places, and agrees within that precision with the measured value set by Nature).
One such example is that of ordinary plane waves in Euclidean space, corresponding to quantum states
of a particle possessing a definite momentum, and the Fourier transformation
of (wave) functions. Strictly speaking such plane waves do not span a Hilbert space since
they lack normalisability. Yet they define ``almost" a Hilbert space (through the theory of
distributions and nested Hilbert spaces), and through the use of Dirac's $\delta$-function
most often one is ``safe" in pretending that the space of states is a Hilbert space.

For the inner product on the space of quantum states, Dirac's ``bra-ket" notation is widely
used and most convenient (in fact it is based on the important concept of the dual of a vector space,
namely the space of linear forms or functions over the vector space taking their values in
the number field over which the vector space is constructed, which
becomes canonically isomorphic to the vector space itself once the vector space is equipped with a nondegenerate inner
product. In that context, the ``bra" corresponds in fact to an element of the dual space,
and the ``ket" to an element of the vector space, with the evaluation of the ``bra" vector
on the ``ket" vector given by the inner product of the two vectors). Let us consider two
vectors of the Hilbert space, $|\psi\rangle$ and $|\chi\rangle$, (``ket" vectors) and
denote their inner product by the ``braket=bra-ket",
\begin{equation}
\langle\psi|\chi\rangle\in\mathbb{C}
\end{equation}
which is thus a complex number. The statement that the inner product is hermitian means
that one has under complex conjugation,
\begin{equation}
\langle\psi|\chi\rangle=\langle\chi|\psi\rangle^*.
\end{equation}
Note that this implies that the bracket of any state with itself is necessarily a real number, 
$\langle\psi|\psi\rangle\in\mathbb{R}$. The statement that the inner product is positive
definite means that one has both the following properties,
\begin{equation}
||\psi||^2\equiv\langle\psi|\psi\rangle\ge 0,\qquad
\langle\psi|\psi\rangle=0\Leftrightarrow |\psi\rangle=0.
\end{equation}
Finally, by inner product over $\mathbb{C}$ one means of course that
$\langle\psi|\chi\rangle$ is a sesquilinear form, namely linear in 
the ket vector $|\chi\rangle$ and antilinear in the bra vector $\langle\psi|$,
\begin{equation}
\langle\sum_\alpha c_\alpha\psi_\alpha|
\sum_\beta d_\beta\chi_\beta\rangle=
\sum_{\alpha,\beta}c^*_\alpha\,d_\beta\,
\langle\psi_\alpha|\chi_\beta\rangle.
\end{equation}
Note that these properties extend to the complex case analogous ones for the ordinary
inner product for a vector space over the real numbers. In the latter case, the
inner product must be linear in both its vector arguments and symmetric under the
exchange of these, and if positive definite it corresponds to a scalar product.
Thus in a Hilbert space over the complex numbers one has, respectively, the properties of sesquilinearity,
hermiticity, and finally positive definiteness.

This much having been said, it does not yet determine how to identify the Hilbert space
to be associated to a given physical system. How is one to choose the space of quantum
states? This is where the second structure comes into action, namely that of the algebraic
properties that must be realised on Hilbert space for the physical observables.
In the classical theory observables are functions defined over phase space for which
one may compute Poisson brackets given the Poisson brackets of the phase space coordinates
$(q^n,p_n)$. In the quantised system, observables are represented by, or associated to
linear operators acting on the quantum states, which therefore no longer commute
with one another in the generic situation. Dirac's proposal for a specific definition
of Bohr's correspondence principle is to state, as a correspondence principle indeed,
that the commutator of two quantum observables is given by the operator associated
to the Poisson bracket of their classical counterparts, up to a factor involving
Planck's constant $\hbar$. More specifically, given two classical observables
$A(q^n,p_n)$ and $B(q^n,p_n)$, let us denote their Poisson bracket as $C(q^n,p_n)$,
$C(q^n,p_n)=\left\{A(q^n,p_n),B(q^n,p_n)\right\}$. At the quantum level, one ought
to associate to these observables quantum operators acting on the space of quantum
states, to be denoted as\footnote{The ``hat" symbol is used to emphasize the
fact that one is dealing with quantum operators. However, at a later stage this notation
will be dropped, since the meaning should become obvious from the context.} $\hat{A}$, $\hat{B}$ and $\hat{C}$,
respectively. The correspondence principle then states that one should have for the equal time commutation
relation of the observables $\hat{A}$ and $\hat{B}$,
\begin{equation}
\left[\hat{A},\hat{B}\right]=i\hbar\,\hat{C}=i\hbar\,\widehat{\left\{A,B\right\}}.
\label{J1.eq:commAB}
\end{equation}
Planck's constant thus enters as a normalisation factor, while the imaginary factor ``$i$"
is required for reasons to which we come hereafter. Incidentally, this normalisation in terms of
Planck's constant $\hbar$ also implies that from now on the action of the system must have
the physical dimension of $\hbar$, while furthermore the absolute numerical normalisation of the
action also implies specific physical properties for the system. This is most readily
seen by considering the commutation relations for the elementary phase space variables.

Thus in particular, associated to the phase
space coordinates $q^n$ and $p_n$, one now has linear operators $\hat{q}^n$ and $\hat{p}_n$
acting on the Hilbert space, which must obey the commutation relations
\begin{equation}
\left[\hat{q}^n,\hat{p}_{n'}\right]=i\hbar\,\delta^n_{n'}.
\label{J1.eq:commqp}
\end{equation}
Note that in the same way as Poisson brackets are defined at equal time, these
commutation relations are defined at equal time, for which we take a specific
reference time $t=t_0$ which is not made explicit in the above relations but must
be kept in mind.

In fact, given that the inner product is hermitian and evaluated over the complex
numbers, one has to specify somewhat further some properties of these quantum observables.
At the classical level a physical observable $A(q^n,p_n)$ is real under complex conjugation,
$A^*(q^n,p_n)=A(q^n,p_n)$. This property should translate at the quantum level into
a corresponding property for the linear operator $\hat{A}$ representing that
observable, known as a self-adjoint property\footnote{Indeed, as is well known, the spectrum of
eigenvalues of a self-adjoint operator is real, as are the values of a real classical observable.}.
To define this concept, let us first
consider an arbitrary linear operator $\hat{A}$ acting on any vector $|\psi\rangle$
of its domain of definition, ${\rm dom}\,A$, in Hilbert space as,
\begin{equation}
|\psi\rangle\in\,{\rm dom}\,A:\qquad
|\psi\rangle\longrightarrow |\psi'\rangle=\hat{A}\,|\psi\rangle\equiv|\hat{A}\psi\rangle,
\end{equation}
where the last form for the transformed vector is for later convenience.
Consider then any other vector $|\chi\rangle$ in Hilbert space and its
inner product with the transformed vector $|\hat{A}\psi\rangle$,
\begin{equation}
\langle\chi|\hat{A}\psi\rangle\equiv \langle\chi|\hat{A}|\psi\rangle.
\end{equation}
The adjoint operator $\hat{A}^\dagger$ of $\hat{A}$ is then defined to be
the operator acting on all the vectors $|\chi\rangle$ of its domain of definition,
${\rm dom}\,\hat{A}^\dagger$, as $\hat{A}^\dagger|\chi\rangle\equiv|\hat{A}^\dagger\chi\rangle$
and such that for any vector $|\psi\rangle\in\,{\rm dom}\,\hat{A}$ we have
\begin{equation}
|\psi\rangle\in\,{\rm dom}\,\hat{A},\quad
|\chi\rangle\in\,{\rm dom}\,\hat{A}^\dagger:\qquad
\langle\hat{A}^\dagger\chi|\psi\rangle=\langle\chi|\hat{A}\psi\rangle.
\end{equation}
In the case of a finite dimensional vector space equipped with an orthonormal
basis for the inner product, the operator $\hat{A}$ is represented by a
matrix (namely the matrix elements of $\hat{A}$ in that basis), and the
adjoint $\hat{A}^\dagger$ of $\hat{A}$ is then represented by the adjoint of that matrix,
namely the complex conjugate of the transposed of the matrix representing $\hat{A}$.
If $n,m=1,2,\ldots,N$ are indices labelling the orthonormalised basis vectors of
a finite $N$ dimensional vector space, the matrix elements of $\hat{A}$ in that
basis are a collection of numbers $A_{nm}$ defining a matrix, of which the
adjoint, $\left(A^\dagger\right)_{nm}=A^*_{mn}$, namely $A^\dagger=\left(A^{\rm T}\right)^*$ as
matrices, gives the matrix elements of the adjoint operator $\hat{A}^\dagger$ in that same basis.

Given this concept of the adjoint of an operator, an operator $\hat{A}$ is said
to be symmetric or hermitian if $\hat{A}$ and $\hat{A}^\dagger$ coincide on the
intersection of their domains of definition,
\begin{equation}
\hat{A}\ \mbox{is symmetric or hermitian if}\
\hat{A}=\hat{A}^\dagger\ \mbox{on}\ {\rm dom}\,\hat{A}\cap{\rm dom}\,\hat{A}^\dagger.
\end{equation}
Indeed, it may be that the two domains are not identical, nor that they would coincide
with the full Hilbert space but are only some vector subspaces of the latter. A self-adjoint
operator $\hat{A}$ is then such that the domains of definition of both $\hat{A}$ and $\hat{A}^\dagger$
are the full Hilbert space ${\cal H}$ while $\hat{A}$ and $\hat{A}^\dagger$ coincide,
\begin{equation}
\hat{A}\ \mbox{is self-adjoint if}\ \hat{A}=\hat{A}^\dagger\ \mbox{and}\
{\rm dom}\,\hat{A}={\cal H}={\rm dom}\,\hat{A}^\dagger.
\end{equation}

Hence in principle one has to identify properly the Hilbert space in such a manner that
quantum observables $\hat{A}$ and $\hat{B}$ (associated to classical real observables) be represented by
self-adjoint operators on ${\cal H}$, $\hat{A}^\dagger=\hat{A}$ and $\hat{B}^\dagger=\hat{B}$.
On the other hand, given two operators $\hat{A}$ and $\hat{B}$, one finds
\begin{equation}
\left[\hat{A},\hat{B}\right]^\dagger=
-[\hat{A}^\dagger,\hat{B}^\dagger],
\end{equation}
since
\begin{equation}
[\hat{A},\hat{B}]^\dagger=\left(\hat{A}\hat{B}-\hat{B}\hat{A}\right)^\dagger=
\hat{B}^\dagger\hat{A}^\dagger-\hat{A}^\dagger\hat{B}^\dagger=[\hat{B}^\dagger,\hat{A}^\dagger].
\end{equation}
Consequently, in the case of self-adjoint operators as must be quantum observables, we conclude
that
\begin{equation}
\left[\hat{A},\hat{B}\right]^\dagger=-[\hat{A},\hat{B}].
\end{equation}
It is this property which explains why it is necessary to include, besides the normalisation
factor specified by Planck's constant $\hbar$, the pure imaginary factor $i$ in the commutation
relations (\ref{J1.eq:commAB}) and (\ref{J1.eq:commqp}).

These concepts having been specified, the correspondence principle thus implies that one should aim to have for all
quantum observables $\hat{A}$, $\hat{B}$ and $\hat{C}$ of which the classical counterparts are such that
$\left\{A,B\right\}=C$, the following equal time commutation relation,
\begin{equation}
[\hat{A},\hat{B}]=i\hbar\hat{C},\quad
\hat{A}^\dagger=\hat{A},\quad\hat{B}^\dagger=\hat{B},\quad\hat{C}^\dagger=\hat{C}.
\end{equation}
In particular for the elementary phase space variables, one must have a realisation
on the Hilbert space ${\cal H}$ of the following equal time canonical commutation relations
\begin{equation}
\left[\hat{q}^n,\hat{p}_{n'}\right]=i\hbar\,\delta^n_{n'},\qquad
\left(\hat{q}^n\right)^\dagger=\hat{q}^n,\quad
\left(\hat{p}_n\right)^\dagger=\hat{p}_n,
\end{equation}
defining an algebra known as {\bf the Heisenberg algebra}.

In conclusion, we thus observe that the Hilbert space ${\cal H}$ of all quantum states of the
system is to be a representation of the algebra of equal commutation relations of the quantum
observables, beginning with the elementary phase space canonical coordinates which must obey a
Heisenberg algebra. As is the case for the concept of classical phase space which combines both a
coordinate parametrisation of that manifold in terms of variables $q^n$ and $p_n$ and
the Poisson bracket structure defined for these and given by the canonical brackets in
the case of canonical phase space coordinates, in the quantum case the notion of the
space of quantum states, namely the Hilbert space of the quantised system, cannot be
dissociated from the algebraic structure of equal time commutation relations that must
be realised in that Hilbert space. Quantising a system consists precisely in the
construction of a Hilbert space for which a given algebra of observables is realised,
namely is a representation of that algebra. We shall briefly come back to this issue hereafter.

The last information available on the classical side of the discussion and for which the
corresponding structure on the quantum side has not yet been introduced is that pertaining
to the dynamics or time evolution of the system generated through the Hamiltonian. Being
an observable, there must correspond a self-adjoint quantum operator $\hat{H}$ to the
classical Hamiltonian $H(q^n,p_n)$, such that $\hat{H}^\dagger=\hat{H}$. At the quantum
level time dependence is thus to be generated by the quantum Hamiltonian $\hat{H}$.
Here there are two physically equivalent ways in which this time dependence may be
represented. The first is by considering that the quantum state describing the system
must evolve in time according to some differential equation in which the Hamiltonian
operator contributes, the latter being defined at the initial time $t=t_0$ at which the
commutation relations are also specified. Consequently, denoting by $|\psi,t\rangle$
the evolving quantum state, its time evolution is governed by {\bf the Schr\"odinger 
equation} in {\bf the Schr\"odinger picture of quantum physics},
\begin{equation}
i\hbar\frac{d|\psi,t\rangle}{dt}=\hat{H}|\psi,t\rangle.
\label{J1.eq:SchrSchr}
\end{equation}
By {\bf Schr\"odinger picture}, one means that the time dependence of the dynamics of the quantum
system is entirely accounted for through a time dependence of the quantum states only, whereas the quantum
operators and observables are defined at the reference time $t=t_0$ at which their
equal time commutation relations have been specified.

Alternatively, because of a reason to which we shall return hereafter, time dependence
of the dynamics of the quantum system may be accounted for by a time dependence of the
operators and observables only, rather than the states, whereas the states are only considered
at the reference time $t=t_0$ at which the equal time commutation relations are
specified. This choice of representation of the time dependence is called 
{\bf the Heisenberg picture of quantum physics}. In this picture the time evolution
equation of these quantum observables is obtained directly from the classical Hamiltonian
equation of motion of an observable through the correspondence principle.
For an observable $\hat{A}$ without any explicit time dependence\footnote{For a classical observable
that carries an explicit time dependence, the general quantum equation of motion reads
$i\hbar d\hat{A}(t)/dt=i\hbar \partial\hat{A}(t)/\partial t + [\hat{A}(t),\hat{H}]$.}, its equal time commutation relation
with the Hamiltonian operator $\hat{H}$ must equal $i\hbar$ multiplying the result of the
corresponding classical Poisson bracket... which is the first order time variation of the
observable, hence leading to {\bf the Schr\"odinger equation} in {\bf the Heisenberg picture
of quantum physics},
\begin{equation}
i\hbar\frac{d\hat{A}(t)}{dt}=\left[\hat{A}(t),\hat{H}\right].
\label{J1.eq:SchrHeis}
\end{equation}
Note that no time dependence is displayed for the Hamiltonian $\hat{H}$. Indeed, its
equation of motion would be $i\hbar\,d\hat{H}/dt=[\hat{H},\hat{H}]=0$, so that this
operator has no time dependence and keeps its value defined at the reference time $t_0$
in any case. Note that when the time evolution parameter $t$ coincides with the physical time,
$\hat{H}$ measures the energy values of the quantum system, implying thus that energy is
conserved even at the quantum level. Later on we shall address the resolution of these Schr\"odinger
equations in terms of the eigenspectrum of the Hamiltonian operator.

This concludes the description of how given a classical dynamics, one may identify a quantum
dynamics associated to it through the correspondence principle. All three structures inherent
to the classical Hamiltonian formalism find their counterparts in the canonical quantisation
of the system. The crucial point of that construction is in fact a construction of the Hilbert space
representation of the algebra of quantum observables.

As a matter of fact, it turns out that it is not possible
to assign to all classical observables a self-adjoint quantum observable while at the same time
obeying all the required commutation relations (a detailed discussion of these
difficulties may be found in Ref.\cite{Gov1.Ali}). The difficulty is
related to the problem of {\bf operator ordering} because variables which at the classical
level commute with one another no longer necessarily do so at the quantum level. Take a single
degree of freedom system with canonical phase coordinates $\hat{q}$ and $\hat{p}$ hence
such that $[\hat{q},\hat{p}]=i\hbar$. Consider then the classical observable $qp$. How is
one to choose a quantum counterpart? Should it be $\hat{q}\hat{p}$, or $\hat{p}\hat{q}$,
or $(\hat{q}\hat{p}+\hat{p}\hat{q})/2$, or yet some other combination of the previous
choices? Clearly there is a potential ambiguity. However in this case it is resolved
by also requiring that the resulting operator be self-adjoint, and would reduce back to
the classical observable when $\hbar\to 0$, namely when $\hat{q}$ and $\hat{p}$ would
commute again. The unique choice meeting these requirements is thus
\begin{equation}
qp\,\longrightarrow\,\frac{1}{2}\left[\hat{q}\hat{p}+\hat{p}\hat{q}\right].
\end{equation}
However, when it comes to higher order monomials in $\hat{q}$ and $\hat{p}$,
the ambiguity remains even when requiring self-adjoint operators, and is even such
that if the correspondence principle can be satisfied for the commutation relations
for all operators bilinear in the quantities $\hat{q}$ and $\hat{p}$, it cannot be
satisfied for all trilinear operators. A specific choice of a subclass of observables
for which the correspondence principle remains satisfied has to be made, based on
yet other considerations, such as those of symmetries which need to be preserved
even at the quantum level (as will be discussed later on, conserved charges related to symmetries
are composite observables constructed from the elementary phase space coordinates).
However, even the latter is not guaranteed, and when it
turns out that there does not exist a quantisation of a system which preserves
its classical symmetries, one has a so-called {\bf quantum anomaly}, namely the
absence in the quantum system of a classical symmetry, the quantum breakdown of a symmetry.

\subsubsection{Illustration: the one dimensional harmonic oscillator}

It is time to illustrate the above general considerations with a simple yet nontrivial
example, for which we shall take the one dimensional harmonic oscillator. We recall that
the classical formulation of that system involves a two dimensional phase space spanned
by the canonical and cartesian coordinates $(q,p)$ with the canonical bracket $\left\{q,p\right\}=1$,
and a dynamics generated by the Hamiltonian
\begin{equation}
H(q,p)=\frac{1}{2m}p^2+\frac{1}{2}m\omega^2q^2.
\end{equation}
It is also of interest to change variables for the description by combining the two real
phase space coordinates into a single complex coordinate,
\begin{equation}
\alpha=\sqrt{\frac{m\omega}{2}}\left[q+\frac{i}{m\omega}p\right],\qquad 
\alpha^*=\sqrt{\frac{m\omega}{2}}\left[q-\frac{i}{m\omega}p\right],
\end{equation}
and conversely
\begin{equation}
q=\frac{1}{\sqrt{2m\omega}}\left[\alpha+\alpha^*\right],\quad
p=-i\frac{m\omega}{\sqrt{2m\omega}}\left[\alpha-\alpha^*\right].
\end{equation}
In terms of this variable and its complex conjugate, the Poisson bracket is
\begin{equation}
\left\{\alpha,\alpha^*\right\}=-i,
\end{equation}
while for the Hamiltonian one finds
\begin{equation}
H=\frac{1}{2}\omega\left[\alpha^* \alpha + \alpha \alpha^*\right],
\end{equation}
without ever having commuted the variables $\alpha$ and $\alpha^*$ with one another
in the calculation.

Applying the correspondence principle of canonical quantisation, the quantised harmonic
oscillator is thus determined by the equal time commutation relation of the Heisenberg algebra
at the reference time $t=t_0$,
\begin{equation}
[\hat{q},\hat{p}]=i\hbar,\qquad\hat{q}^\dagger=\hat{q},\qquad\hat{p}^\dagger=\hat{p},
\end{equation}
with a dynamics governed by the quantum Hamiltonian which we may choose to be
\begin{equation}
\hat{H}=\frac{1}{2m}\hat{p}^2+\frac{1}{2}m\omega^2\hat{q}^2.
\end{equation}
Note that this operator does not suffer an operator ordering ambiguity. The space
of quantum states of this system is thus a representation space of that algebra,
which needs still to be constructed or identified.

Equivalently however, we may also consider the canonical quantisation of the system
based on its description in terms of the complex variable $\alpha$. The correspondence
principle then leads to the equal time commutation relation at the reference time $t=t_0$,
\begin{equation}
[\hat{\alpha},\hat{\alpha}^\dagger]=i\hbar(-i)=\hbar,
\end{equation}
as well as the quantum Hamiltonian
\begin{equation}
\hat{H}=\frac{1}{2}\omega\left[\hat{\alpha}^\dagger \hat{\alpha}\,+\,
\hat{\alpha} \hat{\alpha}^\dagger\right].
\end{equation}
In order to avoid having to carry through all the calculations the $\hbar$ factor
appearing in the above commutation relation, it is better to absorb it in the normalisation
of the operators $\hat{\alpha}$ and $\hat{\alpha}^\dagger$, by dividing each of these operators by
$\sqrt{\hbar}$. Let us thus introduce the quantum operators
\begin{equation}
a=\sqrt{\frac{m\omega}{2\hbar}}\left[\hat{q}+\frac{i}{m\omega}\hat{p}\right],\qquad
a^\dagger=\sqrt{\frac{m\omega}{2\hbar}}\left[\hat{q}-\frac{i}{m\omega}\hat{p}\right],
\end{equation}
and conversely
\begin{equation}
\hat{q}=\sqrt{\frac{\hbar}{2m\omega}}\left[a\,+\,a^\dagger\right],\qquad
\hat{p}=-im\omega\sqrt{\frac{\hbar}{2m\omega}}
\left[a-a^\dagger\right].
\end{equation}
The algebra that the operators $a$ and $a^\dagger$ obey is known as {\bf the Fock algebra},
\begin{equation}
[a,a^\dagger]=\mathbb{I},
\end{equation}
of which the representation in terms of the Fock space is discussed hereafter. In turn,
the quantum Hamiltonian now reads
\begin{eqnarray}
\hat{H} &=& \frac{1}{2}\hbar\omega\left(a^\dagger a+ a a^\dagger\right) \nonumber \\
 &=& \frac{1}{2}\hbar\omega\left(a^\dagger a+ [a,a^\dagger]+a^\dagger a\right) \nonumber \\
 &=& \frac{1}{2}\hbar\omega\left(2a^\dagger a + 1\right) \nonumber \\
 &=& \hbar\omega\left(a^\dagger a+\frac{1}{2}\right).
\end{eqnarray}
Note that the contribution in $\hbar\omega/2$ to this last expression is of a purely
quantum origin, following from the commutator $[a,a^\dagger]=\mathbb{I}$. The reason
why in this calculation we wished to bring the operator $a^\dagger$ to the left of
the operator $a$ will become clear hereafter, when the construction of a representation
of the Fock algebra will have been completed.

\subsubsection{Fock space representation}

In order to identify a Hilbert space providing a realisation of the Fock space algebra,
let us assume there exists some state called the Fock vacuum or ground state (indeed this
state turns out to be the ground state or lowest energy state of the quantum harmonic
oscillator), denoted $|0\rangle$, and such that being acted on with the operator $a$
it is mapped into the null vector of Hilbert space,
\begin{equation}
a|0\rangle=0.
\end{equation}
Furthermore, let us assume at the outset that this state is also normalised, namely
$\langle 0|0\rangle=1$.

Since $|0\rangle$ is annihilated by the operator $a$, the only other possible action
to be considered is that of its adjoint operator, $a^\dagger$, on that state, namely
\begin{equation}
|1\rangle\equiv a^\dagger|0\rangle.
\end{equation}
The question now is to determine whether this new state is really different from $|0\rangle$,
more specifically linearly independent from it and thus defining a new dimension or direction
in Hilbert space independent of that associated to $|0\rangle$ as a basis vector, or rather
whether the state $|1\rangle$ could simply be linearly dependent of $|0\rangle$ with some complex
coefficient $\lambda$ such that
\begin{equation}
|1\rangle=\lambda\,|0\rangle.
\end{equation}
In order to establish that this is excluded through a proof by contradiction, let us assume
it to be the case, and consider now the action of $a$ again on the state $|1\rangle$.
First, independently of the assumption, we have
\begin{equation}
a\,|1\rangle=a\,a^\dagger\,|0\rangle=
\left(aa^\dagger- a^\dagger a+ a^\dagger a\right)\,|0\rangle=
\left([a,a^\dagger]+a^\dagger a\right)|0\rangle=
\left(1+a^\dagger a\right)|0\rangle=|0\rangle,
\end{equation}
in which in the last step of this series of little expressions use has been made of the fact that
the operator $a$ annihilates the vacuum $|0\rangle$. The result $a|1\rangle=|0\rangle$ is thus
valid under all circumstances. However, if in addition we were to have also that
$|1\rangle=\lambda|0\rangle$, it would follow that
\begin{equation}
a|1\rangle=a\left(\lambda|0\rangle\right)=0,
\end{equation}
since $a$ annihilates the state $|0\rangle$. However such a conclusion would be inconsistent
with the result $a|1\rangle=|0\rangle$. Therefore we are led to conclude that indeed the state
$|1\rangle=a^\dagger|0\rangle$ is a state linearly independent from the vacuum $|0\rangle$.

Through a similar discussion in a recursion procedure, it is possible to establish that the
state obtained by acting $n$ times with the operator $a^\dagger$ on the state $|0\rangle$,
$(a^\dagger)^n|0\rangle$, is linearly independent from all the states $(a^\dagger)^m|0\rangle$
with $m=n-1,n-2,\ldots,0$. Consequently, the whole tower of states constructed in this manner out
from the Fock vacuum defines a basis of the Hilbert space which they generate and which
thus provides a representation of the Fock algebra of the operators $a$ and $a^\dagger$.
This space is known as the Fock space representation of the Fock algebra.

In fact, all these states are not only linearly independent for $n=0,1,2,\ldots$ but are
mutually orthogonal or perpendicular. Indeed, this follows from the fact, implicit in the above discussion,
that the constructed Hilbert space is also equipped with an inner product for which the operators $a$ and $a^\dagger$
are adjoint of one another, and such that $\langle 0|0\rangle =1$. For instance, one easily finds
\begin{equation}
\langle 0|1\rangle=\langle 0|a^\dagger|0\rangle=\langle 0|a|0\rangle^*=0.
\end{equation}
In order that they be also normalised, one defines the normalisation of the Fock space
basis vectors, or simply Fock states, as
\begin{equation}
|n\rangle=\frac{1}{\sqrt{n!}}\left(a^\dagger\right)^n\,|0\rangle,\qquad
n=0,1,2,3,\ldots
\end{equation}
Given that choice it does not take much of a calculation using the commutator
$[a,a^\dagger]=\mathbb{I}$ repeatedly to check that one has,
\begin{equation}
a|0\rangle=0,\quad a|n\rangle=\sqrt{n}\,|n-1\rangle,\quad n=1,2,\ldots;\qquad
a^\dagger|n\rangle=\sqrt{n+1}\,|n+1\rangle,\quad n=0,1,2,\ldots,
\end{equation}
from which it also follows that
\begin{equation}
\langle n|m\rangle=\delta_{nm},
\end{equation}
showing that indeed the set of Fock states $\left\{|n\rangle, n=0,1,2,\ldots\right\}$
defines an orthonormalised basis of Fock space. Note that the operators $a$ and $a^\dagger$
thus map between successive Fock states. The latter may be viewed as defining a semi-infinite
ladder, with $a^\dagger$ moving one step upward on that ladder and $a$ one step downward.
The operators $a$ and $a^\dagger$ are thus also known as {\bf the ladder operators}.
However a vocabulary more largely used is that of the creation (for $a^\dagger$) and
annihilation (for $a$) operators since, as will become totally clear hereafter,
they indeed create of annihilate energy quanta of the system, moving between Fock
states differing in a single quantum of excitation in energy.

The action of the ladder operators on the Fock states having been established, it follows that
\begin{equation}
a^\dagger\,a\,|n\rangle=a^\dagger\sqrt{n}|n-1\rangle=n|n\rangle,\qquad
aa^\dagger\,|n\rangle=a\sqrt{n+1}|n+1\rangle=(n+1)|n\rangle,
\end{equation}
hence the Fock algebra is indeed obeyed since it is for each of the Fock basis vectors,
\begin{equation}
[a,a^\dagger]\,|n\rangle=|n\rangle.
\end{equation}
Note that these results also establish that the Fock states are already the eigenstates
of the operator $a^\dagger a$, since $a^\dagger a|n\rangle=n|n\rangle$. Furthermore the
eigenvalue $n$ for the Fock state at level $n$ measures the number of times the creation
or ladder operator $a^\dagger$ has been applied to the Fock vacuum, namely, as shall be
seen later on, the number of quanta present in the system. Hence the operator $N=a^\dagger a$
is often called {\bf the number operator}. Since the Hamiltonian of the system is also
expressed in terms of the number operator, $H=\hbar\omega[N+1/2]$, it follows that the
Fock basis also diagonalises the Hamiltonian of the system, hence immediately providing
the energy spectrum of the quantised one dimensional harmonic oscillator,
\begin{equation}
H|n\rangle=E_n|n\rangle,\qquad
E_n=\hbar\omega\left(n+\frac{1}{2}\right),\quad n=0,1,2,3,\ldots
\end{equation}
Given this result it is useful to represent all these states in a diagram superposed onto
the graph of the potential energy of the system, $V(q)=\frac{1}{2}m\omega^2q^2$. Measured
from the bottom of the harmonic well, the lowest energy state or ground state $|0\rangle$ possesses
an energy $\hbar\omega/2$ which is purely of a quantum origin since that contribution originates
directly from the Fock algebra commutator $[a,a^\dagger]=\mathbb{I}$. This contribution to the
energy is often called {\bf the quantum vacuum energy}. And as quantum excitations of the
system, one has the whole ladder of Fock states $|n\rangle$, of which the energies all differ
for successive states by the quantum $\hbar\omega$. By adding or removing through some coupling
or interaction with the oscillator a certain number of quanta each of energy $\hbar\omega$,
it is possible to bring the system into any of the Fock states. The reason why the spectrum is discrete
also originates in the Fock algebra commutator. The reason why this discreteness is infinite is
because the harmonic potential well has an infinite height. An algebraic reason is that the original
Heisenberg algebra, namely also the Fock algebra itself, may only be represented on an infinite
dimensional vector space. Indeed, if the representation were to be finite dimensional, taking the
trace of either defining commutation relation would lead to an inconsistency of the type $0=1$.

We have thus already learned a great deal and acquired some experience with a Hilbert space
just from this simple system. Let us use the opportunity to develop some further considerations.
Knowing that Fock states define a basis of the full space of quantum states, any state
in that space may be expressed as a mixed state of all Fock states $|n\rangle$ through a linear combination
of which the coefficients are complex numbers $\psi_n$,
\begin{equation}
|\psi\rangle=\sum_{n=0}^\infty\,|n\rangle\,\psi_n.
\end{equation}
Each of the terms $|n\rangle\,\psi_n$ determines the projection of the state $|\psi\rangle$
onto the direction in Hilbert space associated to the Fock state $|n\rangle$, $\psi_n\in\mathbb{C}$
being the component of the state $|\psi\rangle$ with respect to that basis vector. The basis
being orthonormalised, the component itself is obtained by the projection of vectors defined by
the inner product with which the Hilbert space is equipped, namely
\begin{equation}\langle n|\psi\rangle=
\sum_{m=0}^\infty\langle n|m\rangle\,\psi_m=\psi_n.
\end{equation}
Consequently, we may write
\begin{equation}
|\psi\rangle=\sum_{n=0}^\infty\,|n\rangle\langle n|\psi\rangle,
\end{equation}
which is a very useful relation already as such. However, bringing it into the form
\begin{equation}
|\psi\rangle=\left(\sum_{n=0}^\infty|n\rangle\,\langle n|\right)|\psi\rangle,
\label{J1.eq:sumprojector}
\end{equation}
one notices that each of the terms in the brackets on the r.h.s. of this
identity, namely $|n\rangle\langle n|$, stands in fact for an operator which
is nothing else but the projection operator $\mathbb{P}_n=|n\rangle\langle n|$
onto the direction in Hilbert space defined by the Fock state $|n\rangle$, which has the properties
\begin{equation}
\mathbb{P}^2_n=\mathbb{P}_n,\qquad
\mathbb{P}^\dagger_n=\mathbb{P}_n.
\end{equation}
Indeed, when acting on any state $|\psi\rangle$, the projector $\mathbb{P}_n=|n\rangle\langle n|$
produces a new vector which is simply the component of $|\psi\rangle$ in
the direction of $|n\rangle$,
\begin{equation}
\mathbb{P}_n|\psi\rangle=|n\rangle\langle n|\psi\rangle.
\end{equation}
It should thus be clear that when summing each of these projectors $\mathbb{P}_n$
over all independent directions in Hilbert space, one then necessarily recovers the identity operator,
which is indeed what (\ref{J1.eq:sumprojector}) represents,
\begin{equation}
\mathbb{I}=\sum_{n=0}\,|n\rangle\,\langle n|.
\label{J1.eq:spectralone}
\end{equation}
Such a representation of the identity operator is called a {\bf spectral decomposition} (or resolution)
of the identity operator, for a reason which is to become totally clear hereafter. As it turns
out such an identity is extremely useful. In the form of (\ref{J1.eq:sumprojector}) it shows
how it leads directly to the decomposition of any state in the Fock basis. Likewise for an
arbitrary operator $\hat{A}$, we may write directly
\begin{equation}
\hat{A}=\mathbb{I}\,\hat{A}\,\mathbb{I}=
\left(\sum_{n=0}^\infty |n\rangle\langle n|\right)\,\hat{A}\,
\left(\sum_{m=0}^\infty |m\rangle\langle m|\right)=
\sum_{n,m=0}^\infty\,|n\rangle\,
\langle n|\hat{A}|m\rangle\,\langle m|,
\end{equation}
in which the matrix elements $A_{nm}=\langle n|\hat{A}|m\rangle$ of the
(semi-infinite discrete) matrix representing the abstract operator $\hat{A}$ in the
Fock basis appear naturally and explicitly. In particular for an operator $\hat{\Lambda}$
which is diagonalised in the Fock basis, and for which the matrix representation is thus
diagonal with its eigenvalues $\left\{\lambda_n, n=0,1,2,\ldots\right\}$ on the diagonal,
the above double sum reduces to a single sum, $\Lambda_{nm}=\langle n|\hat{\Lambda}|m\rangle=\lambda_n\delta_{nm}$, so that
\begin{equation}
\hat{\Lambda}=\sum_{n=0}^\infty\,|n\rangle\,\lambda_n\,\langle n|.
\end{equation}
Such a decomposition of a diagonal abstract operator in the basis of its eigenvectors
is called {\bf the spectral decomposition} (or resolution) of the operator, with the spectrum of its
eigenvalues $\lambda_n$ indeed appearing in between the product of the ket, $|n\rangle$,
and bra, $\langle n|$, eigenvectors, or multiplying each of the projection operators
$\mathbb{P}_n=|n\rangle\langle n|$ associated to these eigenvectors. The case of
the identity operator (\ref{J1.eq:spectralone}) is the particular situation when all
these eigenvalues reduce to unity. As another example, one has the Hamiltonian
operator in the Fock state basis which diagonalises it,
\begin{equation}
\hat{H}=\sum_{n=0}^\infty\,|n\rangle\,E_n\,\langle n|,\qquad
E_n=\hbar\omega\left(n+\frac{1}{2}\right).
\end{equation}

The above discussion relied mostly on a purely algebraic and abstract approach, and managed
to identify most straightforwardly the energy spectrum of the harmonic oscillator. However once
the Fock state basis is singled out, abstract operators may also be represented in terms of
matrices of which the entries are the matrix elements of the operator in that basis. This
then enables a matrix representation of quantum physics, which is essentially how Heisenberg
first conceived of the rules of quantum mechanics. As an illustration, knowing how the
ladder operators act on the Fock states the values for their matrix elements are readily
identified as follows, in the order of increasing $n=0,1,2,\ldots$ values for the Fock states,
\begin{equation}
a:\qquad\left(\begin{array}{c c c c c}
0 & \sqrt{1} & 0 & 0 & \cdots \\
0 & 0 & \sqrt{2} & 0 & \cdots\\
0 & 0 & 0 & \sqrt{3} & \cdots \\
\vdots & \vdots & \vdots & \vdots 
\end{array}
\right),\qquad
a^\dagger:\qquad\left(\begin{array}{c c c c c}
0 & 0 & 0 & 0 & \cdots \\
\sqrt{1} & 0 & 0 & 0 & \cdots\\
0 & \sqrt{2} & 0 & 0 & \cdots \\
\vdots & \vdots & \vdots & \vdots 
\end{array}
\right),
\end{equation}
and as a consequence, we also have the matrix representations of the generators $\hat{q}$ and $\hat{p}$
of the Heisenberg algebra,
\begin{equation}
\hat{q}:\quad
\sqrt{\frac{\hbar}{2m\omega}}\left(\begin{array}{c c c c c}
0 & \sqrt{1} & 0 & 0 & \cdots \\
\sqrt{1} & 0 & \sqrt{2} & 0 & \cdots\\
0 & \sqrt{2} & 0 & \sqrt{3} & \cdots \\
\vdots & \vdots & \vdots & \vdots 
\end{array}
\right),\quad
\hat{p}:\quad
-im\omega\sqrt{\frac{\hbar}{2m\omega}}\left(\begin{array}{c c c c c}
0 & \sqrt{1} & 0 & 0 & \cdots \\
-\sqrt{1} & 0 & \sqrt{2} & 0 & \cdots\\
0 & -\sqrt{2} & 0 & \sqrt{3} & \cdots \\
\vdots & \vdots & \vdots & \vdots 
\end{array}
\right).
\end{equation}
Based on these matrices, once again it is possible to check for the commutation relations
whether for the Heisenberg algebra, $[\hat{q},\hat{p}]=i\hbar$, or the Fock algebra,
$[a,a^\dagger]=\mathbb{I}$.

\subsection{Quantum evolution}

As discussed previously, once the Hilbert space appropriate to a given system has been identified or constructed
as a representation of the algebra of equal time commutation relations, quantum dynamics is generated
by the quantum Hamiltonian through the Schr\"odinger equation, whether in the Schr\"odinger or the
Heisenberg picture. Hereafter we first discuss certain considerations in relation to the solution to either
of these forms of the Schr\"odinger equation, to be followed by a demonstration that if the spectrum of the
quantum Hamiltonian has been determined, in fact all that there is to know about the dynamics of the quantum
system is available explicitly. Determining the energy spectrum amounts to solving the full quantum dynamics.

\subsubsection{The Schr\"odinger and Heisenberg pictures}

As discussed previously, in the Schr\"odinger picture time dependence is totally accounted for through the
time dependence of the quantum states, $|\psi,t\rangle$, which have to obey the Schr\"odinger equation
\begin{equation}
i\hbar\frac{d|\psi,t\rangle}{dt}=\hat{H}\,|\psi,t\rangle,
\end{equation}
whereas operators are time independent and are considered at the reference time $t_0$ at which
the quantisation programme based on the equal time commutation relations is being developed.
Being first order in the time derivative, the general solution to this equation requires a single
integration constant or boundary condition, say the value for the state at the reference time $t=t_0$
at which the canonical quantisation and the equal time commutation relations are specified, $|\psi,t_0\rangle$.
The solution then reads
\begin{equation}
|\psi,t\rangle=U(t,t_0)\,|\psi,t_0\rangle,
\end{equation}
where $U(t,t_0)$ is the operator defined by
\begin{equation}
U(t,t_0)=e^{-\frac{i}{\hbar}(t-t_0)\hat{H}}.
\end{equation}
This operator is known as {\bf the quantum evolution operator} or also {\bf the propagator}\footnote{Not
to be confused with the Feynman propagator of quantum field theory.} of the quantum system.
Being defined through the exponential of the Hamiltonian operator through the usual power series
expansion,
\begin{equation}
e^{-\frac{i}{\hbar}(t-t_0)\hat{H}}=
\sum_{n=0}^\infty\frac{1}{n!}\left(-\frac{i}{\hbar}(t-t_0)\hat{H}\right)^n,
\end{equation}
there are certain conditions that may have to be met to ensure the convergence of such an
expression. Furthermore, it is crucial that the Hamiltonian $\hat{H}$ also be self-adjoint
to guarantee unitarity, namely a time evolution preserving the probabilities of quantum physics.
In fact, the quantum evolution operator must obey two important properties,
\begin{eqnarray}
&{\rm Convolution}&:\qquad
U(t_3,t_2)\,U(t_2,t_1)= U(t_3,t_1),\nonumber\\
&{\rm Unitarity}&:\qquad
U^\dagger(t_2,t_1)=U^{-1}(t_2,t_1)\left(=U(t_1,t_2)\right).
\end{eqnarray}
In particular, the property of unitarity of the operator $U(t_2,t_1)$ is indeed required
to preserve the probability of a state under time evolution,
\begin{equation}
\langle\psi,t|\psi,t\rangle=\langle\psi,t_0|U^\dagger(t,t_0)\,U(t,t_0)|\psi,t_0\rangle=
\langle\psi,t_0|\psi,t_0\rangle .
\end{equation}

In the Heisenberg picture however, dynamics of the quantum system is totally accounted for 
through the time dependence of the quantum operators, which have to obey the Schr\"odinger
equation\footnote{In the case of observables that carry no explicit time dependence.}
\begin{equation}
i\hbar\frac{d\hat{A}(t)}{dt}=\left[\hat{A}(t),\hat{H}\right],
\end{equation}
whereas quantum states are time independent and are considered at the reference time $t=t_0$ at
which the canonical quantisation programme is being developed. But since the Schr\"odinger and
Heisenberg pictures of quantum physics are physically equivalent, how are they related? To answer
this, let us for instance consider the expectation value in a state $|\psi,t\rangle$ of a given
observable $\hat{A}(t_0)$ defined in the Schr\"odinger picture,
$\langle\psi,t|\hat{A}(t_0)|\psi,t\rangle$, the observable being assumed not to carry any explicit time
dependence (more generally one could consider expectation values of
products of observables). Using the solution to the Schr\"odinger equation, and assuming the
state $|\psi,t\rangle$ to have been normalised, $\langle\psi,t|\psi,t\rangle=1$, we then have
for the expectation value of the observable $\hat{A}$ as a function of time $t$,
\begin{equation}
\langle\hat{A}\rangle(t)\equiv\frac{\langle\psi,t|\hat{A}(t)|\psi,t\rangle}
{\langle\psi,t|\psi,t\rangle}=\langle\psi,t_0|U^\dagger(t,t_0)\,\hat{A}(t_0)\,U(t,t_0)|\psi,t_0\rangle=
\langle\psi,t_0|\hat{A}(t)|\psi,t_0\rangle,
\end{equation}
in which we have introduced
\begin{equation}
\hat{A}(t)=U^\dagger(t,t_0)\,\hat{A}(t_0)\,U(t,t_0).
\label{J1.eq:Heisenbergpicture}
\end{equation}
Hence, this definition establishes the relation between the Schr\"odinger and Heisenberg pictures
of quantum physics. In particular, this definition also provides the solution to the Schr\"odinger
equation in the Heisenberg picture in terms of the quantum evolution operator $U(t,t_0)$, the
operator $\hat{A}(t_0)$ then playing the r\^ole of an integration constant. Indeed, from the above
definition of $\hat{A}(t)$ it readily follows that we have
\begin{equation}
i\hbar\frac{d\hat{A}(t)}{dt}=\left[\hat{A}(t),\hat{H}\right],
\end{equation}
which is indeed the relevant Schr\"odinger equation when the quantum observable $\hat{A}$
does not possess any explicit time dependence. The quantum evolution operator is thus central
in solving the Schr\"odinger equation whatever the picture of quantum physics being used.

\subsubsection{Diagonalisation of time evolution}

Let us now assume that the eigenspectrum of the Hamiltonian operator $\hat{H}$ has been
determined (which, in practice, is an extremely difficult problem for any system with the slightest
relevance to physical reality),
\begin{equation}
\hat{H}\,|E_m\rangle=E_m\,|E_m\rangle,
\end{equation}
where the notation is meant to be schematic. In general the index $m$ stands for
a multi-index some of which components could even take values in a continuous
rather than a discrete set. Furthermore we assume all the eigenstates $|E_m\rangle$
to have been orthonormalised,
\begin{equation}
\langle E_m|E_{m'}\rangle=\delta_{mm'}.
\end{equation}
Consequently, one has the spectral decomposition of the identity operator
\begin{equation}
\mathbb{I}=\sum_m\,|E_m\rangle\,\langle E_m|,
\end{equation}
with in particular, when applied onto any diagonal operator in the basis $|E_m\rangle$,
a similar spectral decomposition, such as for instance
\begin{equation}
\hat{H}=\sum_m\,|E_m\rangle\,E_m\,\langle E_m|,
\end{equation}
\begin{equation}
U(t_2,t_1)=\sum_m\,|E_m\rangle\,e^{-\frac{i}{\hbar}(t_2-t_1)\,E_m}\,\langle E_m|.
\end{equation}

Given this last decomposition of the evolution operator, the solution to the Schr\"odinger
equation whether in the Schr\"odinger or the Heisenberg pictures is (in the latter case again
for an operator that carries no explicit time dependence),
\begin{equation}
|\psi,t\rangle=\sum_m\,|E_m\rangle\,e^{-\frac{i}{\hbar}(t-t_0)\,E_m}\,\langle E_m|\psi,t_0\rangle,
\end{equation}
\begin{equation}
\hat{A}(t)=\sum_{m,m'}\,|E_m\rangle\,e^{\frac{i}{\hbar}(t-t_0)E_m}\,
\langle E_m|\hat{A}(t_0)|E_{m'}\rangle\,e^{-\frac{i}{\hbar}(t-t_0)E_{m'}}\,
\langle E_{m'}|.
\end{equation}
Consequently, if the eigenspectrum of the Hamiltonian operator is completely known,
the entire dynamical time evolution of the quantum system is also determined.

\subsubsection{Illustration: the one dimensional harmonic oscillator}

Since the energy spectrum of the harmonic oscillator is
\begin{equation}
E_n=\hbar\omega\left(n+\frac{1}{2}\right),\qquad n=0,1,2,\ldots,
\end{equation}
the above general discussion translates into the following simple terms.
Based on the spectral decomposition of the unit operator
in terms of the Fock states $|n\rangle$,
\begin{equation}
\mathbb{I}=\sum_{n=0}^\infty\,|n\rangle\,\langle n|,
\end{equation}
one simply obtains for the quantum evolution operator
\begin{equation}
U(t_2,t_1)=\sum_{n=0}^\infty\,|n\rangle\,e^{-\frac{i}{\hbar}(t_2-t_1)\hbar\omega(n+1/2)}\,
\langle n|=
e^{-\frac{1}{2}i\omega(t_2-t_1)}\,\sum_{n=0}^\infty\,|n\rangle\,
e^{-in\omega(t_2-t_1)}\,\langle n|.
\end{equation}
Based on that spectral decomposition, it follows that for quantum states in the
Schr\"odinger picture one has for the solutions to the Schr\"odinger equation,
\begin{equation}
|\psi,t\rangle=U(t,t_0)\,|\psi,t_0\rangle=e^{-\frac{1}{2}i\omega(t-t_0)}\,
\sum_{n=0}^\infty\,|n\rangle\,e^{-in\omega(t-t_0)}\,\langle n|\psi,t_0\rangle.
\end{equation}

For what concerns the Heisenberg picture, let us consider as observables the
ladder operators as well as $\hat{q}$ and $\hat{p}$. Given the definition
(\ref{J1.eq:Heisenbergpicture}) of the time dependence of operators in the Heisenberg
picture, one needs to use one of the Baker--Campbell--Hausdorff formulae,
\begin{equation}
e^A\,B\,e^{-A}=B+\left[A,B\right]+\frac{1}{2!}\left[A,\left[A,B\right]\right]+\ldots
+\frac{1}{n!}\left[A,\left[A,\left[\cdots,\left[A,\left[A,B\right]\right]\cdots\right]\right]\right.+\ldots
\end{equation}
Applying this expression to
\begin{equation}
a(t)=e^{\frac{i}{\hbar}(t-t_0)\hat{H}}\,a(t_0)\,e^{-\frac{i}{\hbar}(t-t_0)\hat{H}},
\end{equation}
and using the fact that
\begin{equation}
\left[\frac{i}{\hbar}(t-t_0)\hat{H},a(t_0)\right]=
i\omega(t-t_0)\left[a^\dagger a,a\right]=-i\omega(t-t_0),
\end{equation}
it follows that
\begin{equation}
a(t)=a(t_0)\,e^{-i\omega(t-t_0)},
\end{equation}
and thus also
\begin{equation}
a^\dagger(t)=a^\dagger(t_0)\,e^{i\omega(t-t_0)}.
\end{equation}
Note that these time dependencies are precisely those of the classical solutions as well,
in terms of the coefficients denoted $\alpha(t)$ and $\alpha^*(t)$ as introduced previously but differing from 
$a(t)$ and $a^\dagger(t)$ only by a normalisation factor of $\sqrt{\hbar}$. Furthermore,
the time dependence of the position and conjugate momentum operators then also follows,
\begin{equation}
\hat{q}(t)=\sqrt{\frac{\hbar}{2m\omega}}
\left[a(t_0)\,e^{-i\omega(t-t_0)}\,+\,a(t_0)\,e^{i\omega(t-t_0)}\right],
\end{equation}
\begin{equation}
\hat{p}(t)=-im\omega\sqrt{\frac{\hbar}{2m\omega}}
\left[a(t_0)\,e^{-i\omega(t-t_0)}\,-\,a(t_0)\,e^{i\omega(t-t_0)}\right].
\end{equation}
Once again these time dependencies coincide with those of the classical solutions to
the Hamiltonian equations of motion. Of course this is as expected given that the
Schr\"odinger equation for a quantum  observable in the Heisenberg picture is
in correspondence, through  canonical quantisation, with the Hamiltonian equation of motion for
the classical counterpart of that observable. In particular, the configuration space operator
$\hat{q}(t)$ in the Heisenberg picture obeys precisely, as an operator now, the original
Euler--Lagrange equation of motion of the system,
\begin{equation}
\left[\frac{d^2}{dt^2}\,+\,\omega^2\right]\,\hat{q}(t)=0.
\end{equation}
This observation serves as a basis for a heuristic argument showing that
any theory of relativistic quantum point particles is necessarily also a
theory of relativistic quantum fields \cite{Gov1.Gov1}.

\subsection{Representations of the Heisenberg algebra I}

Within the context of the harmonic oscillator, starting from the Heisenberg algebra,
$[\hat{q},\hat{p}]=i\hbar$, of the phase space observables, we introduced the
creation and annihilation operators $a^\dagger$ and $a$ in terms of which 
an abstract Hilbert space representation of the Fock algebra was constructed,
as well as the Fock basis identified within it, which enabled directly the
diagonalisation of the Hamiltonian of the harmonic oscillator. Hence the
Fock space representation of the Fock algebra already provides in fact
an abstract representation of the Heisenberg algebra as well, through the relations
\begin{equation}
\hat{q}=\sqrt{\frac{\hbar}{2m\omega}}\left[a+a^\dagger\right],\qquad
\hat{p}=-im\omega\sqrt{\frac{\hbar}{2m\omega}}\left[a-a^\dagger\right].
\end{equation}
Furthermore, given the Fock basis in that Hilbert space, operators acquire a matrix
representation which obeys the same commutation relations once again. In other words,
there also exists a matrix representation, albeit in terms of semi-infinite matrices, 
of the same abstract algebraic structure defining the space of quantum states of the
harmonic oscillator. This raises the issue of finding all possible representations of
the Heisenberg algebra, beginning with a single degree of freedom system. For a finite
number $N$ of degrees of freedom, the Heisenberg algebra is the $N$-fold tensor product
of the Heisenberg algebra for a single degree of freedom, hence so is its representation
space. Let us thus restrict to the Heisenberg algebra for a single degree of freedom.

In the case that the classical phase space $(q,p)$ is simply Euclidean (and thus
by extension for any Euclidean configuration space of any dimension $N$ with cartesian coordinates $q^n$),
there exists a famous result due to von Neumann and Stone stating that up to unitary transformations
in Hilbert space (indeed quantum states are defined up to an arbitrary overall unitary
transformation which does not affect physical observations of the quantum dynamics),
there exists a single representation of the Heisenberg algebra (a derivation of this result may be
found for example in Refs.\cite{Gov1.GovBook,Gov1.Villa}). In other words, the two
representations that have already been constructed are but two different realisations of
a common underlying abstract Hilbert space realisation of the Heisenberg algebra.

However, let us just state here that if the configuration space possesses a nontrivial
topology such that there exist noncontractible cycles within it (namely, when its first homotopy group is nontrivial),
then there exists in fact an infinity of unitarily inequivalent representations of the Heisenberg
algebra, labelled by U(1) holonomies around the noncontractible cycles \cite{Gov1.Villa}. Furthermore, the
discussion presented hereafter may also be extended to curved configuration space manifolds,
but only the case of Euclidean geometry will be detailed here.

\subsubsection{Configuration and momentum space representations}

Let us thus consider as configuration space the real line $\mathbb{R}$, with $q\in\mathbb{R}$,
and its momentum conjugate $p$ also spanning that range of values, $p\in\mathbb{R}$. Given
this configuration and phase space Euclidean geometry, one may then establish the
existence of two representations of the Heisenberg algebra in terms of
complex functions of $q$ in one case and of $p$ in the other case, hence known
as {\bf configuration} and {\bf momentum space wave function representations} of the Heisenberg
algebra. These two possibilities correspond to what is usually introduced to define
quantum mechanics in a first course on the subject. Here these considerations
follow a discussion of the representation theory of the abstract Heisenberg algebra, which
arises whatever dynamical system is being quantised. Note well that our discussion thus
does not assume implicitly the dynamics of the harmonic oscillator. The discussion to
be developed hereafter is totally independent from any dynamical consideration, but is
purely kinematical in character.

\vspace{10pt}

\noindent
a) {\sl The configuration space wave function representation}

\vspace{10pt}

Let us assume there exists in Hilbert space a basis of position eigenstates, namely
states $|q\rangle$ labelled by the eigenvalues of the configuration space or position
operator $\hat{q}$ with a spectrum spanning all of $\mathbb{R}$,
\begin{equation}
\hat{q}\,|q\rangle=q\,|q\rangle,\qquad q\in\,\mathbb{R}.
\end{equation}
Furthermore, let us assume that these states are normalised, while they may be assumed
to be orthogonal since $\hat{q}$ is self-adjoint\footnote{Indeed, a self-adjoint operator is
necessarily diagonalisable and such that its eigenvectors for distinct eigenvalues are
orthogonal. These results are well known for finite dimensional matrices, and extend to operators on a Hilbert space.},
$\hat{q}^\dagger=\hat{q}$. Consequently, we have the inner products
\begin{equation}
\langle q|q'\rangle=\delta(q-q'),
\label{J1.eq:normqq}
\end{equation}
where $\delta(q-q')$ is the Dirac $\delta$-function. The main property of the $\delta$-``function"
(strictly speaking it is a distribution, and its evaluation has to be understood inside an
integral where it is multiplied with a test function for the purpose of evaluating the
integral) is that given any function $f(x)$ of a single variable, the $\delta$-function
$\delta(x-a)$ is such that
\begin{equation}
\int_{-\infty}^{+\infty}\,dx\,f(x)\delta(x-a)=f(a).
\end{equation}
In other words, the $\delta$-function is analogous to the Kronecker $\delta_{nm}$ symbol
in which the set of integers would have been closed into the set of real numbers by having
grown ever more dense on the real line. The $\delta$-function thus vanishes whenever its argument is nonvanishing, whereas
it takes an infinite value when its argument vanishes but in such a manner that one has a finite
value thus normalised to unity when the $\delta$-function is evaluated inside
the following integral\footnote{Even though the $\delta$-function
is infinite for a vanishing argument, the latter value is a point of zero measure in the integral so that it remains possible
that the integral be finite. One may also introduce different approximations to the $\delta$-function
through a limit procedure, such as in the Gaussian representation
$\delta(x-a)=\lim_{\epsilon\to 0}\frac{1}{\sqrt{2\pi\epsilon}}\,e^{-\frac{1}{2\epsilon}(x-a)^2}$.}
(again in the same way that when summing the Kronecker $\delta_{nm}$ symbol
over all the range of values of one of its indices it returns the unit value),
\begin{equation}
\int_{-\infty}^{+\infty}dx\,\delta(x-a)=1.
\end{equation}
Given these remarks, it should clear that the condition (\ref{J1.eq:normqq}) indeed
expresses the orthonormality of the basis $|q\rangle$ of the position eigenstates of $\hat{q}$.

Given this basis, once again one may consider the sum over all basis vectors of the
associated projection operators $|q\rangle\langle q|$, which is to reproduce the
identity operator,
\begin{equation}
\mathbb{I}=\int_{-\infty}^{+\infty}\,dq\,|q\rangle\,\langle q|.
\label{J1.eq:spectral2}
\end{equation}
That this expression is indeed consistent may easily be checked by applying it
onto any of the position eigenstates, $|q\rangle=\mathbb{I}|q\rangle$,
\begin{equation}
\mathbb{I}|q\rangle=\int_{-\infty}^{+\infty}dq'\,|q'\rangle\,\langle q'|q\rangle=
\int_{-\infty}^{+\infty}dq'\,|q'\rangle\,\delta(q-q')=|q\rangle,
\end{equation}
indeed as it should. Being true for all basis vectors $|q\rangle$, by linearity
in Hilbert space it is true for any quantum state, hence (\ref{J1.eq:spectral2})
does apply.

But then for any quantum state $|\psi\rangle$ we have
\begin{equation}
|\psi\rangle=\mathbb{I}|\psi\rangle=
\int_{-\infty}^{+\infty}\,dq\,|q\rangle\,\langle q|\psi\rangle=
\int_{-\infty}^{+\infty}\,dq\,|q\rangle\,\psi(q).
\end{equation}
This identity thus provides the decomposition of the state $|\psi\rangle$ in the basis
$|q\rangle$ in terms of its components $\langle q|\psi\rangle$ which define
{\bf the configuration space wave function} of the state $|\psi\rangle$,
\begin{equation}
\psi(q)=\langle q|\psi\rangle ,
\end{equation}
indeed a complex valued quantity.

As a consequence, it also becomes possible to determine the representation of the
operators $\hat{q}$ and $\hat{p}$ acting on any state $|\psi\rangle$ through their matrix 
element in the configuration space basis $|q\rangle$. A simple analysis using the
Heisenberg algebra\footnote{See for instance Ref.\cite{Gov1.GovBook},
even though there exist alternative derivations of these results, all of which have as
starting point the Heisenberg algebra commutator.} then finds
\begin{equation}
\langle q|\hat{q}|\psi\rangle=q\,\langle q|\psi\rangle=q\,\psi(q),\qquad
\langle q|\hat{p}|\psi\rangle=-i\hbar\frac{d\psi(q)}{dq}.
\end{equation}
In other words, when using the configuration space wave function representation $\psi(q)$
of quantum states, the action of the quantum operators $\hat{q}$ and $\hat{p}$ is through multiplication
by $q$ in the first case, and differentiation with respect to $q$ as well as multiplication
by $(-i\hbar)$ in the second case, of the wave function $\psi(q)$. That these operations
now acting on functions (which is indeed a set spanning a vector space over the complex numbers)
define a representation of the abstract Heisenberg algebra follows from their commutator
as differential operators,
\begin{equation}
\left[q,-i\hbar\frac{d}{dq}\right]=i\hbar.
\end{equation}

Furthermore the inner product of quantum states, $\langle \psi|\chi\rangle$, then also translates
into an integral expression in terms of the associated configuration space wave functions, which
indeed defines an inner product over the space of functions possessing all the required properties
for a Hilbert space. Namely, using once again the spectral decomposition of the identity operator
(\ref{J1.eq:spectral2}), it follows
\begin{equation}
\langle\psi|\chi\rangle=\langle\psi|\mathbb{I}|\chi\rangle=
\langle\psi|\int_{-\infty}^{+\infty}dq|q\rangle\langle q|\chi\rangle=
\int_{-\infty}^{+\infty}dq\,\langle\psi|q\rangle\,\langle q|\chi\rangle=
\int_{-\infty}^{+\infty}dq\,\psi^*(q)\,\chi(q),
\label{J1.eq:innerproduct1}
\end{equation}
with $\psi(q)=\langle q|\psi\rangle$ and $\chi(q)=\langle q|\chi\rangle$. The abstract Hilbert
space of the Heisenberg algebra is thereby represented in terms of the space $L^2(\mathbb{R},dq)$
of square integrable configuration space wave functions,
\begin{equation}
\int_{-\infty}^{+\infty}dq\,|\psi(q)|^2\,<\,\infty,
\end{equation}
so that they may be divided by their finite norm and be of norm unity for the inner product
(\ref{J1.eq:innerproduct1}), while the abstract operators $\hat{q}$ and $\hat{p}$ are then
represented by the functional operators as specified above.

\vspace{10pt}

\noindent
b) {\sl The momentum space wave function representation}

\vspace{10pt}

It should be quite obvious that a story similar to the one above
also applies to a momentum space wave function
representation of the Heisenberg algebra. For instance, under the exchange of both operators
$\hat{q}$ and $\hat{p}$ and a change of sign in $\hbar$ (a ``duality" transformation in a certain
sense), the Heisenberg algebra remains invariant. Consequently all the above considerations
and results translate into corresponding ones for a momentum space wave function representation.

For that purpose, let us again assume that, since the operator $\hat{p}$ is self-adjoint, there
exists a basis of momentum eigenstates $|p\rangle$ of which the eigenspectrum is the real line
which is also the range of the classical conjugate momentum variable in phase space,
\begin{equation}
\hat{p}\,|p\rangle=p\,|p\rangle,\qquad p\in\mathbb{R}.
\end{equation}
Even though this leaves open still the phase of the states $|p\rangle$, their normalisation
may be specified once again through their inner products, which should be proportional
to the Dirac $\delta$-function in momentum space this time, and for which we choose again
a normalisation which is that of an orthonormalised basis,
\begin{equation}
\langle p|p'\rangle=\delta(p-p').
\end{equation}
Consequently the spectral decomposition of the identity operator reads
\begin{equation}
\mathbb{I}=\int_{-\infty}^{+\infty} dp\,|p\rangle\,\langle p|,
\end{equation}
as is confirmed by applying this identity onto any of the the momentum
eigenstates $|p\rangle$ as was discussed above for the
configuration space eigenbasis. In particular, abstract quantum states $|\psi\rangle$
are then represented in terms of a complex wave function over momentum space\footnote{The tilde
above the wave function symbol in momentum space is included to avoid confusing that wave
function with the configuration space one.}, $\widetilde{\psi}(p)$,
which specifies the components of the state $|\psi\rangle$ in the momentum eigenbasis,
\begin{equation}
|\psi\rangle=\int_{-\infty}^{+\infty}dp\,|p\rangle\,\langle p|\psi\rangle=
\int_{-\infty}^{+\infty} dp\,|p\rangle\,\widetilde{\psi}(p),\qquad
\widetilde{\psi}(p)=\langle p|\psi\rangle.
\end{equation}

Furthermore the abstract operators $\hat{q}$ and $\hat{p}$ now also acquire realisations
in terms of functional operators acting on the momentum wave function, through
\begin{equation}
\langle p|\hat{q}|\psi\rangle = +i\hbar\frac{d\widetilde{\psi}(p)}{dp},\qquad\qquad
\langle p|\hat{p}|\psi\rangle = p\,\widetilde{\psi}(p).
\end{equation}
Namely $\hat{q}$ is now represented by the derivative of the wave function with respect
to $p$ and multiplied by $(i\hbar)$ while $\hat{p}$ is realised simply as multiplication
of the wave function by $p$. In particular, one may check that these two functional
operators do indeed obey the Heisenberg algebra on the vector space of momentum wave functions
$\widetilde{\psi}(p)$,
\begin{equation}
\left[i\hbar\frac{d}{dp},p\right]=i\hbar.
\end{equation}

The vector space of such wave functions is equipped with an inner product which possesses
all the required properties of sesquilinearity, hermiticity and positive definiteness.
By making use once again of the spectral decomposition of the identity operator in momentum space,
the inner product of any two states $|\psi\rangle$ and $|\chi\rangle$ represented
by their momentum wave functions $\widetilde{\psi}(p)$ and $\widetilde{\chi}(p)$ is simply
\begin{equation}
\langle\psi|\chi\rangle=\int_{-\infty}^{+\infty} dp\,\langle\psi|p\rangle\,\langle p|\chi\rangle
=\int_{-\infty}^{+\infty} dp \,\widetilde{\psi}^*(p)\,\widetilde{\chi}(p).
\end{equation}
In particular, and to be precise, the actual Hilbert space (in the strict mathematical sense)
consists of all those square integrable wave functions, namely those of finite norm so that
the associated states may be normalised to unity,
\begin{equation}||\psi||^2=\langle\psi|\psi\rangle=
\int_{-\infty}^{+\infty} dp|\widetilde{\psi}(p)|^2<\infty.
\end{equation}

\vspace{10pt}

\noindent
c) {\sl Change of basis}

\vspace{10pt}

Given the statement that, up to unitary transformations (namely in the case of each of the above
two wave function representations, the fact that the position or momentum eigenstates, hence
also the wave functions are only defined up to local phase factors), there exists a single
representation of the abstract Heisenberg algebra, we know that the Hilbert spaces
realised by the above two wave function representations are in fact identical. In other
words, the two bases of vectors, $|q\rangle$ and $|p\rangle$, that have been identified
provide different bases within the same abstract Hilbert space as a vector space.
Since both are orthonormalised bases, there should exist a unitary transformation relating
these two bases and transforming the two classes of wave function representations for a
same abstract quantum state $|\psi\rangle$. By analogy with the situation for an ordinary
finite dimensional vector space over the real numbers, it is clear that knowing the
decomposition of one set of basis vectors in terms of the other basis is all that is
required to determine the unitary transformation. But such a decomposition amounts to
determining the projections of one set of basis vectors onto those of the other set,
namely in the present case determine the quantities
\begin{equation}
\langle q|p\rangle,\qquad \langle p|q\rangle=\langle q|p\rangle^* .
\end{equation}
For instance, $\langle q|p\rangle$ stands for the configuration space wave function
of the momentum eigenstate $|p\rangle$, namely a function of $q$ for a fixed value of $p$.
In order to determine this function, let us establish a differential equation for it
by considering the matrix element
\begin{equation}
\langle q|\hat{p}|p\rangle = p\,\langle q|p\rangle,
\end{equation}
which indeed produces the sought for function $\langle q|p\rangle$. On the other hand
since that matrix element also defines the realisation of the abstract operator
$\hat{p}$ in the configuration space representation, one has
\begin{equation}
\langle q|\hat{p}|p\rangle=-i\hbar\frac{d\langle q|p\rangle}{dq}.
\end{equation}
We have thus established the differential equation
\begin{equation}
\frac{d\langle q|p\rangle}{dq}=\frac{i}{\hbar}\,p\,\langle q|p\rangle,
\end{equation}
of which the solution is
\begin{equation}
\langle q|p\rangle=N\,e^{\frac{i}{\hbar}qp},
\end{equation}
$N$ being some complex normalisation factor. However one should have,
for instance,
\begin{equation}
\delta(q-q')=\langle q|q'\rangle=\int_{-\infty}^{+\infty}\,dp\,
\langle q|p\rangle\,\langle p|q'\rangle=
|N|^2\,\int_{-\infty}^{+\infty}\,dp\,e^{\frac{i}{\hbar}p(q-q')}=
2\pi\hbar|N|^2\,\delta(q-q').
\end{equation}
In conclusion, up to an arbitrary phase factor set to unity here, we have
$N=(2\pi\hbar)^{-1/2}$, and finally
\begin{equation}
\langle q|p\rangle=\frac{1}{\sqrt{2\pi\hbar}}\,e^{\frac{i}{\hbar} qp},\qquad
\langle p|q\rangle=\langle q| p\rangle^*=\frac{1}{\sqrt{2\pi\hbar}}\,e^{-\frac{i}{\hbar}qp}.
\end{equation}

Having established the change of bases for the position and momentum eigenstates, let us
turn to the unitary transformation of the wave functions themselves for an arbitrary state $|\psi\rangle$.
Using once again the spectral decompositions of the identity operator, one finds
\begin{equation}
\widetilde{\psi}(p)=\langle p|\psi\rangle=
\int_{-\infty}^{+\infty}dq\,\langle p|q\rangle\,\langle q|\psi\rangle=
\int_{-\infty}^{+\infty}\frac{dq}{\sqrt{2\pi\hbar}}\,e^{-\frac{i}{\hbar}qp}\,\psi(q),
\end{equation}
\begin{equation}
\psi(q)=\langle q|\psi\rangle=
\int_{-\infty}^{+\infty}dp\,\langle q|p\rangle\langle p|\psi\rangle=
\int_{-\infty}^{+\infty}\frac{dp}{\sqrt{2\pi\hbar}}\,e^{\frac{i}{\hbar}qp}\,\widetilde{\psi}(p).
\end{equation}
In these expressions one recognises the Fourier transformation formulae of a complex function.
Hence configuration and momentum wave functions are related through the above Fourier transforms,
while this connection between the two finds its origin in the abstract Heisenberg algebra
which underlies the whole discussion. In the pure imaginary exponential factors for
$\langle q|p\rangle$ and $\langle p|q\rangle$ one also recognises the ordinary ``plane wave"
factors, which have to do with the behaviour of quantum states under translations  in phase space
in either $q$ or $p$, namely a symmetry group. We shall thus come back to this point when discussing
symmetries and the (first) Noether theorem.

\vspace{10pt}

\noindent
\underline{\bf Remark}: {\bf The Heisenberg uncertainty relation}

\vspace{10pt}

As already mentioned, given an operator $\hat{A}(t_0)$ and a quantum state $|\psi,t\rangle$
(both, say, in the Schr\"odinger picture), the expectation value of that observable is defined as
\begin{equation}
\langle\hat{A}\rangle(t)=\frac{\langle\psi,t|\hat{A}(t_0)|\psi,t\rangle}
{\langle\psi,t|\psi,t\rangle}.
\end{equation}
In the case of the Heisenberg operators $\hat{q}$ and $\hat{p}$, let us then introduce
the following quantities. First the expectation values or mean values for both the
position and the conjugate momentum of the state,
\begin{equation}
\bar{x}(t)=\langle\hat{x}\rangle(t),\qquad
\bar{p}(t)=\langle\hat{p}\rangle(t),
\end{equation}
and next the mean values for the variations from these means,
\begin{equation}
\left(\Delta x\right)^2(t)=\langle\left(\hat{x}-\bar{x}\right)^2\rangle(t),\qquad
\left(\Delta p\right)^2(t)=\langle\left(\hat{p}-\bar{p}\right)^2\rangle(t),
\end{equation}
with $\Delta x(t)>0$ and $\Delta p(t)>0$. Heisenberg's uncertainty relation in this case
states that one always has
\begin{equation}
\Delta x\,\Delta p\ge\,\frac{1}{2}\hbar,
\end{equation}
for whatever quantum state $|\psi,t\rangle$, the inequality being saturated only for
specific types of states. This inequality is in direct relation with the Heisenberg
commutator $[\hat{q},\hat{p}]=i\hbar$. Thus any other pair of canonically conjugated
observables with the same commutation relation will also obey that uncertainty
relation, which may also be extended into a more general form given any commutator. As a consequence
in the context of quantum physics canonically conjugated variables may no longer
be known both to arbitrarily good precision, the product of the intrinsic uncertainties inherent
to their quantum noncommutativity being bounded below essentially by Planck's constant.

\subsubsection{The nonrelativistic quantum particle}

Let us apply the previous general discussion now to the nonrelativistic particle of mass $m$
subjected to conservative forces of total potential energy $V(\vec{r}\,)$, of which the Lagrange
function is
\begin{equation}
L=\frac{1}{2}m\dot{\vec{r}}\,^2\,-\,V(\vec{r}\,)
\end{equation}
(a generalisation to an arbitrary number of distinct particles is straightforward).
The Hamiltonian of the system is
\begin{equation}
H=\frac{1}{2m}\vec{p}\,^2\,+\,V(\vec{r}\,),
\end{equation}
with the canonical phase space variables $(\vec{r},\vec{p}\,)=(x^i,p_j)$ ($i,j=1,2,3$)
of which the Poisson brackets are $\left\{x^i,p_j\right\}=\delta^i_j$.

Hence at the quantum level we simply have the Heisenberg algebra
\begin{equation}
\left[\hat{x}^i,\hat{p}_j\right]=i\hbar\,\delta^i_j,\qquad
\left(\hat{x}^i\right)^\dagger=\hat{x}^i,\quad \left(\hat{p}_i\right)^\dagger=\hat{p}_i,
\end{equation}
as well as the quantum Hamiltonian
\begin{equation}
\hat{H}=\frac{1}{2m}\hat{\vec{p}}\,^2\,+\,V(\hat{\vec{r}}\,),
\end{equation}
which is in direct correspondence with the classical Hamiltonian and
is obviously also self-adjoint if the position and momentum operators are
themselves self-adjoint, $\hat{H}^\dagger=\hat{H}$.

The Hilbert space of this system is simply the 3-fold tensor product of the
Hilbert space of the single degree of freedom Heisenberg algebra constructed
above. Diagonalisation of the Hamiltonian over that space depends on the choice
of the potential energy, namely the forces to which the particle is subjected.
In the case of harmonic forces for which $V(\vec{r}\,)$ is some given but otherwise
arbitrary positive definite quadratic polynomial of the cartesian coordinates $x^i$,
a purely algebraic solution in terms of the Fock space representation of the
Heisenberg algebra is readily established. However in a general case for which algebraic
methods are not available, one approach to solving the eigenvalue problem is by
considering the Schr\"odinger equation of wave quantum mechanics. Namely, consider
the abstract Schr\"odinger equation in the Schr\"odinger picture,
\begin{equation}
\hat{H}\,|\psi,t\rangle\,=\,i\hbar\,\frac{d|\psi,t\rangle}{dt},
\end{equation}
and project it onto the configuration space eigenstates, leading to the
Schr\"odinger equation for the configuration space wave function,
$\psi(t,\vec{r}\,)=\langle\vec{r}\,|\psi,t\rangle$,
of the quantum particle,
\begin{equation}
\left[-\frac{\hbar^2}{2m}\vec{\nabla}^2\,+\,V(\vec{r}\,)\right]\,\psi(t,\vec{r}\,)=
i\hbar\,\frac{\partial\psi(t,\vec{r}\,)}{\partial t},
\end{equation}
in which the correspondences $\hat{\vec{r}}\rightarrow \vec{r}$
and $\hat{\vec{p}}\rightarrow -i\hbar\vec{\nabla}$ have been applied. For a given potential
energy this differential equation lends itself to methods of a more analytical
or even numerical character. Note that by considering an expansion of the
general state into the energy eigenbasis of which the spectrum is to be found
and which may include components which are both discrete and continuous (such as
the spectrum of the hydrogen atom with its discrete but yet infinite bound state
spectrum and its infinite continuous spectrum of scattering or unbound states),
as was done in the Fock basis in the case of the one dimensional harmonic oscillator,
it suffices in fact to consider the energy eigenvalue problem for the Schr\"odinger
equation, namely the so-called stationary Schr\"odinger equation,
\begin{equation}
\left[-\frac{\hbar^2}{2m}\vec{\nabla}\,^2\,+\,V(\vec{r}\,)\right]\,\psi_E(\vec{r}\,)=
E\,\psi_E(\vec{r}\,).
\end{equation}
Corresponding to any of the eigenvalues $E$ of this differential eigenvalue problem,
the solution to the original Schr\"odinger equation is then, up to an arbitrary
constant phase factor both in time and space,
\begin{equation}
\psi_E(t,\vec{r}\,)=\psi_E(\vec{r}\,)\,e^{-\frac{i}{\hbar}E\,(t-t_0)},
\end{equation}
showing that this state is indeed stationary since its only time dependence
is through a simple phase factor linear in time and with as coefficient the
energy measured in units of $\hbar$, namely an angular frequency. The general
solution to the full original Schr\"odinger equation is then constructed through
the most general possible linear combination of all stationary states. Indeed
given the energy eigenspectrum of the abstract Hamiltonian operator,
$\hat{H}|E_m\rangle=E_m|E_m\rangle$, we had established in the general case that
the general solution to the abstract Schr\"odinger equation in the Schr\"odinger
picture is of the form,
\begin{equation}
|\psi,t\rangle=\sum_m\,|E_m\rangle\,e^{-\frac{i}{\hbar}(t-t_0)E_m}\,
\langle E_m|\psi,t_0\rangle.
\end{equation}
It suffices to project this relation onto the position eigenstates $|\vec{r}\,\rangle$
to obtain the same statement in terms of the configuration space wave function
of the general quantum state of the system,
\begin{equation}
\psi(t,\vec{r}\,)=\langle\vec{r}\,|\psi,t\rangle=
\sum_m\,\psi_{E_m}(\vec{r}\,)\,e^{-\frac{i}{\hbar}E_m(t-t_0)}\,
\langle E_m|\psi,t_0\rangle.
\end{equation}

\vspace{10pt}

\noindent
\underline{\bf Remark}

\vspace{10pt}

\noindent
It is also possible to write the Schr\"odinger equation in the
Schr\"odinger picture in the momentum space wave function representation of the
Heisenberg algebra. It should be quite clear that this equation reads,
\begin{equation}
\left[\frac{1}{2m}\vec{p}\,^2\,+\,V\left(i\hbar\vec{\nabla}\,\right)\right]\,
\widetilde{\psi}(t,\vec{p}\,)=i\hbar\frac{\partial\widetilde{\psi}(t,\vec{r}\,)}
{\partial t}.
\end{equation}
Depending on the considered system and the issues to be solved, one of these different
representations of the same abstract Schr\"odinger equation may be more convenient
to use than the others.

\vspace{10pt}

\noindent
\underline{\bf Application}: {\bf The free nonrelativistic particle}

\vspace{10pt}

\noindent
In the case of the free particle the potential energy is vanishing
(or an arbitrary constant, leading to an arbitrary constant phase redefinition of the quantum
states). It is best to consider the Schr\"odinger
equation (in the Schr\"odinger picture) in the momentum representation since
it is obvious that the energy eigenstates are then simply the momentum eigenstates,
$|\vec{p}\,\rangle$, with
\begin{equation}
\hat{H}\,|\vec{p}\,\rangle=E(\vec{p}\,)\,|\vec{p}\,\rangle,\qquad
E(\vec{p}\,)=\frac{1}{2m}\vec{p}\,^2.
\end{equation}
Consequently the stationary solutions to the configuration space Schr\"odinger
equation are
\begin{equation}
\psi_{\vec{p}}(t,\vec{r}\,)=\langle\vec{r}\,|\vec{p},t\rangle=\frac{1}{(2\pi\hbar)^{3/2}}\,
e^{\frac{i}{\hbar}\vec{r}\cdot\vec{p}}\,e^{-\frac{i}{\hbar}(t-t_0)\,E(\vec{p}\,)}.
\end{equation}

It is also possible to consider the Schr\"odinger equation in the Heisenberg picture.
It should be quite clear that the solution to that equation in the case of the
position and momentum operators is of the form
\begin{equation}
\hat{\vec{r}}(t)=\hat{\vec{r}}(t_0)+\frac{1}{m}\hat{\vec{p}}(t_0)\,(t-t_0),\qquad
\hat{\vec{p}}(t)=\hat{\vec{p}}(t_0),
\end{equation}
which are indeed the operator solutions in direct correspondence with their
classical counterparts in the Hamiltonian formulation of the same system.
Once again we notice that when the full quantum dynamics may be solved, the
Heisenberg picture of that dynamics is in direct correspondence with the
classical solutions to the Hamiltonian first order equations of motion for the
observables.

\subsubsection{The one dimensional harmonic oscillator}

In the case of this system the configuration space Schr\"odinger equation (in the
Schr\"odinger picture) reads, given the potential energy $V(q)=m\omega^2 q^2/2$,
\begin{equation}
\left[-\frac{\hbar^2}{2m}\frac{\partial^2}{\partial q^2}\,+\,\frac{1}{2}m\omega^2 q^2\right]\,
\psi(t,q)=i\hbar\frac{\partial\psi(t,q)}{\partial t}.
\end{equation}
Even the Schr\"odinger equation for stationary states,
\begin{equation}
\left[-\frac{\hbar^2}{2m}\frac{d^2}{dq^2}\,+\,\frac{1}{2}m\omega^2\,q^2\right]
\psi_n(q)=E_n\,\psi_n(q),
\end{equation}
where an index $n$ distinguishing energy eigenvalues has already been introduced,
is not the most appealing. Solving this differential eigenvalue problem is not
readily achieved, and has produced historically a certain class of special functions, namely
the Hermite polynomials.

However, let us show how the knowledge of the purely algebraic solution based on
Fock space techniques allows a direct resolution of the above differential equation.
First, we already know that the spectrum of energy eigenvalues is given by
\begin{equation}
E_n=\hbar\omega\left(n+\frac{1}{2}\right),\qquad
n=0,1,2,\ldots
\end{equation}
To each of these eigenvalues there thus corresponds an eigen-wave function $\psi_n(q)=\langle q|n\rangle$
for the associated stationary state $\psi_n(t,q)=e^{-\frac{i}{\hbar}(t-t_0)E_n}\,\psi_n(q)$,
which is nothing else but the configuration space wave function of the Fock state $|n\rangle$.
In other words, solving the problem of determining these eigenfunctions $\psi_n(q)$ amounts
also to establishing the change of basis between the Fock basis of the abstract Hilbert space
providing the representation (up to unitary equivalence) and the configuration space
basis of the position eigenstates $|q\rangle$.

In order to identify the functions $\psi_n(q)$, let us return to the defining property of
the Fock states, beginning with the Fock vacuum which is annihilated by the operator $a$,
$a|0\rangle=0$. Hence, projecting that relation onto a position eigenstate $|q\rangle$, we have
\begin{equation}
\langle q|a|0\rangle=0.
\end{equation}
However, in terms of the abstract operators we have
\begin{equation}
a=\sqrt{\frac{m\omega}{2\hbar}}\left[\hat{q}+\frac{i}{m\omega}\hat{p}\right],
\end{equation}
so that when acting on configuration space wave functions the abstract operator $a$ is
realised by the functional operator
\begin{equation}
a:\qquad \sqrt{\frac{m\omega}{2\hbar}}\left[q+\frac{\hbar}{m\omega}\frac{d}{dq}\right].
\end{equation}
Consequently the above property for the Fock vacuum, $\langle q|a|0\rangle=0$, translates into
the following differential equation for the configuration space wave function of the
ground state, $\psi_0(q)=\langle q|0\rangle$,
\begin{equation}
\left[q+\frac{\hbar}{m\omega}\frac{d}{dq}\right]\,\psi_0(q)=0,
\end{equation}
of which the solution is
\begin{equation}
\psi_0(q)=N\,e^{-\frac{m\omega}{2\hbar}\,q^2},
\end{equation}
$N$ being some complex normalisation constant. The latter is fixed by the normalisation
condition of the Fock vacuum,
\begin{equation}
1=\langle 0|0\rangle=\int_{-\infty}^{+\infty}dq\,\langle 0|q\rangle\,\langle q|0\rangle=
\int_{-\infty}^{+\infty}\,dq\,|\langle q|0\rangle|^2=
|N|^2\,\int_{-\infty}^{+\infty}\,dq\,e^{-\frac{m\omega}{\hbar}\,q^2}=
|N|^2\sqrt{\frac{\pi\hbar}{m\omega}}.
\end{equation}
Hence, up to an arbitrary choice of phase set to unity once again ({\it i.e.\/}, up a
unitary transformation in Hilbert space), one has finally the configuration space
wave function of the Fock vacuum,
\begin{equation}
\psi_0(q)=\langle q|0\rangle =\left(\frac{m\omega}{\pi\hbar}\right)^{1/4}\,
e^{-\frac{m\omega}{2\hbar}\,q^2}.
\end{equation}

In order to also determine the wave functions for the excited Fock states,
\begin{equation}
\psi_n(q)=\langle q|n\rangle=\frac{1}{\sqrt{n!}}\,
\langle q|\left(a^\dagger\right)^n|0\rangle,
\end{equation}
we now need to consider the functional operator realisation of the abstract creation operator,
$a^\dagger$, in the configuration space representation of the Heisenberg algebra. From the
above discussion in the case of the annihilation operator $a$, it is clear that we have the
correspondence,
\begin{equation}
a^\dagger:\qquad
\sqrt{\frac{m\omega}{2\hbar}}\left[q-\frac{\hbar}{m\omega}\frac{d}{dq}\right].
\end{equation}
Consequently,
\begin{equation}
\psi_n(q)=\frac{1}{\sqrt{n!}}\,\left(\frac{m\omega}{2\hbar}\right)^{n/2}\,
\left[q-\frac{\hbar}{m\omega}\frac{d}{dq}\right]^n\,
\left(\frac{m\omega}{\pi\hbar}\right)^{1/4}\,
e^{-\frac{m\omega}{2\hbar}\,q^2},
\end{equation}
in which the function $\psi_0(q)=\langle q|0\rangle$ has already been substituted by its solution
established previously. This expression is reminiscent of one possible definition of the Hermite polynomials, namely
\begin{equation}
H_n(x)=e^{\frac{1}{2}x^2}\,\left[x-\frac{d}{dx}\right]^n\,e^{-\frac{1}{2}x^2},\quad n=0,1,2,\ldots
\end{equation}
Hence, by an appropriate rescaling of the coordinate $q$, the above expression for $\psi_n(q)$
may be brought into the form of this definition of the Hermite polynomials. All factors
combine to give the following final expression for the configuration space wave functions of the
energy eigenstates of the harmonic oscillator,
\begin{equation}
\psi_n(q)=\left(\frac{m\omega}{\pi\hbar}\right)^{1/4}\,
\frac{1}{\sqrt{2^n\,n!}}\,e^{-\frac{m\omega}{2\hbar}q^2}\,
H_n\left(q\sqrt{\frac{m\omega}{\hbar}}\right).
\end{equation}
As indicated already previously in the general case, from the knowledge of these stationary
solutions one may construct the expression for the general solution to the Schr\"odinger
equation (in the Schr\"odinger picture) in the configuration space representation of the
Heisenberg algebra. Incidentally, since the Hamiltonian is quadratic in both $p$ and $q$,
the solution in the momentum space wave function representation is again constructed
in terms of the Hermite polynomials evaluated for a rescaling of the conjugate momentum variable $p$.
The Fourier transformation of the above products of the Hermite polynomials with
the Gaussian factor included produces again similar products of the same Hermite polynomial with
a common Gaussian factor, but of course now as a function of $p$ rather than $q$ including
some appropriate dimensionful scaling parameters as displayed for instance in the above explicit expression
for the configuration space solutions.

\subsection{The path integral representation and quantisation}

It is well known that besides the canonical quantisation path,
there is another royal avenue towards the quantisation of a classical
system whose dynamics is defined through some action and the variational
principle, namely the so-called path integral or functional integral 
formulation of quantum mechanics. Here we shall discuss how,
starting from the canonical quantisation of any such system following
the approach outlined in the previous Sections, it is possible to set up
integral representations for matrix elements of quantum operators, which
acquire the interpretation of functional integrals over phase space.
When reducing from these integrals the conjugate momentum degrees of
freedom, one recovers a functional integral over configuration space
in which the original classical action expressed in terms of the
Lagrange function plays again a central r\^ole. Further remarks as to
quantisation directly through the functional integral are made
at the end of this discussion. It should already be clear
that these two approaches are complementary, each with its own
advantages and difficulties both with respect to an intuitive 
understanding of the physics that they both encode as well as to the
calculational advantages of one compared to the other. However, when properly 
implemented, they represent in complementary ways an identical
physical content.

The procedure for constructing an integral representation for matrix
elements of operators, starting from canonical quantisation, follows
essentially always the same avenue, based on the insertion of
complete sets of states in terms of which the unit operator possess
a spectral resolution. Here, we shall illustrate this feature for
the configuration and momentum space representations of the Heisenberg algebra,
even though more general cases may be envisaged as well. In a later
Section such an illustration will be provided in terms of so-called coherent states,
to be introduced hereafter. Furthermore,
we shall consider configuration space matrix elements of the evolution
operator for a given quantum system, namely the propagator
$\langle q_f|U(t_f,t_i)|q_i\rangle$ of the system
(in configuration space\footnote{It would be an excellent exercise
to establish a path integral representation for the momentum space
matrix elements of the same operator.}). Indeed, the physical meaning of this
quantity is that it measures the overlap with the position eigenstate $|q_f\rangle$
of the time evolved position eigenstate $|q_i\rangle$ over the time interval $(t_f-t_i)$,
namely the probability amplitude for finding the system, initially localised at $q=q_i$
at time $t=t_i$, at the position $q=q_f$ at time $t=t_f$: this is indeed the probability amplitude for
propagating the system in configuration space for a given time interval.

The quantum evolution operator may also be expressed as the product of such
operators representing the evolution of the system through a succession of time slices,
\begin{equation}
U(t_f,t_i)=e^{-\frac{i}{\hbar}(t_f-t_i)\hat{H}}=
\left[e^{-\frac{i}{\hbar}\epsilon\hat{H}}\right]^N=
\lim_{N\rightarrow\infty}\left[1-\frac{i}{\hbar}\epsilon\hat{H}\right]^N,
\end{equation}
with
\begin{equation}
\epsilon=\frac{t_f-t_i}{N}=\frac{\Delta t}{N},\qquad \Delta t=t_f-t_i,
\end{equation}
while $N$ is some arbitrary positive
integer specifying an equally spaced slicing of the finite time
interval $(t_f-t_i)$. In what follows, the $n$ index for the degrees
of freedom $(q^n,p_n)$ is suppressed, to keep expressions as
transparent as possible. Given this time sliced form of the evolution
operator, the idea now is to insert twice the spectral resolution of the
unit operator $\mathbb{I}$, once in terms of the position eigenstates,
and once in terms of the momentum eigenstates, and this in between each of
the $N$ factors that appear in the above $N$ factorised form for
$U(t_f,t_i)$, as follows,
\begin{equation}
\mathbb{I}=\int_{-\infty}^{+\infty}\,dp_\alpha\,\int_{-\infty}^{+\infty}\,
dq_{\alpha+1}\,|q_{\alpha+1}\rangle\langle q_{\alpha+1}|p_\alpha\rangle\langle p_\alpha|
,\qquad \alpha=0,1,2,\ldots,N-2.
\end{equation}
Setting then $q_f=q_{\alpha=N}$ and $q_i=q_{\alpha=0}$, a straightforward
substitution into the considered matrix element leads to the expression
(a substitution of the unit operator as
$\mathbb{I}=\int_{-\infty}^{+\infty} dp|p\rangle\langle p|$ is also
performed to the right of the external final state $\langle q_f|$, leading
to one more integration over the $p_\alpha$'s than over the $q_\alpha$'s),
\begin{equation}
\langle q_f|U(t_f,t_i)|q_i\rangle=\int_{-\infty}^{+\infty}\prod_{\alpha=1}^{N-1}
dq_\alpha\,\prod_{\alpha=0}^{N-1}dp_\alpha\,
\prod_{\alpha=0}^{N-1}\left[\langle q_{\alpha+1}|p_\alpha\rangle
\langle p_\alpha|e^{-\frac{i}{\hbar}\epsilon\hat{H}}|q_\alpha\rangle\right].
\end{equation}
Using then the value for the matrix element $\langle q|p\rangle$ given previously, 
this quantity finally reduces to,
\begin{equation}
\langle q_f|U(t_f,t_i)|q_i\rangle = \lim_{N\rightarrow\infty}\int_{-\infty}^{+\infty}\prod_{\alpha=1}^{N-1}
dq_\alpha\,\prod_{\alpha=0}^{N-1}\frac{dp_\alpha}{2\pi\hbar}\,
{\rm exp}\left\{\frac{i}{\hbar}\sum_{\alpha=0}^{N-1}\epsilon
\left[\frac{q_{\alpha+1}-q_\alpha}{\epsilon}p_\alpha-h_\alpha\right]\right\},
\label{J1.eq:PI1}
\end{equation}
with the Hamiltonian matrix elements
\begin{equation}
h_\alpha=\frac{\langle p_\alpha|\hat{H}|q_\alpha\rangle}
{\langle p_\alpha|q_\alpha\rangle }.
\end{equation}
Clearly, the discretised integral representation (\ref{J1.eq:PI1}) of
the configuration space propagator corresponds to a specific construction
of the otherwise formal expression for the phase space path integral
or functional integral corresponding to that quantity, namely the
following integral over the space of functions $q(t)$ and $p(t)$,
\begin{equation}
\langle q_f|U(t_f,t_i)|q_i\rangle=
\int_{q(t_i)=q_i}^{q(t_f)=q_f}
\left[{\cal D}q\frac{{\cal D}p}{2\pi\hbar}\right]\,
e^{\frac{i}{\hbar}S[q,p]},
\label{J1.eq:formalPI1}
\end{equation}
in which the phase space action is that of the first-order Hamiltonian 
formulation of the system, namely
\begin{equation}
S[q,p]=\int_{t_i}^{t_f}dt\left[\dot{q}p-H(q,p)\right],
\end{equation}
which is that associated to the choice of boundary conditions corresponding
to the configuration space propagator when imposing the variational
principle in a strong sense, namely with the induced boundary terms also
required to vanish through the boundary conditions $q(t_{i,f})=q_{i,f}$. 
Note that contrary to what the formal
expression (\ref{J1.eq:formalPI1}) may lead one to believe, the integration
measure is not quite the phase space Liouville measure, since in fact
there is always one more $p_\alpha$ integration than the number of $q_\alpha$ 
integrations. One should always keep this remark in mind when developing
formal arguments based on the formal expression (\ref{J1.eq:formalPI1})
of the functional integral.

Considering the momentum space matrix elements of the same
operator, a similar analysis leads to an analogous specific discretised
expression, namely
\begin{equation}
\langle p_f|U(t_f,t_i)|p_i\rangle=
\lim_{N\to\infty}\int_{-\infty}^{+\infty}\prod_{\alpha=1}^{N-1}dp_\alpha\,
\prod_{\alpha=0}^{N-1}\frac{dq_\alpha}{2\pi\hbar}\,
{\rm exp}\,\left\{\frac{i}{\hbar}\sum_{\alpha=0}^{N-1}\epsilon
\left[-q_\alpha\frac{p_{\alpha+1}-p_\alpha}{\epsilon}-h_\alpha\right]\right\},
\end{equation}
with $h_\alpha={\langle q_\alpha|\hat{H}|p_\alpha\rangle}/{\langle q_\alpha|p_\alpha\rangle}$,
corresponding to the formal quantity,
\begin{equation}
\langle p_f|U(t_f,t_i)|p_i\rangle=
\int_{p(t_i)=p_i}^{p(t_f)=p_f}
\left[\frac{{\cal D}q}{2\pi\hbar}{\cal D}p\right]\,
e^{\frac{i}{\hbar}S[q,p]},
\label{J1.eq:formalPI2}
\end{equation}
where the appropriate Hamiltonian first-order action now reads
\begin{equation}
S[q,p]=\int_{t_i}^{t_f}\,dt\left[-q\dot{p}-H(q,p)\right],
\end{equation}
being this time associated to the choice of boundary conditions
$p(t_{i,f})=p_{i,f}$ as opposed to $q(t_{i,f})=q_{i,f}$ for the
propagator in configuration space. Note that the same remark as above
concerning the phase space Liouville measure applies here as well.

In the particular situation that the Hamiltonian is such that the
matrix elements $h_\alpha$ are quadratic in the momenta,
\begin{equation}
h_\alpha=\frac{p^2_\alpha}{2m}+V(q_\alpha),
\end{equation}
which is the case when the quantum Hamiltonian is of the form
$\hat{H}=\hat{p}^2/2m+V(\hat{q})$,
the integration over momentum space may be completed explicitly
in the above discretised expressions\footnote{Using the following Gaussian integral,
$\int_{-\infty}^{+\infty} dx e^{-\alpha x^2}=\sqrt{\pi/\alpha}$, valid within the
complex plane through analytic continuation from the region with ${\rm Re}\,\alpha>0$.},
thereby leading to the configuration space functional integral representation,
\begin{equation}
\langle q_f|U(t_f,t_i)|q_i\rangle=\lim_{N\rightarrow\infty}
\left(\frac{m}{2i\pi\hbar\epsilon}\right)^{N/2}\,
\int_{-\infty}^{+\infty}\prod_{\alpha=1}^{N-1}dq_\alpha\,
{\rm exp}\left\{\frac{i}{\hbar}\sum_{\alpha=0}^{N-1}\epsilon
\left[\frac{1}{2}m\left(\frac{q_{\alpha+1}-q_\alpha}{\epsilon}\right)^2
-V(q_\alpha)\right]\right\},
\label{J1.eq:PI2}
\end{equation}
or at the formal level,
\begin{equation}
\langle q_f|U(t_f,t_i)|q_i\rangle
=\int_{q(t_i)=q_i}^{q(t_f)=q_f}
\left[{\cal D}q\right]\,e^{\frac{i}{\hbar}S[q]},
\label{J1.eq:formalPI3}
\end{equation}
with
\begin{equation}
S[q]=\int_{t_i}^{t_f}\,dt\,L(q,\dot{q}), \qquad
L(q,\dot{q})=\frac{1}{2}m\dot{q}^2-V(q).
\end{equation}
The above explicit discretised representation of this latter formal functional 
integral coincides exactly with the explicit construction performed by 
Feynman when he first developed the path integral quantisation approach \cite{Feyn}.

Hence, we have come back full circle. Starting from the action principle
defined within the Lagrangian formulation of dynamics, the canonical
Hamiltonian formulation of the same dynamics on phase space has been
constructed, allowing for the canonical operator quantisation of the
associated algebraic and geometric structures, for which operator
matrix elements may be given a functional integral representation
on phase space or configuration space, in which the classical Hamiltonian
or Lagrangian action functionals reappear on equal terms. The concept
which is central to this whole construction is that of the action,
through one of the many forms by which it contributes whether for the
classical or the quantum dynamics.

Having chosen to follow the operator quantisation path, once a specific
choice of operator ordering has been made, in principle the functional
integral representation acquires a totally unambiguous and well defined
discretised expression, which defines in an exact manner otherwise
ill defined formal path integral expressions whose actual meaning always
still needs to be specified properly. Nonetheless, as we have indicated, 
difficulties lie at the operator level precisely in the choice of operator 
ordering required so as to obtain a consistent unitary quantum theory.

Had one taken the functional integral path towards quantisation,
whether from the Lagrangian or Hamiltonian classical actions,
the difficulty of a proper construction of the quantised system
then lies hidden in the necessity of giving a precise definition and
meaning, through some discretisation procedure or otherwise, to the
formal and thus ill defined functional integrals such as those in 
(\ref{J1.eq:formalPI1}), (\ref{J1.eq:formalPI2}) and (\ref{J1.eq:formalPI3}). 
As a matter of fact, the arbitrariness which exists at this level in the 
choice of discretisation procedure and functional integration measure
(whether over configuration, momentum or phase space) is in direct
correspondence with the arbitrariness which exists on the operator
side of this relationship in terms of the choice of operator ordering.
Taking either path towards quantisation, for appropriate choices on
both sides which are in correspondence, the same dynamical quantum system 
is being represented in a complementary manner. It is extremely
fruitful to constantly keep in one's mind these equivalent
representations of a quantum dynamics when properly implemented,
in particular in a manner that should ensure its quantum unitarity.

As a final illustration, consider the free nonrelativistic particle, with $V(q)=0$.
Given the exact expression in (\ref{J1.eq:PI2}), the remaining Gaussian integrations
may then all be completed, leading {\it in fine\/} to the matrix element,
\begin{equation}
\langle q_f|U(t_f,t_i)|q_i\rangle=
\left(\frac{m}{2i\pi\hbar\Delta t}\right)^{1/2}\,
e^{\frac{i}{\hbar}\frac{m}{2\Delta t}\left(q_f-q_i\right)^2}=
\left(\frac{m}{2i\pi\hbar\Delta t}\right)^{1/2}\,e^{\frac{i}{\hbar}S_c},
\end{equation}
with for the classical action $S_c$,
\begin{equation}
q(t)=q_i+\frac{q_f-q_i}{\Delta t}(t-t_i),\quad
\dot{q}(t)=\frac{q_f-q_i}{\Delta t},\quad
S_c=\int_{t_i}^{t_f}dt\,\frac{1}{2}m\dot{q}^2=\frac{m}{2\Delta t}\left(q_f-q_i\right)^2.
\end{equation}
As a matter of fact, that the classical action appears as a phase factor in the overall
path integral and this matrix element is no accident, but may be understood through a saddle point
evaluation of the path integral\footnote{Such a saddle point evaluation is equivalent to the classical
limit $\hbar\rightarrow 0$, which is such that only the classical trajectories, which extremise the classical
action, end up contributing to the path integral. From that point of view, this very fact is a justification
{\it a posteriori\/} for the variational principle of classical dynamics.} which happens to be exact in the present case, and more generally
for any Lagrange function which is quadratic in $q$ and $\dot{q}$, as is also the case for the
harmonic oscillator. That the above expression for this matrix element is correct may also be checked
directly from the wave function representations of the Heisenberg algebra, and using the momentum representation
in which the Hamiltonian, hence the quantum evolution operator $U(t_f,t_i)$, is diagonal.
This is left as a useful exercise to the reader.

A few more remarks are in order. First, it is straightforward to extend to an arbitrary number $N$ of degrees
of freedom $(q^n,p_n)$ both the above discretised and the formal expressions for the relevant path integral representations
of matrix elements. Second, by proper consideration of representations of the Heisenberg algebra on an arbitrary
configuration space\footnote{These representations are not being discussed in these notes.}, possibly with nontrivial
curvature and when curvilinear coordinates are used, even in the Euclidean case, it is possible to extend the analysis
to such situations as well. When configuration space possesses nontrivial topology, extra features of a purely
quantum character then also come into play, having to do with the first homotopy group of the configuration manifold
and its U(1) holonomy representations \cite{Gov1.Villa}. Finally, in spite of appearances from its formal representation,
the precise construction of the path integral over phase space is not invariant under canonical transformations
of phase space---which indeed leave the phase space Liouville measure invariant---, since the integration measure
in the path integral is not exactly the Liouville measure.

\clearpage

\subsection{Representations of the Heisenberg algebra II: Coherent states}

So far for a single degree of freedom system, we have constructed essentially three different though
unitarily equivalent representations of the Heisenberg algebra,
\begin{equation}\left[\hat{q},\hat{p}\right]=i\hbar,\qquad
\hat{q}^\dagger=\hat{q},\qquad \hat{p}^\dagger=\hat{p},
\end{equation}
namely the configuration space and momentum space wave function representations in terms of square integrable
wave functions on the real line, $q,p\in\mathbb{R}$, and in terms of the Fock algebra associated to the
Heisenberg algebra provided a parameter with the physical dimensions of $(m\omega)$ is available, as is the case for the harmonic
oscillator. Once the Fock space representation is achieved, it is possible to identify yet another
realisation of the Heisenberg algebra, in terms of so-called canonical coherent states. The notion of coherent state
extends much beyond the simple Heisenberg algebra, but retains most of the properties of the canonical
coherent states discussed hereafter. Note also that the discussion will be restricted to a single degree of
freedom, but may readily be extended to many degrees of freedom through a simple tensor product of Hilbert spaces.

In order to identify a Fock algebra associated to the above Heisenberg algebra, one needs to introduce
an extra real and positive parameter, to be denoted $\lambda>0$ hereafter, having the physical dimensions of
a mass multiplying an angular frequency (after all, this is the situation for the harmonic oscillator, but here
the discussion does not presume a specific dynamics; it remains of a purely kinematical character). Given such a
parameter, consider then the operators,
\begin{equation}
a=\sqrt{\frac{\lambda}{2\hbar}}\left[\hat{q}+\frac{i}{\lambda}\hat{p}\right],\qquad
a^\dagger=\sqrt{\frac{\lambda}{2\hbar}}\left[\hat{q}-\frac{i}{\lambda}\hat{p}\right],
\end{equation}
and their inverse relations,
\begin{equation}
\hat{q}=\sqrt{\frac{\hbar}{2\lambda}}\left[a\,+\,a^\dagger\right],\qquad
\hat{p}=-i\lambda\sqrt{\frac{\hbar}{2\lambda}}\left[a\,-\,a^\dagger\right].
\end{equation}
It is clear that the operators $(a,a^\dagger)$ span a Fock algebra over the Hilbert space
associated to the original Heisenberg algebra,
\begin{equation}
\left[a,a^\dagger\right]=\mathbb{I}.
\end{equation}

Consequently, the associated Fock states span a discrete basis of this Hilbert space. Consider
the Fock vacuum state $|\Omega\rangle$ associated to the choice of $\lambda$, defined by the properties
\begin{equation}
a|\Omega\rangle=0,\qquad \langle\Omega|\Omega\rangle=1,
\end{equation}
as well as the Fock states $|n\rangle$ ($n=0,1,2,\ldots$),
\begin{equation}
|n\rangle=\frac{1}{\sqrt{n!}}\left(a^\dagger\right)^n\,|\Omega\rangle,\qquad
\langle n|m\rangle=\delta_{n,m},\qquad
\mathbb{I}=\sum_{n=0}^{+\infty}\,|n\rangle\,\langle n|.
\end{equation}
As we know one has the following actions of the ladder, or creation and annihilation operators,
\begin{equation}
a|n\rangle=\sqrt{n}\,|n-1\rangle,\qquad
a^\dagger|n\rangle=\sqrt{n+1}\,|n+1\rangle,\qquad
N=a^\dagger a:\quad N|n\rangle=n\,|n\rangle.
\end{equation}
In particular
\begin{equation}
|\psi\rangle=\sum_{n=0}^{+\infty}\,|n\rangle\,\psi_n,\qquad\qquad \psi_n=\langle n|\psi\rangle,
\end{equation}
while the change of basis from the Fock state basis to the configuration space eigenbasis, say, is specified
by the matrix elements
\begin{equation}
\langle q|n\rangle=\left(\frac{\lambda}{\pi\hbar}\right)^{1/4}\,
\frac{1}{\sqrt{2^n n!}}\,e^{-\frac{\lambda}{2\hbar}q^2}\,H_n\left(q\sqrt{\frac{\lambda}{\hbar}}\right),
\end{equation}
$H_n(x)$ being the Hermite polynomials of order $n\in\mathbb{N}$. These functions thus provide a complete
and infinite discrete basis in $L^2(\mathbb{R},dx)$.

\vspace{10pt}

\noindent
\underline{\bf Remark}

\vspace{10pt}

\noindent
It is of interest to consider this discussion in the specific case of
the harmonic oscillator once again, with
\begin{equation}
L(q,\dot{q})=\frac{1}{2}m\dot{q}^2-\frac{1}{2}m\omega^2 q^2,\qquad
H(q,p)=\frac{1}{2m}p^2\,+\,\frac{1}{2}m\omega^2 q^2.
\end{equation}
At the quantum level one then finds
\begin{eqnarray}
\hat{H}&=&\frac{1}{2m}\hat{p}^2+\frac{1}{2}m\omega^2\hat{q}^2 \nonumber\\
 &=&\frac{\hbar}{2m\lambda}\left[(m\omega)^2+\lambda^2\right]
\left[a^\dagger a+\frac{1}{2}\right]\,+\,
\frac{\hbar}{4m\lambda}\left[(m\omega)^2-\lambda^2\right]
\left[{a^\dagger}^2+a^2\right].
\end{eqnarray}
Hence the choice of Fock algebra which diagonalises the Hamiltonian of the system corresponds
to the value $\lambda=m\omega$, readily leading to the energy spectrum,
\begin{equation}
\hat{H}=\hbar\omega\left[a^\dagger a+\frac{1}{2}\right],\qquad
\hat{H}|n\rangle=E_n\,|n\rangle,\quad
E_n=\hbar\omega\left(n+\frac{1}{2}\right).
\end{equation}

\subsubsection{Phase space or canonical coherent states}

Given the Fock algebra constructed as indicated above, the associated phase space or canonical
coherent states are defined according to the following relations. Given any point $(q,p)$ in phase space,
namely any classical state, there corresponds to it a quantum state in Hilbert space parametrised by
these two coordinates and obtained from the exponentiated action of the Heisenberg algebra, or the Fock algebra,
on the Fock vacuum\footnote{In fact, most properties characteristic of coherent states discussed hereafter remain applicable
whatever the choice of normalised ``fiducial" quantum state chosen in place of $|\Omega\rangle$.}, $|\Omega\rangle$,
\begin{equation}
|q,p\rangle=e^{-\frac{i}{\hbar}\left(q\hat{p}-p\hat{q}\right)}\,|\Omega\rangle=
e^{-\frac{1}{2}|z|^2}\,e^{z a^\dagger}\,|\Omega\rangle=|z\rangle,\qquad
|z\rangle=e^{-\frac{1}{2}|z|^2}\sum_{n=0}^{+\infty}\frac{z^n}{\sqrt{n!}}\,|n\rangle,
\label{J1.eq:CS1}
\end{equation}
with
\begin{equation}
z=\sqrt{\frac{\lambda}{2\hbar}}\left[q+\frac{i}{\lambda}p\right],\qquad
\bar{z}=\sqrt{\frac{\lambda}{2\hbar}}\left[q-\frac{i}{\lambda}p\right].
\end{equation}
Note how this complex parameter labelling phase space is in direct correspondence with the relations
defining the creation and annihilation operators in terms of the position and momentum operators $\hat{q}$ and $\hat{p}$.
It is the very last relation on the r.h.s. of (\ref{J1.eq:CS1}) which explains the name given to these states.
Indeed, each of the Fock quantum states $|n\rangle$ are involved but with relative phases for the coefficients
which define their combination which are coherent, namely given by the successive powers of the unique and common complex parameter $z$.
In fact, coherent states play a central r\^ole in quantum optics, for instance, or any other field where coherence
effects are at play, as is the case for laser physics and the optical coherence properties of laser beams.
Coherent states also play an important r\^ole in quantum field theory, as models for the quantum states corresponding
to classical fields, such as classical electric and magnetic fields.

The different expressions above for these coherent states follow from applying the following Baker--Campbell--Hausdorff
formula valid for any two operators $A$ and $B$,
\begin{equation}
e^A\,e^B=e^{A+B+\frac{1}{2}[A,B]+\frac{1}{12}[A,[A,B]]-\frac{1}{12}[B,[A,B]]+\cdots}.
\end{equation}
In particular if $[A,B]$ commutes with both $A$ and $B$, this formula reduces to
\begin{equation}
e^{A+B}=e^{-\frac{1}{2}[A,B]}\,e^A\,e^B,
\end{equation}
which is the relevant situation in the present case.

A first noteworthy property of these coherent states is that they are eigenstates of the
annihilation operator,
\begin{equation}
a|z\rangle=z\,|z\rangle,\quad a^n|z\rangle=z^n\,|z\rangle.
\end{equation}
Also, the overlap of any such coherent state with a Fock state is a pure monomial in $z$,
except for a common Gaussian factor,
\begin{equation}
\langle n|z\rangle=\frac{1}{\sqrt{n!}}\langle\Omega|a^n|z\rangle=
\frac{1}{\sqrt{n!}}\,e^{-\frac{1}{2}|z|^2}\,z^n.
\end{equation}
Since it will turn out that coherent states define yet another (overcomplete) basis of Hilbert space,
this fact shows that in that basis Fock states are represented by simple monomials in $z$, as
compared to Hermite polynomials in configuration (or momentum) space, a much welcome simpler situation.

Even though the above coherent states are normalised, $\langle z|z\rangle=1$, their general overlaps
are nonvanishing,
\begin{equation}
\langle z_1|z_2\rangle=e^{-\frac{1}{2}|z_1|^2-\frac{1}{2}|z_2|^2+\bar{z}_1 z_2},\qquad
\langle z|z\rangle=1,\qquad \langle z_1|z_2\rangle\ne 0,
\end{equation}
or in terms of the real $(q,p)$ parametrisation,
\begin{equation}
\langle q_2,p_2|q_1,p_1\rangle={\rm exp}\,\left\{
-\frac{\lambda}{4\hbar}\left[\left(q_2-q_1\right)^2+\frac{1}{\lambda^2}\left(p_2-p_1\right)^2\right]\,+\,
\frac{i}{2\hbar}\left(q_2p_1-q_1p_2\right)\right\}.
\end{equation}

Hence even though coherent states span the entire Hilbert space they are not linearly independent, namely they
provide an overcomplete basis of the space of states. The fact that they generate the whole space of
quantum states follows from the {\bf overcompleteness relation}, or resolution of the unit operator
in terms of coherent states.
\begin{equation}
\mathbb{I}=\int_{\mathbb{R}^2}\frac{dq dp}{2\pi\hbar}\,|q,p\rangle\,\langle q,p|=
\int_{\mathbb{C}}\frac{dz d\bar{z}}{\pi}\,|z\rangle\,\langle z|\qquad
[dzd\bar{z}=d{\rm Re}\,z\,d{\rm Im}\,z].
\end{equation}
This relation may be verified by computing explicitly all its matrix elements in the Fock state
basis, for instance. Indeed, this resolution of the unit operator is a nontrivial property of coherent states
which is crucial for the relevance of these states to quantum physics in general.

Given the resolution of the unit operator in terms of coherent states, and as was the case for the
configuration and momentum space wave function representations of quantum states, coherent states lead
to (anti)holomorphic (also called Bargmann) wave function representations of quantum states, namely in
terms of functions of the variable $\bar{z}$ only, again up to an overall and common Gaussian factor
${\rm exp}\,(-|z|^2/2)$,
\begin{equation}
\psi(z)=\langle z|\psi\rangle=e^{-\frac{1}{2}|z|^2}\,\varphi(\bar{z}),\qquad
|\psi\rangle=\int_{\mathbb{C}}\frac{dz d\bar{z}}{\pi}\,|z\rangle\,\psi(z)=
\int_{\mathbb{C}}\frac{dz d\bar{z}}{\pi}\,e^{-\frac{1}{2}|z|^2}\,
|z\rangle\,\varphi(\bar{z}).
\end{equation}
Consequently, one has the following representations for the annihilation and creation
operators,
\begin{eqnarray}
a:\quad \langle z|a|\psi\rangle &=& \left[\partial_{\bar{z}}+\frac{1}{2}z\right]\,\psi(z)=
e^{-\frac{1}{2}|z|^2}\,\partial_{\bar{z}}\varphi(\bar{z}), \nonumber\\
a^\dagger:\quad \langle z|a^\dagger|\psi\rangle &=& \bar{z}\,\psi(z)=
e^{-\frac{1}{2}|z|^2}\,\bar{z}\,\varphi(\bar{z}).
\end{eqnarray}

Coherent states still possess other noteworthy properties. For instance
the diagonal matrix elements for both $\hat{q}$ and $\hat{p}$ are both
sharp in phase space,
\begin{equation}
\bar{q}=\langle q,p|\hat{q}|q,p\rangle=q,\qquad
\bar{p}=\langle q,p|\hat{p}|q,p\rangle=p.
\end{equation}
In view of Heisenberg's uncertainty principle, such a property is remarkable indeed.
It needs to be emphasized though, that it is true only provided the diagonal matrix elements
in the coherent state basis are considered; non diagonal matrix elements of these same two
operators do not possess that property. Nevertheless, when considering the Heisenberg
uncertainty relation itself for any coherent state $|z\rangle$, one in fact finds
that this relation is exactly saturated whatever the value for $z$,
\begin{eqnarray}
\left(\Delta q\right)^2=\langle q,p|\left(\hat{q}-\bar{q}\right)^2|q,p\rangle=\frac{\hbar}{2\lambda}&,&
\left(\Delta p\right)^2=\langle q,p|\left(\hat{p}-\bar{p}\right)^2|q,p\rangle=\frac{1}{2}\hbar\lambda,\\
\Delta q\,\Delta p &=& \frac{1}{2}\hbar.
\end{eqnarray}
Consequently, coherent states are quantum states which are closest to being ordinary classical states,
sharing quite a number of properties of classical states, being in particular eigenstates of $a$ with
as eigenvalue $z$, and in one-to-one correspondence with classical states in phase space.

Yet another property characteristic of coherent states in general is that the above canonical coherent
states are stable under time evolution. Consider as time evolution generating
Hamiltonian the operator $\hat{H}=\hbar\omega(a^\dagger a+1/2)$.
A straightforward application of the following Baker--Campbell--Hausdorff formula,
\begin{equation}
e^A e^B e^{-A}=e^{B+[A,B]+\frac{1}{2!}[A,[A,B]]+\frac{1}{3!}[A,[A,[A,B]]]+\cdots},
\end{equation}
then finds
\begin{equation}
e^{-\frac{i}{\hbar}\hat{H}t}\,|z\rangle=e^{-\frac{1}{2}i\omega t}\,|z e^{-i\omega t}\rangle.
\end{equation}
This remarkable property implies that in the case of the harmonic oscillator,
given the one-to-one correspondence between classical states in classical phase space
and quantum coherent states in quantum Hilbert space, the time evolution of the latter in quantum space is in one-to-one
correspondence with the classical trajectory in phase space. Other dynamical
systems for which coherent states may be constructed in a similar manner also share this remarkable property.
The canonical coherent states discussed here are the coherent states associated to the Weyl--Heisenberg group
and to the harmonic oscillator.

In fact, having established this time dependence for the coherent states of the harmonic oscillator, and using the overlap of
phase space coherent states as given above, it follows that the phase space coherent state matrix elements of the quantum evolution
operator of the harmonic oscillator are given by, with $\Delta t=t_f-t_i$ and of course the choice $\lambda=m\omega$,
\begin{equation}
\begin{array}{rl}
\langle q_f,p_f|&U(t_f,t_i)|q_i,p_i\rangle=
e^{-\frac{1}{2}i\omega\Delta t}\,e^{\frac{i}{2\hbar}\left(q_fp_i-q_ip_f\right)e^{-i\omega\Delta t}}\times\\
 & \\
& \times\ {\rm exp}\,\left\{-\frac{m\omega}{4\hbar}\left[q^2_f+q^2_i-2q_fq_ie^{-i\omega\Delta t}\right]\,-\,
\frac{1}{4\hbar m\omega}\left[p^2_f+p^2_i-2p_fp_ie^{-i\omega\Delta t}\right]\right\}.
\end{array}
\end{equation}
The small time, $\Delta t\to 0$, behaviour of this quantity is thus the basic building block for the
construction of a path integral representation of such phase space coherent state matrix elements,
to be described hereafter.

As a final remark, let us consider an application of coherent states to diagonalise the harmonic
oscillator coupled to an external dipole field leading to a shifted
vacuum state,
\begin{equation}
L=\frac{1}{2}m\dot{q}^2-\frac{1}{2}m\omega^2 q^2 - \alpha q=
\frac{1}{2}m\dot{Q}^2-\frac{1}{2}mQ^2+\frac{\alpha^2}{2m\omega^2},\qquad
Q(t)=q(t)+\frac{\alpha}{m\omega}.
\end{equation}
Here the variable $Q(t)$ is such that it vanishes at the minimum of the total potential
energy $V(q)=m\omega^2q^2/2+\alpha q$. In terms of the annihilation operators associated
to $Q$ ({\it i.e.\/}, $A$) and $q$ ({\it i.e.\/}, $a$), which, given the above
definition of $Q$ in terms of $q$, are related as 
\begin{equation}
A=a+\sqrt{\frac{m\omega}{2\hbar}}\frac{\alpha}{m\omega}=a+\frac{\alpha}{\sqrt{2\hbar m\omega}},
\end{equation}
it obviously follows that the actual ground state of the shifted oscillator is a specific
coherent state of the ground state of the unshifted oscillator, when $\alpha=0$, since one has
\begin{equation}
A|\Omega_\alpha\rangle=0 \Longrightarrow a|\Omega_\alpha\rangle=-\frac{\alpha}{\sqrt{2\hbar m\omega}}|\Omega_\alpha\rangle.
\end{equation}
Since coherent states are eigenstates of the annihilation operator $a$, up to an arbitrary phase factor,
the unique normalised solution to the latter condition is thus given by the following coherent state,
\begin{equation}
|\Omega_\alpha\rangle=|z(\alpha)\rangle=e^{-\frac{1}{2}|z(\alpha)|^2}\,e^{z(\alpha) a^\dagger}\,|\Omega\rangle,\quad
z(\alpha)=-\frac{\alpha}{\sqrt{2\hbar m\omega}}.
\end{equation}
This shows that when $\alpha\ne 0$, the true vacuum of the system is the vacuum of the original system filled
with a ``condensate" of a coherent infinite number of quanta of the original system. When $\alpha\ne 0$, the point $q=0$
is no longer the minimum of the potential energy but becomes unstable and decays into the true ground state at $Q=0$,
which at the quantum level corresponds to the nonperturbative coherent vacuum state $|\Omega_\alpha\rangle=|z(\alpha)\rangle$.

\subsubsection{The coherent state phase space path integral}

Given the resolution of the unit operator in terms of the phase space coherent states,
\begin{equation}
\mathbb{I}=\int_{\mathbb{R}^2}\frac{dq dp}{2\pi\hbar}\,|q,p\rangle\,\langle q,p|,
\end{equation}
in a manner analogous to the one discussed previously for the construction of path
integral representations of configuration or momentum space matrix elements of the quantum evolution operator,
one finds the following formal coherent state phase space path integral representation of the
quantum evolution operator,
\begin{equation}
\langle q_f,p_f|U(t_f,t_i)|q_i,p_i\rangle=
\int_{(q_i,p_i)}^{(q_f,p_f)}\left[\frac{{\cal D}q{\cal D}p}{2\pi\hbar}\right]\,
e^{\frac{i}{\hbar}\int_{t_i}^{t_f}dt\left[\frac{1}{2}\left(\dot{q}p-q\dot{p}\right)-H(q,p)\right]},
\end{equation}
corresponding to the following exact and explicit construction,
\begin{equation}
\begin{array}{rl}
&\langle q_f,p_f|U(t_f,t_i)|q_i,p_i\rangle=\lim_{N\to\infty}
\int_{-\infty}^{+\infty}\prod_{\alpha=1}^{N-1}\frac{dq_\alpha dp_\alpha}{2\pi\hbar} \times\\
 & \\
&\times\ {\rm exp}\,\left\{\frac{i}{\hbar}\sum_{\alpha=0}^{N-1}\epsilon
\left(\frac{1}{2}\left[\frac{q_{\alpha+1}-q_\alpha}{\epsilon}p_\alpha\,-\,
q_\alpha\frac{p_{\alpha+1}-p_\alpha}{\epsilon}\right]\,-\,h_\alpha\,+
\,\frac{1}{4}i\epsilon\left[\lambda\left(\frac{q_{\alpha+1}-q_\alpha}{\epsilon}\right)^2\,+\,
\frac{1}{\lambda}\left(\frac{p_{\alpha+1}-p_\alpha}{\epsilon}\right)^2\right]\right)\right\},
\end{array}
\end{equation}
where
\begin{equation}
h_\alpha=\frac{\langle q_{\alpha+1},p_{\alpha+1}|\hat{H}|q_\alpha,p_\alpha\rangle}
{\langle q_{\alpha+1},p_{\alpha+1}|q_\alpha,p_\alpha\rangle}.
\end{equation}

Note that in contradistinction to the previous path integral representations, in the
coherent state phase space one all phase space variables are treated identically; there are
as many of the $q$ as of the $p$ type, all integrated over with the Liouville measure in
phase space. Nevertheless, the path integral remains non invariant under phase space canonical
transformations, nor is it invariant under changes of coordinates in configuration space.

\subsection{The nonrelativistic charged particle in a background electromagnetic field}

{}From a previous discussion of that system, we know that the classical Hamiltonian for a charged nonrelativistic
particle subjected to an electromagnetic field, described by the scalar and vector gauge potentials
$\Phi(t,\vec{r}\,)$ and $\vec{A}(t,\vec{r}\,)$, as well as to conservative forces of total
potential energy $V(\vec{r}\,)$, is given by
\begin{equation}
H=\frac{1}{2m}\left[\vec{p}-q\vec{A}(t,\vec{r}\,)\right]^2+q\Phi(t,\vec{r}\,)+V(\vec{r}\,),
\end{equation}
the cartesian coordinates of the phase space variables $(\vec{r},\vec{p}\,)$ being canonically
conjugate with the canonical Poisson brackets $\left\{x^i,p_j\right\}=\delta^i_j$ ($i,j=1,2,3$).
Here $m$ and $q$ denote of course the mass and charge of the particle, respectively.

According to the correspondence principle of canonical quantisation, the quantum Hamiltonian
of this system is then
\begin{equation}
\hat{H}=\frac{1}{2m}\left[\vec{\hat{p}}-q\vec{A}(t,\vec{\hat{r}}\,)\right]^2
+q\Phi(t,\vec{\hat{r}}\,)+V(\vec{\hat{r}}\,),
\end{equation}
in which now the quantum operators $\vec{\hat{r}}$ and $\vec{\hat{p}}$ have the following
commutation relations for their cartesian coordinates, $[\hat{x}^i,\hat{p}_j]=i\hbar\delta^i_j$,
it being understood that these are also (preferably) self-adjoint operators
on the space of quantum states, thus defining a 3-fold tensor product (over the cartesian components)
of the Heisenberg algebra over $\mathbb{R}$.

Since the physical space in which the particle is moving is assumed to be the Euclidean space
$\mathbb{R}^3$, there exists essentially a unique representation of that algebra, given for
instance in terms of the configuration space wave function representation of states,
$\psi(t,\vec{r}\,)=\langle\vec{r}\,|\psi,t\rangle$, in which case the operators $\vec{\hat{r}}$
and $\vec{\hat{p}}$ have the following functional representations
\begin{equation}
\langle\vec{r}\,|\vec{\hat{r}}\,|\psi,t\rangle=\vec{r}\,\psi(t,\vec{r}\,),\qquad
\langle\vec{r}\,|\vec{\hat{p}}\,|\psi,t\rangle=-i\hbar\vec{\nabla}\psi(t,\vec{r}\,).
\end{equation}
By direct substitution into the above abstract quantum Hamiltonian, the Schr\"odinger equation
in configuration space then reads,
\begin{equation}
\left\{-\frac{\hbar^2}{2m}\left[\vec{\nabla}-i\frac{q}{\hbar}\vec{A}(t,\vec{r}\,)\right]^2\,+\,
q\Phi(t,\vec{r}\,)\,+\,V(\vec{r}\,)\right\}\,\psi(t,\vec{r}\,)=i\hbar\frac{\partial\psi(t,\vec{r}\,)}{\partial t}.
\end{equation}

It is also possible to express the same quantum dynamics in the following form,
\begin{equation}
\left\{-\frac{\hbar^2}{2m}\left[\vec{\nabla}-i\frac{q}{\hbar}\vec{A}(t,\vec{r}\,)\right]^2
\,+\,V(\vec{r}\,)\right\}\,\psi(t,\vec{r}\,)=
i\hbar\left[\frac{\partial }{\partial t}+i\frac{q}{\hbar}\Phi(t,\vec{r}\,)\right]\,\psi(t,\vec{r}\,).
\end{equation}
This form of the Schr\"odinger equation is most relevant to study the issue of its possible invariance
under the gauge transformations of the electromagnetic potentials,
\begin{equation}
\Phi'(t,\vec{r}\,)=\Phi(t,\vec{r}\,)-\partial_t\chi(t,\vec{r}\,),\qquad
\vec{A}'(t,\vec{r}\,)=\vec{A}(t,\vec{r}\,)+\vec{\nabla}\chi(t,\vec{r}\,),
\end{equation}
$\chi(t,\vec{r}\,)$ being an arbitrary function of time and space.
It is clear that since these potentials contribute to the Schr\"odinger equation in
combination with time or space derivatives, if there is any chance to identify these
transformations also as a symmetry of the Schr\"odinger equation, the quantum wave function
has to transform accordingly with a phase factor. After but only a little reflection,
one quickly comes to the conclusion that the appropriate transformation of the quantum wave function
in configuration space is
\begin{equation}
\psi'(t,\vec{r}\,)=e^{i\frac{q}{\hbar}\chi(t,\vec{r}\,)}\,\psi(t,\vec{r}\,).
\end{equation}
Indeed, one then finds that each of the terms involving the time or space derivatives and the
scalar and vector gauge potentials transforms as follows,
\begin{eqnarray}
\left[\vec{\nabla}-i\frac{q}{\hbar}\vec{A}'(t,\vec{r}\,)\right]\,\psi'(t,\vec{r}\,) &=&
e^{i\frac{q}{\hbar}\chi(t,\vec{r}\,)}\,\left[\vec{\nabla}-i\frac{q}{\hbar}\vec{A}(t,\vec{r}\,)\right]\,
\psi(t,\vec{r}\,),\nonumber \\
\left[\partial_t+i\frac{q}{\hbar}\Phi'(t,\vec{r}\,)\right]\,\psi'(t,\vec{r}\,) &=&
e^{i\frac{q}{\hbar}\chi(t,\vec{r}\,)}\,\left[\partial_t+i\frac{q}{\hbar}\Phi(t,\vec{r}\,)\right]\,
\psi(t,\vec{r}\,).
\label{J1.eq:gaugetransf}
\end{eqnarray}
Consequently, both sides of the Schr\"odinger equation vary with the same phase factor as
the wave function, thus leaving the equation {\it in fine\/} invariant indeed under the local gauge
symmetry of the electromagnetic interaction.

Note that the actual transformation associated to the symmetry is that through the phase
factor. Indeed, introducing
\begin{equation}
U(t,\vec{r}\,)=e^{i\frac{q}{\hbar}\chi(t,\vec{r}\,)},
\end{equation}
one may write
\begin{eqnarray}
\psi'(t,\vec{r}\,) &=& U(t,\vec{r}\,)\,\psi(t,\vec{r}\,),\nonumber \\
\Phi'(t,\vec{r}\,) &=& U(t,\vec{r}\,)\,\Phi(t,\vec{r}\,)\,U^{-1}(t,\vec{r}\,)\,-\,
i\frac{\hbar}{q}\,U(t,\vec{r}\,)\,\partial_t\,U^{-1}(t,\vec{r}\,), \nonumber \\
\vec{A}'(t,\vec{r}\,) &=& U(t,\vec{r}\,)\,\vec{A}(t,\vec{r}\,)\,U^{-1}(t,\vec{r}\,)\,+\,
i\frac{\hbar}{q}\,U(t,\vec{r}\,)\,\vec{\nabla}\,U^{-1}(t,\vec{r}\,),
\end{eqnarray}
a form which readily extends to nonabelian symmetries for which $U(t,\vec{r}\,)$ then stands
for elements of some nonabelian symmetry group. In the present case the symmetry group is thus
that of phase transformations, or rotations of the unit circle in the complex plane, or
simply the group U(1) of $1\times 1$ unitary matrices such that $U^\dagger=U^{-1}$, namely
complex numbers of unit norm, thus pure phases. Hence the electromagnetic interaction is
intimately connected the local gauge symmetry based on the group U(1).

In fact in the above expressions (\ref{J1.eq:gaugetransf}) lies hidden one of the two secrets
necessary to construct in general Yang-Mills theories of the abelian and nonabelian type.
Note that for a constant phase transformation, $\chi(t,\vec{r}\,)=\chi_0$, the wave function
$\psi(t,\vec{r}\,)$ and its time and space derivatives transform in a similar fashion, simply
being multiplied by the same phase factor. However when the symmetry is made local, namely when
$\chi(t,\vec{r}\,)$ becomes an arbitrary function of time and space, the ordinary time and space
derivatives of the wave function do no longer transform in the same covariant way under the
phase symmetry as does the wave function itself. There is always one more contribution,
\begin{eqnarray}
\partial_t\,e^{i\frac{q}{\hbar}\chi(t,\vec{r}\,)}\,\psi(t,\vec{r}\,) &=& 
e^{i\frac{q}{\hbar}\chi(t,\vec{r}\,)}\left[\partial_t\,\psi(t,\vec{r}\,)\,+\,
i\frac{q}{\hbar}\partial_t\chi(t,\vec{r}\,)\psi(t,\vec{r}\,)\right],\nonumber \\
\vec{\nabla}\,e^{i\frac{q}{\hbar}\chi(t,\vec{r}\,)}\,\psi(t,\vec{r}\,) &=&
e^{i\frac{q}{\hbar}\chi(t,\vec{r}\,)}\left[\vec{\nabla}\psi(t,\vec{r}\,)
\,+\,i\frac{q}{\hbar}\vec{\nabla}\chi(t,\vec{r}\,)\psi(t,\vec{r}\,)\right].
\end{eqnarray}
These additional terms are precisely those that are cancelled by the transformation of the
gauge potentials in the combinations
\begin{equation}
\left[\partial_t-i\frac{q}{\hbar}\Phi(t,\vec{r}\,)\right]\psi(t,\vec{r}\,),\qquad
\left[\vec{\nabla}+i\frac{q}{\hbar}\vec{A}(t,\vec{r}\,)\right]\psi(t,\vec{r}\,).
\end{equation}
These types of extended or generalised derivatives are known as {\bf covariant derivatives},
since when acting on an object transforming covariantly under a symmetry they produce again
a quantity covariant for the same transformations, in contradistinction to the ordinary
derivatives.

Incidentally, there is a difference in sign in the contributions of the electromagnetic
gauge potentials to the time and space covariant derivatives. This difference in sign is
directly related to the opposite sign in the time and space contributions to the spacetime
metric, $(ct)^2-\vec{r}\,^2$, of Minkowski spacetime in special relativity. Indeed, the
electromagnetic interaction is also explicitly invariant under the symmetry group of that
geometry, namely the Lorentz and Poincar\'e groups. Thus the germs of two of the most important
insights of XX$^{\rm th}$ century physics, namely special relativity and the fundamental r\^ole
of the gauge symmetry principle as governing all fundamental interactions, are already
present in the Schr\"odinger equation of the nonrelativistic charged particle in a background
electromagnetic field.

\subsection{The two dimensional spherically symmetric harmonic oscillator}

The Lagrange function of the spherically symmetric harmonic oscillator of angular frequency $\omega$
and mass $m$, in two dimensional Euclidean space, is
\begin{equation}
L=\frac{1}{2}m\left(\dot{x}^2_1+\dot{x}^2_2\right)-\frac{1}{2}m\omega^2\left(x^2_1+x^2_2\right),
\end{equation}
$x_1$ and $x_2$ being the cartesian coordinates representing the two degrees of freedom of this dynamics.
As such, this system is literally the sum of two independent one dimensional harmonic oscillators.
One may thus simply take over all the results from previous discussions of the classical and quantum
harmonic oscillator, and add an index $\alpha=1,2$ to the variables, and finally sum over these, which at
the quantum level means taking the ordinary tensor product of the two spaces of quantum states.

Consequently, we know that the Hamiltonian of the system is simply,
\begin{equation}
H=\frac{1}{2m}\left(p^2_1+p^2_2\right)+\frac{1}{2}m\omega^2\left(x^2_1+x^2_2\right),
\end{equation}
$p_1$ and $p_2$ being of course the momenta canonically conjugate to the coordinates $x_1$
and $x_2$, respectively. At the quantum level, one has the two pairs of operators $\hat{x}_1$ and $\hat{p}_1$, on the one hand,
and $\hat{x}_2$ and $\hat{p}_2$ on the other hand, with each pair obeying the Heisenberg algebra,
$[\hat{x}_\alpha,\hat{p}_\beta]=i\hbar\delta_{\alpha\beta}$ $(\alpha,\beta=1,2)$.
Introducing the associated creation and annihilation operators, we thus have,
\begin{eqnarray}
a_1 &=& \sqrt{\frac{m\omega}{2\hbar}}\left[\hat{x}_1+\frac{i}{m\omega}\hat{p}_1\right],\qquad
a^\dagger_1=\sqrt{\frac{m\omega}{2\hbar}}\left[\hat{x}_1-\frac{i}{m\omega}\hat{p}_1\right],\nonumber \\
a_2 &=& \sqrt{\frac{m\omega}{2\hbar}}\left[\hat{x}_2+\frac{i}{m\omega}\hat{p}_2\right],\qquad
a^\dagger_2=\sqrt{\frac{m\omega}{2\hbar}}\left[\hat{x}_2-\frac{i}{m\omega}\hat{p}_2\right],
\end{eqnarray}
obeying the tensor product Fock algebra,
\begin{equation}
[a_\alpha,a^\dagger_\beta]=\delta_{\alpha\beta}\mathbb{I}.
\end{equation}
Conversely,
\begin{eqnarray}
\hat{x}_1 &=& \sqrt{\frac{\hbar}{2m\omega}}\left[a_1+a^\dagger_1\right],\qquad
\hat{p}_1=-im\omega\sqrt{\frac{\hbar}{2m\omega}}\left[a_1-a^\dagger_1\right],\nonumber \\
\hat{x}_2 &=& \sqrt{\frac{\hbar}{2m\omega}}\left[a_2+a^\dagger_2\right],\qquad
\hat{p}_2=-im\omega\sqrt{\frac{\hbar}{2m\omega}}\left[a_2-a^\dagger_2\right].
\end{eqnarray}
Obviously, the quantum Hamiltonian then reads
\begin{equation}
\hat{H}=\hbar\omega\left(a^\dagger_1 a_1+\frac{1}{2}\right)\,+\,
\hbar\omega\left(a^\dagger_2 a_2+\frac{1}{2}\right)=
\hbar\omega\left(a^\dagger_1 a_1+a^\dagger_2 a_2+1\right).
\end{equation}
Introducing the orthonormalised Fock state basis
\begin{equation}
|n_1,n_2\rangle=\frac{1}{\sqrt{n_1!\,n_2!}}\left(a^\dagger_1\right)^{n_1}\,
\left(a^\dagger_2\right)^{n_2}\,|0,0\rangle,\qquad
\langle n_1,n_2|n'_1,n'_2\rangle=\delta_{n_1n'_1}\delta_{n_2n'_2},
\end{equation}
it is clear that these states also diagonalise the Hamiltonian with the following energy spectrum,
\begin{equation}
\hat{H}|n_1,n_2\rangle=E(n_1,n_2)\,|n_1,n_2\rangle,\qquad
E(n_1,n_2)=\hbar\omega\left(n_1+n_2+1\right).
\end{equation}
However, this spectrum is degenerate at each level except for the ground state at $(n_1,n_2)=(0,0)$.
Indeed, at a given energy level $E(n_1,n_2)=\hbar\omega(N+1)$ with $N=0,1,2\ldots$, we have as many states
as there are partitions of the natural number $N$ in two natural numbers $n_1$ and $n_2$, namely
$(N+1)$ states. There must exist an actual solid and sound explanation for this fact. It cannot
be just a mere numerical coincidence. The first thought that comes to one's mind is that of a symmetry.

Certainly, the system is invariant under rotations in the plane, and indeed it possesses a conserved
angular-momentum (perpendicular to the plane),
\begin{equation}
L_3=x_1p_2-x_2p_1.
\end{equation}
At the quantum level and expressed in terms of the creation and annihilation operators, one has the
operator,
\begin{equation}
\hat{L}_3=-i\hbar\left(a^\dagger_1 a_2\,-\,a^\dagger_2 a_1\right).
\end{equation}
From this expression, even though the operators $\hat{H}$ and $\hat{L}_3$ commute, the Fock space
basis $|n_1,n_2\rangle$ does not diagonalise them both. We need to find another basis of Hilbert
space which diagonalises these two commuting operators.

For this purpose, let us introduce an helicity basis as follows,
\begin{eqnarray}
a_\pm &=& \frac{1}{\sqrt{2}}\left[a_1\mp i a_2\right],\qquad
a^\dagger_\pm=\frac{1}{\sqrt{2}}\left[a^\dagger_1\pm i a^\dagger_2\right],\nonumber\\
a_1 &=& \frac{1}{\sqrt{2}}\left[a_++a_-\right],\qquad
a^\dagger_1=\frac{1}{\sqrt{2}}\left[a^\dagger_++a^\dagger_-\right],\nonumber \\
a_2 &=& \frac{i}{\sqrt{2}}\left[a_+-a_-\right],\qquad
a^\dagger_2=-\frac{i}{\sqrt{2}}\left[a^\dagger_+-a^\dagger_-\right],
\end{eqnarray}
such that
\begin{equation}
[a_\pm,a^\dagger_\pm]=\mathbb{I},\qquad
[a_\pm,a^\dagger_\mp]=0.
\end{equation}
These redefinitions in turn imply
\begin{eqnarray}
\hat{x}_1 &=&\frac{1}{2}\sqrt{\frac{\hbar}{m\omega}}\left[a_++a_-+a^\dagger_++a^\dagger_-\right],\qquad
\hat{p}_1=-i\frac{m\omega}{2}\sqrt{\frac{\hbar}{m\omega}}\left[a_++a_--a^\dagger_+-a^\dagger_-\right],
\nonumber \\
\hat{x}_2 &=&\frac{i}{\sqrt{2}}\sqrt{\frac{\hbar}{m\omega}}\left[a_+-a_--a^\dagger_++a^\dagger_-\right],\qquad
\hat{p}_2=\frac{m\omega}{2}\sqrt{\frac{\hbar}{m\omega}}\left[a_+-a_-+a^\dagger_+-a^\dagger_-\right],
\end{eqnarray}
hence
\begin{equation}
\hat{x}_\pm=\hat{x}_1\pm i \hat{x}_2=\sqrt{\frac{\hbar}{m\omega}}\left[a_\mp+a^\dagger_\pm\right],\qquad
\hat{p}_\pm=\frac{1}{2}\left[\hat{p}_1\mp i\hat{p}_2\right]=
-\frac{im\omega}{2}\sqrt{\frac{\hbar}{m\omega}}
\left[a_\pm - a^\dagger_\mp\right],
\end{equation}
which are such that
\begin{equation}
[\hat{x}_\pm,\hat{p}_\pm]=i\hbar,\qquad
[\hat{x}_\pm,\hat{p}_\mp]=0.
\end{equation}
One then also finds
\begin{eqnarray}
\hat{H} &=& \hbar\omega\left(a^\dagger_+a_+ + a^\dagger_- a_-+1\right), \nonumber \\
\hat{L}_3 &=& \hbar\left(a^\dagger_+a_+ - a^\dagger_- a_-\right).
\end{eqnarray}
Hence by considering rather the orthonormalised Fock space helicity basis
\begin{equation}
|n_+,n_-\rangle=\frac{1}{\sqrt{n_+!\,n_-!}}\,\left(a^\dagger_+\right)^{n_+}\,
\left(a^\dagger_-\right)^{n_-}\,|0,0\rangle,\qquad
\langle n_+,n_-|n'_+,n'_-\rangle=\delta_{n_+n'_+}\,\delta_{n_- n'_-},
\end{equation}
based on the alternative Fock algebra $(a_\pm,a^\dagger_\pm)$, one has indeed diagonalised
both operators,
\begin{eqnarray}
\hat{H}|n_+,n_-\rangle &=& E(n_+,n_-)|n_+,n_-\rangle,\qquad
E(n_+,n_-)=\hbar\omega(n_+ + n_-+1),\nonumber \\
\hat{L}_3|n_+,n_-\rangle &=& \hbar(n_+-n_-)|n_+,n_-\rangle.
\end{eqnarray}
The previously described degeneracy of energy levels is again observed in terms of $N=n_++n_-$.
However the angular-momentum operator does not map states belonging to a same level into
one another. Hence, the degeneracy cannot be related to that symmetry under SO(2)=U(1) rotations in the
plane. A larger symmetry must be at play. Note that the quanta of the $a^\dagger_+$ type carry a 
unit $(+\hbar)$ of angular-momentum, whereas those of type $a^\dagger_-$ carry a unit
$(-\hbar)$ of angular-momentum. The combinations of the degrees of freedom which defined these
quantities are indeed associated to the helicities $(+1)$ and $(-1)$ of the oscillating modes of the system.

Both operators $\hat{H}$ and $\hat{L}_3$ may be written in the form,
\begin{equation}
\hat{H}=\hbar\omega\left(\begin{array}{c c}
	a^\dagger_+ & a^\dagger_-
			\end{array}\right)
\left(\begin{array}{c c}
1 & 0 \\
0 & 1 \end{array}\right)\left(\begin{array}{c}
	a_+ \\ a_- \end{array}\right)+\hbar\omega,\qquad
\hat{L}_3=\hbar\left(\begin{array}{c c}
	a^\dagger_+ & a^\dagger_-
			\end{array}\right)
\left(\begin{array}{c c}
1 & 0 \\
0 & -1 \end{array}\right)\left(\begin{array}{c}
	a_+ \\ a_- \end{array}\right).
\end{equation}
From that point of view, it should be quite clear that any SU(2) unitary transformation, namely
a $2\times 2$ matrix $U$ over $\mathbb{C}$ which is unitary, $U^\dagger=U^{-1}$, and of unit
determinant, ${\rm det}\,U=1$, and acting on the pair of annihilation operators as
\begin{equation}
\left(\begin{array}{c}
	a_+ \\ a_- \end{array}\right)\ \longrightarrow
\left(\begin{array}{c}
	a'_+ \\ a'_- \end{array}\right)= U
\left(\begin{array}{c}
	a_+ \\ a_- \end{array}\right),
\end{equation}
leaves the Hamiltonian invariant. In other words, there is a larger symmetry present in the
system than simply the SO(2)=U(1) symmetry of rotations which on the pair of annihilation operators act as
\begin{equation}
U=\left(\begin{array}{c c}
	e^{i\alpha} & 0 \\
	0 & e^{-i\alpha} \end{array}\right),
\end{equation}
$\alpha$ being a rotation angle. Note that the latter remark also shows that the symmetry
SU(2) includes as a subgroup the SO(2)=U(1) symmetry under rotations in the plane. This larger
SU(2) symmetry is an example of what is called a {\bf dynamical symmetry}, namely a symmetry
of the Hamiltonian formulation of the system which is not one of its Lagrangian formulation. Such
a situation is indeed possible {\it a priori\/} since the Hamiltonian formulation involves both the
configuration space coordinates and their conjugate momenta which may therefore mix under
a symmetry transformation, a situation which is not possible in the Lagrangian formulation
involving the configuration space coordinates only.

Let us see whether we may introduce now, based on the Fock algebra generators $a^\dagger_\pm$ and
$a_\pm$, operators which map states into one another within a same energy level, thus still commuting
with the Hamiltonian, and then determine their algebra. The obvious candidates are,
\begin{equation}
T_+=a^\dagger_+ a_-,\qquad
T_-=a^\dagger_- a_+,\qquad
T_3=\frac{1}{2}\left(a^\dagger_+a_+ - a^\dagger_- a_-\right)=\frac{1}{2\hbar}\hat{L}_3.
\end{equation}
It takes but only a little calculation to obtain
\begin{equation}
[T_+,T_-]=2T_3,\qquad [T_3,T_\pm]=\pm T_\pm,\qquad
[\hat{H},T_\pm]=0,\qquad [\hat{H},T_3]=0.
\end{equation}
Introducing now
\begin{equation}
T_1=\frac{1}{2}\left(T_++T_-\right),\qquad
T_2=-\frac{i}{2}\left(T_+-T_-\right),\qquad
T_\pm=T_1\pm i T_2,
\end{equation}
the same algebra reads
\begin{equation}
[T_i,T_j]=i\epsilon_{ijk}\,T_k,\qquad
i,j,k=1,2,3,
\end{equation}
$\epsilon_{ijk}$ being the totally antisymmetric invariant tensor in three dimensions with $\epsilon_{123}=+1$.
The last set of commutation relations is directly identified with the SU(2) algebra, or also the SO(3) algebra
which as algebras are identical (but not as groups since SO(3)=SU(2)$/\mathbb{Z}_2$ as a quotient of groups).
One famous finite dimensional representation of that algebra is given by the $2\times 2$
Pauli matrices $T_i=\frac{1}{2}\sigma_i$, with
\begin{equation}
\sigma_1=\left(\begin{array}{c c}
	0 & 1 \\ 1 & 0 \end{array}\right),\qquad
\sigma_2=\left(\begin{array}{c c}
	0 & -i \\ i & 0 \end{array}\right),\qquad
\sigma_3=\left(\begin{array}{c c}
	1 & 0 \\ 0 & -1 \end{array}\right).
\end{equation}

Let us now turn to the action of the operators $T_\pm$ on the Fock basis states $|n_+,n_-\rangle$. Since for
a single harmonic oscillator one has
\begin{equation}
a|n\rangle=\sqrt{n}|n-1\rangle,\qquad
a^\dagger|n\rangle=\sqrt{n+1}|n+1\rangle,
\end{equation}
it follows that
\begin{eqnarray}
T_+|n_+,n_-\rangle &=& \sqrt{(n_++1)n_-}\ |n_++1,n_--1\rangle, \nonumber \\
T_-|n_+,n_-\rangle &=& \sqrt{n_+(n_-+1)}\ |n_+-1,n_-+1\rangle, \nonumber \\
T_3|n_+,n_-\rangle &=& \frac{1}{2}(n_+-n_-)\,|n_+,n_-\rangle,
\end{eqnarray}
showing that indeed it is this algebra of operators which accounts for the degeneracy of
the energy levels. Let us introduce the following representation for the helicity occupation
numbers $n_\pm$,
\begin{equation}
n_++n_-=N=2j,\qquad n_+-n_-=2m,\qquad \frac{1}{2}\left(n_+-n_-\right)=m,
\end{equation}
hence
\begin{equation}
n_+=j+m,\qquad n_-=j-m,
\end{equation}
where $j$ thus takes positive integer or half-integer values according to whether the energy level $N$
is even or odd, while $m$ takes values between $j$ and $(-j)$ in unit steps, $-j\le m\le j$ ($m$ thus measures
the $T_3$ eigenvalue, or half the angular-momentum eigenvalue $L_3$ in units of $\hbar$). The above actions
then read
\begin{equation}
T_+|j,m\rangle=\sqrt{(j-m)(j+m+1)}\,|j,m+1\rangle,\qquad
T_-|j,m\rangle=\sqrt{(j+m)(j-m+1)}\,|j,m-1\rangle,
\end{equation}
with $|j,m\rangle=|n_+,n_-\rangle$ given the above correspondence between these integer or half-integer
variables.

In conclusion, for a given energy level $N=0,1,2,\ldots$, one obtains a certain SU(2) representation
as a finite $N+1=2j+1$ dimensional subspace of the complete Hilbert space of quantum states of this system.
This representation is characterised by the single integer or half-integer number $j$, known as
the {\bf spin} of that representation, whereas the states within that representation of given spin
$j$ are distinguished by their $T_3$ eigenvalue $m$ lying between $(-j)$ and $j$ in integer steps, with
the above matrix elements for the action of the two other operators $T_\pm$ of the SU(2) algebra.
As a matter of fact, we have in this manner recovered {\bf all} finite dimensional representations
of the symmetry algebra SU(2), which is also the algebra of the symmetry group SO(3) of rotations
in three dimensional Euclidean space. The above formula for the action of $T_\pm$ and $T_3$
in a given spin $j$ representation are valid as such in full generality, independently of the
system in which these symmetries may be realised. Any rotationally invariant system in three
dimensions will find its quantum states organised according to these spin representations of
SU(2). But it is matter of experiment to determine which spin values are realised for a specific
physical system. For example, that the electron has spin 1/2 may only be determined experimentally.

As an illustration, consider the value $N=1$ or $j=1/2$, namely the first excited state of the present
system. It is thus doubly degenerated, with
\begin{equation}
T_3 \left|j=\frac{1}{2},m=\frac{1}{2}\right\rangle=
\frac{1}{2}\left|j=\frac{1}{2},m=\frac{1}{2}\right\rangle,\qquad
T_3\left|j=\frac{1}{2},m=-\frac{1}{2}\right\rangle=
-\frac{1}{2}\left|j=\frac{1}{2},m=-\frac{1}{2}\right\rangle,
\end{equation}
\begin{equation}
T_+\left|j=\frac{1}{2},m=\frac{1}{2}\right\rangle=0,\qquad
T_+\left|j=\frac{1}{2},m=-\frac{1}{2}\right\rangle=
\left|j=\frac{1}{2},m=\frac{1}{2}\right\rangle,
\end{equation}
\begin{equation}
T_-\left|j=\frac{1}{2},m=\frac{1}{2}\right\rangle=
\left|j=\frac{1}{2},m=-\frac{1}{2}\right\rangle,\qquad
T_-\left|j=\frac{1}{2},m=-\frac{1}{2}\right\rangle=0,
\end{equation}
and the associations $\left|j=\frac{1}{2},m=\frac{1}{2}\right\rangle=\left|n_+=1,n_-=0\right\rangle$,
$\left|j=\frac{1}{2},m=-\frac{1}{2}\right\rangle=\left|n_+=0,n_-=1\right\rangle$.
In other words, in this two dimensional basis, the matrix representation of the operators $T_3$
and $T_\pm$ is given precisely by the Pauli matrices. Clearly, the ground state of the system,
$|j=0,m=0\rangle=|n_+=0,n_-=0\rangle$, corresponds to a trivial representation
of the SU(2) algebra for which $T_3=0$ and $T_\pm=0$.

\section{Symmetries and the First Noether Theorem}
\label{Gov1.Sec5}

\subsection{Motivation}

Symmetries play a central r\^ole nowadays in fundamental and more applied physics,
not only as means for solving otherwise analytically intractable problems but more
importantly, as providing profound insight into properties of interactions and of
conservation laws. Indeed, when it comes to a continuous symmetry associated to
the notion of a Lie group, Noether's first theorem establishes a direct relation
between the existence of such continuous symmetries and that of conserved quantities,
often also called {\bf Noether charges}.
Furthermore within the Hamiltonian formulation the algebra of Poisson brackets of these
Noether charges proves
to be identical to the abstract algebra of the Lie symmetry group of which the
charges are then the generators. In other words the abstract algebra of the Lie
symmetry group is then realised on the phase space of the system through the
conserved quantities and their Poisson brackets. When canonical quantisation may
proceed in a manner consistent with the Lie algebra of symmetries, the associated
quantum operators then generate, as linear transformations, the symmetry algebra on
the quantum space of states. Consequently, the latter Hilbert space then provides
a linear representation of the Lie algebra. It is here that the whole of
Lie algebra representation theory becomes most relevant, with in particular a classification
of the possible quantum representations of a given classical symmetry. As an example,
if a system is invariant under SO(3) rotations in three Euclidean space dimensions,
the Noether charges correspond to the total angular-momentum vector of that system,
of which the Poisson brackets are isomorphic to the Lie algebra $so(3)$ of that
Lie group. When quantised, one then is able to classify the quantum space of states
in terms of representations of $so(3)$, namely also $su(2)$, since the algebras of
SO(3) and SU(2) are identical. The representations of SU(2) are labelled in terms of
the spin value $j$ taking an integer or half-integer positive value (these
representations were introduced in the context of the two dimensional spherically
symmetric harmonic oscillator in the previous Section).
This fundamental result based on Noether's first theorem is especially important
when the symmetry group is a compact Lie group, since all finite dimensional and unitary
representations of all compact Lie groups have been classified, one of the
towering achievements of pure algebra in XX$^{\rm th}$ century mathematics.
But this is not the purpose of the present notes.

The situation just described is also the generic one, when it comes to continuous 
Lie symmetries, namely a group of transformations defined in terms of a
collection of continuous parameters. A system may also be invariant under a discrete
symmetry of which the group elements depend on a collection of parameters taking
only a discrete set of values. As an example, consider a square in the plane.
Its symmetry group is a specific subgroup of O(2) or SO(2), namely those rotations
by $\pm\pi/2$ radians modulo a reflection in the origin of one of the coordinate axes.
There are no conserved quantities, or Noether charges associated to a discrete
symmetry, even though such a symmetry does impose some restrictions or relations
between physical quantities, of relevance especially when it comes to the
quantum world. Famous symmetries of that type are those of space parity, $P$,
time reversal, $T$, and charge conjugation, $C$, the latter corresponding to the
exchange of particles with their antiparticles.

In these notes we shall not discuss discrete symmetries, but focus only on Noether's
first theorem. There exist still one or two more Noether theorems, in relation with
gauge symmetries. As was illustrated through the example of the electromagnetic 
interaction in previous Sections, a gauge symmetry is a continuous
Lie symmetry of which the continuous parameters themselves may be continuous
functions of time or even spacetime (the latter in a field theory context). As such,
when as a particular case one takes for these function parameters just constant
values, one has what is also often called a {\bf global symmetry}. This is
the situation addressed above, corresponding to Noether's first theorem, leading
to conserved Noether charges. However, in the eventuality of a local or gauged
symmetry, Noether's second and third theorems imply even further restrictions
on the Noether charges and other quantities that are involved in the gauge
symmetry transformations, which are of relevance to the quantisation of gauge
invariant systems. In particular, Noether's second theorem implies that for
configurations of the system which are ``physical" in the sense of being indeed
gauge invariant, the Noether charges of the gauge symmetries have to vanish altogether.
Given the previous remark that Noether charges are the generators of the symmetries
of which they are the conserved quantities, indeed they would need to vanish in the
case of gauge invariant configurations. In these notes we shall not discuss Noether's
second and third theorem (for those interested, a discussion may be found in Ref.\cite{Gov1.GovBook}).

This much having been said, it is necessary to specify what is
exactly meant by a symmetry. Clearly it must consist in a transformation of the configuration
space variables, $q^n$, of the system, and possibly even in combination with a transformation
in the time variable\footnote{Why not for instance redefine time by an arbitrary shift
of $t_0$?}, $t$, of the form
\begin{equation}
t'=t'(t),\qquad {q^n}'(t')={q^n}'\left(q^n,t\right).
\end{equation}
In the case of a discrete symmetry the functions $t'(t)$ and ${q^n}'\left(q^n,t\right)$
would depend on a collection of parameters taking values in a discrete set ({\it viz.\/},
the example of the symmetries of the square). In the case of a continuous symmetry forming
a Lie group, the symmetry parameters take their values in a continuous set. For instance,
the two or three dimensional spherically symmetric harmonic oscillator is invariant
under all space rotations. These form the Lie groups SO(2) or SO(3), respectively, and
their elements may be characterised in terms of angular variables, the rotation angle in the
plane in the first case, or the three Euler angles in the second case, each of these angular
variables taking values in continuous albeit finite intervals. So far these precisions only
concern the transformation, but when does it qualify as a symmetry?

Contrary may be to everyday's usage of the word which is taken to mean that something
is symmetric under a transformation if it is left invariant under that transformation (such
as the square of the above example), in the context of physics and the dynamics of systems,
what one means by the concept of symmetry is not that a particular configuration of the system is
left invariant by the transformation (say, only when a set of particles occupies
a preselected collection of points in space does there exist a particular symmetry), but rather
{\bf that the space of solutions to its equations of motion is left invariant under the
symmetry transformation}. In other words, a symmetry is a transformation which maps any 
given solution to the equations of motion into another solution to the same equations of
motion. Often, one says that a symmetry leaves the equations of motion form invariant.
This means that when expressing the equations of motion in terms of the not-yet-transformed
variables or the transformed ones, the functional relations between these variables in
each case are identical, {\it i.e.\/}, that they have the same form. Given the above representation of such a
transformation, this means that when expressed in terms of the variables carrying a
prime, $t'$ and ${q^n}'$, the equations they obey are the same as those for the variables not carrying
that prime, namely $t$ and $q^n$. One may easily imagine examples for oneself's, and a few
will be discussed hereafter.

Since our discussion is rooted in the variational principle, how then does the action of
a system transform under a symmetry? Since the equations of motion are form invariant,
necessarily under a symmetry the action may only change by a total time derivative.
Indeed as is well known actions that differ only by a total time derivative share
identical equations of motion. Hence when transformations of the above class
define a symmetry of a dynamics, the action of the system must transform according to
\begin{equation}
S[{q^n}']=\int\,dt'\,L\left({q^n}',\frac{d{q^n}'}{dt'}\right)=
\int\,dt\left[L\left(q^n,\frac{dq^n}{dt}\right)+\frac{d\Lambda(q^n,t)}{dt}\right],
\end{equation}
where $\Lambda(q^n,t)$ is some function implicitly defined through the transformation of
the action under the symmetry. In particular, it depends on the parameters of the symmetry
group, say the rotation angles in the case of a rotational symmetry in space. Using then
the composition law for differentials,
\begin{equation}
dt'=dt\,\frac{dt'}{dt},
\end{equation}
a system is invariant under a symmetry transformation if its Lagrange function changes
according to
\begin{equation}
\frac{dt'}{dt}\,L\left({q^n}',\frac{dt}{dt'}\,\frac{d{q^n}'}{dt}\right)=
L\left(q^n,\frac{dq^n}{dt}\right)\,+\,
\frac{d\Lambda\left(q^n,t\right)}{dt}.
\label{J1.eq:transf}
\end{equation}
Note that in the case of a field theory where the action is given as the spacetime
integral of the Lagrangian density ${\cal L}$, the above total time derivative may
be replaced by a total surface term, namely a total spacetime divergence.
In the presence of nontrivial topology in space
such contributions may be physically relevant for some symmetries, leading at the
quantum level to quantisation rules on some parameters. A noteworthy example of
this are supersymmetric field theories; under a supersymmetry transformation,
which exchanges bosons and fermions (states of integer and half-integer spin), the
Lagrangian density varies precisely always by a total surface term.

The discussion so far, including the property (\ref{J1.eq:transf}), applies whether the
symmetry is a continuous or a discrete one. Let us now restrict to a set of transformations
forming a Lie group, namely a continuous symmetry.

\subsection{Linearisation of a Lie symmetry group}

In the case of a continuous group of symmetry transformations, at least for the component
of the group connected to the identity transformation, one may consider infinitesimal
transformations, namely transformations in a linearised form, simply by linearising
the transformations in the parameters of the group. Hence let us now consider transformations
of the following form,
\begin{equation}
t'=t+\delta t(t),\qquad
{q^n}'=q^n+\delta q^n\left(q^n,t\right),\qquad
\Lambda=\delta\Lambda\left(q^n,t\right),
\end{equation}
with $\delta t(t)$ and $\delta q^n(q^n,t)$ some infinitesimal (or linearised) variations
of the variables $t$ and $q^n$ possessing the indicated dependencies on $t$ and $q^n$.
Since in the absence of a transformation no total time derivative term is induced in the
action, the function $\Lambda(q^n,t)$ has no zeroth order contribution, while the function
$\delta\Lambda(q^n,t)$ is determined from the knowledge of $\Lambda(q^n,t)$ and its dependence
on the group parameters, given a symmetry.

All that is required now is to substitute these variations in the variables in the
fundamental identity (\ref{J1.eq:transf}), and expand the latter to first order in the
quantities $\delta t$, $\delta q^n$ and $\delta\Lambda$. It takes only a little calculation
detailed hereafter to establish that for linearised continuous symmetries one has the following
fundamental Noether identity,
\begin{equation}
\left[\delta q^n-\delta t\,\dot{q}^n\right]\,
\left[\frac{\partial L}{\partial q^n}\,-\,\frac{d}{dt}\,
\frac{\partial L}{\partial\dot{q}^n}\right]\ +\
\frac{d}{dt}\left[\delta t\left(L-\dot{q}^n\frac{\partial L}{\partial\dot{q}^n}\right)\,+\,
\delta q^n\frac{\partial L}{\partial\dot{q}^n}\,-\,\delta\Lambda\right]=0,
\label{J1.eq:FundNoether}
\end{equation}
where once again the implicit summation rule over repeated indices is to be understood.

In order to establish this result, let us first note that we have, to first order in the
variation $\delta t(t)$,
\begin{equation}
t'=t+\delta t(t),\qquad
t=t'-\delta t(t),\qquad
\frac{dt'}{dt}=1+\frac{d\delta t(t)}{dt},\qquad
\frac{dt}{dt'}=1-\frac{d\delta t(t)}{dt}.
\end{equation}
Consequently, again to first order the identity (\ref{J1.eq:transf}) reads
\begin{equation}
\left[1+\frac{d\delta t(t)}{dt}\right]\,
L\left(q^n+\delta q^n,\left(1-\frac{d\delta t(t)}{dt}\right)
\left(\frac{dq^n}{dt}+\frac{d\delta q^n}{dt}\right)\right)\,-\,
L\left(q^n,\frac{dq^n}{dt}\right)=\frac{d\delta\Lambda}{dt}.
\end{equation}
Expanded to first order we thus have
\begin{equation}
\frac{d\delta t(t)}{dt}L(q^n,\dot{q}^n)+\delta q^n
\frac{\partial L}{\partial q^n}+\left[\frac{d\delta q^n}{dt}-
\frac{d\delta t}{dt}\dot{q}^n\right]\frac{\partial L}{\partial\dot{q}^n}
-\frac{d\delta\Lambda}{dt}=0.
\end{equation}
Let us now bring as many terms as possible in the form of a total time derivative,
\begin{eqnarray}
&&\ \ \frac{d}{dt}\left[\delta t L+\delta q^n\frac{\partial L}{\partial\dot{q}^n}
-\delta t\,\dot{q}^n\frac{\partial L}{\partial\dot{q}^n}-\delta\Lambda\right] \nonumber \\
&&-\delta t \frac{dL}{dt}+\delta q^n\frac{\partial L}{\partial q^n}-
\delta q^n\frac{d}{dt}\frac{\partial L}{\partial\dot{q}^n}+
\delta t\, \frac{d}{dt}\left[\dot{q}^n\frac{\partial L}{\partial\dot{q}^n}\right]=0.
\end{eqnarray}
Making explicit the contributions of the last line one finds
\begin{equation}
-\delta t\,\dot{q}^n\frac{\partial L}{\partial q^n}-
\delta t\,\ddot{q}^n\frac{\partial L}{\partial\dot{q}^n}+
\delta q^n\frac{\partial L}{\partial q^n}-
\delta q^n\frac{d}{dt}\frac{\partial L}{\partial\dot{q}^n}
+\delta t\,\ddot{q}^n\frac{\partial L}{\partial\dot{q}^n}
+\delta t\,\dot{q}^n\frac{d}{dt}\frac{\partial L}{\partial\dot{q}^n},
\end{equation}
in which some terms cancel explicitly. As a consequence, the fundamental identity
(\ref{J1.eq:FundNoether}) is established.

What is truly remarkable about this result is that the Euler--Lagrange equations of
motion appear explicitly in it, whereas all remaining terms contribute only through
a total time derivative. Thus in particular, this implies that given any solution to
the equations of motion there exists a quantity constructed out of the degrees of freedom
of the system and their generalised velocities of which the total time derivative
vanishes identically. In other words, as a consequence of the property (\ref{J1.eq:transf})
expressing the existence of a symmetry, in the case of a continuous symmetry there
exist quantities which are conserved or constants of motion for any solution to the
equations of motion of the system. In itself this is quite a remarkable result.
However the identity (\ref{J1.eq:FundNoether}) states something even stronger, namely
that there exist specific combinations of the Euler--Lagrange equations of motion which
reduce to total time derivatives. Let us make these remarks more specific by
introducing now the collection of parameters of such Lie group symmetries.

\subsection{The first Noether theorem: global Lie symmetry group}

Let us now make explicit the parameters\footnote{Again, in the case of a rotational symmetry in
space, say, these parameters would correspond to the angles parametrising any such rotation.}
$\alpha_a$ of the Lie group of which the
transformations acting on the configurations $q^n$ of the system and the time variable $t$
define a symmetry of its dynamics in the sense of (\ref{J1.eq:transf}),
\begin{equation}
\delta t(t)=\alpha_a\,\chi^a(t),\qquad
\delta q^n\left(q^n,t\right)=\alpha_a\,\phi^{an}\left(q^n,t\right),\qquad
\delta\Lambda\left(q^n,t\right)=\alpha_a\,\Lambda^a\left(q^n,t\right).
\end{equation}
Here $\chi^a(t)$, $\phi^{an}(q^n,t)$ and $\Lambda^a(q^n,t)$ are specific
functions of the indicated variables, obtained from the functions
$t'(t)$, ${q^n}'(q^n,t)$ and $\Lambda(q^n,t)$ by expanding these to first order
in the group parameters $\alpha_a$.

A direct substitution of these expansions in terms of the group parameters $\alpha_a$
into the fundamental Noether identity (\ref{J1.eq:FundNoether}) clearly leads to 
{\bf the First Noether identity},
\begin{equation}
\frac{d\gamma^a}{dt}=
\left[\phi^{an}\,-\,\chi^a\,\dot{q}^n\right]\,
\left[\frac{d}{dt}\frac{\partial L}{\partial\dot{q}^n}\,
-\,\frac{\partial L}{\partial q^n}\right],
\end{equation}
in which the {\bf Noether charges} are given by
\begin{equation}
\gamma^a=\phi^{an}\,\frac{\partial L}{\partial\dot{q}^n}\,-\,
\chi^a\left(\dot{q}^n\frac{\partial L}{\partial\dot{q}^n}\,-\,L\right)\,-\,\Lambda^a.
\label{J1.eq:NoetherCharges}
\end{equation}
These identities, one for each independent value of the index $a$, namely each
independent Lie group parameter hence generator, thus provide the statement of
the first Noether theorem. If the Lie group $G$ is of dimension $N_G$, there exist
$N_G$ independent linear combinations of the Euler--Lagrange equations of motion
which are total time derivatives. As a corollary, it also follows that given any
solution to the equations of motion there exist $N_G$ quantities which are conserved,
namely constants of motion of the system.

This conclusion is remarkable indeed. Lie symmetries are statements of a purely algebraic
and geometric character, yet when they apply to the dynamics of a system they imply
restrictions so powerful that independently of the explicit knowledge of any solution
in analytic form (which is often indeed impossible), one nevertheless knows it to be true for a fact
that there exists a collection of conserved quantities in direct correspondence with each
of the independent symmetry generators of the Lie group. To each of the Lie group generated
symmetries, parametrised by the variable $\alpha_a$, there corresponds a conserved charge,
$\gamma^a$, the Noether charge. This fundamental result gives a profound insight into
the very reason for the possibility of conserved quantities. Furthermore as is discussed
hereafter, the expressions for the Noether charges $\gamma^a$ readily extend to phase space
and the Hamiltonian formulation of the dynamics. Within that framework, it turns out that
the Noether charges possess an algebra of Poisson brackets such that, on the one hand
this algebra is isomorphic to the abstract algebra of Lie brackets of the Lie symmetry
group $G$ of which they are the conserved charges, and on the other hand, the Noether
charges generate through their Poisson brackets with phase space observables the
infinitesimal (or linearised) symmetry transformations of the phase space coordinates and observables
under the (connected component of the) Lie symmetry group $G$. From that point of view, and as already mentioned earlier,
when a quantisation of the system preserves the structure of the algebra of Noether charges
the same conclusions extend to the Hilbert space of quantum states, with in particular
conservation of the Noether charges. Quantum states may then be classified in terms of
linear representations of the symmetry group $G$. Their eigenvalues for a maximal abelian
subgroup of $G$ (namely, a subset of which all Noether charges commute with one another and may
thus be diagonalised simultaneously) then define conserved {\bf quantum numbers} for the
quantised system. As an example the conservation of the electric charge immediately comes
to mind. Indeed in that case the electric charge is but the Noether charge associated to the
U(1) phase symmetry of the electromagnetic interaction. As we know this symmetry is not only
a global one, hence leading to a Noether charge which is the electric charge that matter
degrees of freedom carry when they couple to the electromagnetic field, but is also a
local gauged symmetry with further consequences.

As mentioned already in the case of a local Lie symmetry group, further statements following
from the fundamental identity (\ref{J1.eq:transf}) imply two more Noether theorems. One of the
consequences of these additional theorems is then that given any solution to the Euler--Lagrange
equations of motion, some of which are actually constraint equations rather than genuine
dynamical equations of motion, the Noether charges $\gamma^a$ are not only conserved but
vanish identically, $\gamma^a=0$. Indeed any such solution should thus also be gauge invariant,
namely not transform at all under the symmetry, which is possible only provided these charges
all vanish. Such configurations, within the context of gauge invariant dynamics, are called
{\bf physical configurations} or {\bf physical states}. In the case of the electromagnetic
interaction, this condition would translate in an identically vanishing total electric charge
of a physical system interacting with the electromagnetic field, thus including also the
latter which extends through all of space and time.

\subsection{The Noether charge algebra}

Given the expression for the Noether charges,
\begin{equation}
\gamma^a=\phi^{an}\,\frac{\partial L}{\partial\dot{q}^n}\,-\,
\chi^a\left(\dot{q}^n\frac{\partial L}{\partial\dot{q}^n}\,-\,L\right)\,-\,\Lambda^a,
\end{equation}
it is clear that these quantities are readily defined over phase space as the following
observables,
\begin{equation}
\gamma^a(q^n,p_n)=\phi^{an}(q^n,t)p_n-\chi^a(t)\,H_0(q^n,p_n)\,-\,\Lambda^a(q^n,t),
\end{equation}
where $H_0(q^n,p_n)=\dot{q}^n\,p_n-L(q^n,\dot{q}^n)$ is the canonical Hamiltonian.
Note that even though for a given solution to the equations of motion (whether in
Lagrangian or Hamiltonian form) these quantities are conserved with a time independent
value, their kinematical expression as phase space observables may carry an explicit
time dependence through the functions $\chi^a(t)$, $\phi^{an}(q^n,t)$ and $\Lambda^a(q^n,t)$,
depending on the Lie symmetry group $G$ under consideration.

Given this observation, the immediate question which arises is to determine the algebra
of Poisson brackets of the Noether charges over phase space. It is indeed possible to
give an explicit answer to that question through a careful analysis of the consequences
following from the fundamental identity (\ref{J1.eq:transf}) within the Hamiltonian
framework. The details of the argument are not discussed in these notes.

In order to describe the result for the algebra of Poisson brackets of the Noether
charges, let us first consider the abstract Lie symmetry group $G$ and the algebra of
its abstract generators $T^a$. Being a continuous Lie group of transformations, any of its
elements in the component connected to the identity transformation may be written as
\begin{equation}
g[\alpha]=e^{i\alpha_a\,T^a},
\end{equation}
$\alpha_a$ being the coordinate parameters of the Lie group as a differential manifold,
and $T^a$ the abstract generators of the Lie algebra associated to the Lie group $G$.
The presence of the factor $i$ is a physicist's convention, since within a quantum context
one needs such transformations to be also unitary ones in order to preserve quantum
probabilities of quantum states under the symmetry transformations, hence one needs self-adjoint
generators $T^a$ if that factor $i$ is included, ${T^a}^\dagger=T^a$. Now, given the
group composition rules for group elements $g[\alpha]$ and $g[\beta]$, it follows that
for the linearised or infinitesimal form of these group transformations, the algebra of
generators $T^a$ must possess Lie brackets of the form\footnote{A Lie algebra is defined
in terms of a Lie bracket, in a manner similar to that in which a phase space is defined
in terms of Poisson brackets. For all practical purposes in these notes one may think of
the Lie bracket as the ordinary commutator of abstract operators, or even matrices, since
one is implicitly interested in Hilbert space realisations of the abstract Lie group
symmetry and its algebra.}
\begin{equation}
[T^a,T^b]=i{C^{ab}}_c\,T^c,
\end{equation}
where the constant coefficients ${C^{ab}}_c$ (the summation convention is again implicit)
are known as {\bf the structure constants} or structure coefficients of the Lie algebra of the Lie group $G$.
In the case of a compact Lie group $G$, such as SU(2) or more generally the unitary
groups SU(N), these structure coefficients are real numbers. Note that they are also
antisymmetric in their first two indices. Again in the case of compact simple semi-simple
compact Lie groups, by introducing the positive definite Killing form on the algebra
which then defines an hermitian metric on that vector space, it becomes possible to raise
and lower the indices $a,b,c$ (which transform in the adjoint representation of the algebra
and group), in which case the structure coefficients $C^{abc}$ or $C_{abc}$ are totally
antisymmetric in all three indices.

As an illustration, consider the group\footnote{The reader is invited to consider the case
of rotations in the two dimensional plane with symmetry group SO(2), and derive the
same considerations, to conclude that the algebra is that of the group U(1) of phase
transformations in the complex plane, with a single generator hence an abelian algebra
and thus a vanishing structure constant.} SO(3)
of all orientation preserving rotations in three dimensional Euclidean space. Rather than using for instance the
Euler angle parametrisation of that group, it should be clear that any element of $G=$SO(3)
may be obtained through the composition of the three independent rotations around the three
cartesian coordinate axes $x^i$ ($i=1,2,3$) with an arbitrary angle $\theta_i$ in the range
$0\le\theta_i\le 2\pi$. Any such transformation thus corresponds to an element of the form
\begin{equation}
g_i[\theta_i]=e^{i\theta_i\,T^i}\qquad
[\mbox{no summation over the index}\ i]
\end{equation}
with a generator $T^i$ (the index $i$ is not summed over in this expression). In order to
identify the algebra of SO(3), let us use the defining representation of SO(3) in terms of
$3\times 3$ orthogonal matrices of unit determinant acting on three component vectors. It is
well known how the matrix representations of each of the above elements $g_i[\theta_i]$
are expressed, for example for the rotation of angle $\theta_3$ around the third axis $i=3$,
\begin{equation}
g_3[\theta_3]:\qquad
\left(\begin{array}{c c c}
\cos\theta_3 & \sin\theta_3 & 0 \\
-\sin\theta_3 & \cos\theta_3 & 0 \\
0 & 0 & 1 
\end{array}\right).
\end{equation}
Expanding such expressions to first order in the angles $\theta_i$, the matrix representations
of the generators $T^i$ are readily identified through $g_i[\theta_i]=\mathbb{I}+i\theta_i T^i+\cdots$,
leading to the following matrix elements
\begin{equation}
{\left(T^i\right)^j}_k=-i\epsilon^{ijk},\qquad i,j,k=1,2,3,
\end{equation}
$\epsilon^{ijk}$ being the usual totally antisymmetric invariant tensor in three dimensional
Euclidean space, with $\epsilon^{123}=+1$. Simple matrix multiplication then finds that the algebra
of commutators of these SO(3) generators, defining the Lie brackets of the abstract Lie algebra $so(3)$
of the finite compact Lie group SO(3), is
\begin{equation}
\left[T^i,T^j\right]=i\,\epsilon^{ijk}\,T^k,
\end{equation}
where the summation of the repeated index $k$ is implicit and to be understood. Hence the structure
constants of the Lie algebra $so(3)$ are indeed real and totally antisymmetric structure coefficients,
given by the antisymmetric tensor $\epsilon^{ijk}$. It is left as an exercise for the reader to
repeat the analysis for the compact Lie group SU(2) of unitary $2\times 2$ matrices of unit determinant
over the complex numbers, to discover that the algebra $su(2)$ of that Lie group in fact coincides
with that of $so(3)$, $su(2)=so(3)$. Incidentally, this is precisely what has been observed when
finding the symmetry reason behind the degeneracies of the energy spectrum of the two dimensional
spherically symmetric harmonic oscillator. This equivalence between a unitary group Lie algebra,
$su(2)$, and an orthogonal group Lie algebra, $so(3)$, is specific to this particular case.

Having thus introduced the concepts of generators and structure constants of the Lie algebra
associated to a Lie symmetry group, let us consider the issue of the Poisson brackets of the
Noether charges $\gamma^a$. First, in the case that no total time derivative term $\Lambda(q^n,t)$
is induced in the action through the symmetry transformation, it may be shown that the algebra
of these Poisson brackets is always given as
\begin{equation}
\left\{\gamma^a,\gamma^b\right\}={C^{ab}}_c\,\gamma^c
\label{J1.eq:Noetheralgebra1}
\end{equation}
(the summation convention is implicit), precisely in terms of the structure constants ${C^{ab}}_c$
of the Lie algebra of the symmetry group $G$. Except for a factor $i$ which may easily be introduced
by considering the Poisson brackets of the pure imaginary quantities $i\gamma^a$, this algebra of
Poisson brackets thus coincides with the Lie bracket of the abstract Lie algebra of $G$. Note that at
the quantum level, through the correspondence principle, the Noether charge operators $\hat{\gamma}^a$
should then obey the algebra of commutation relations
\begin{equation}
\left[\hat{\gamma}^a,\hat{\gamma}^b\right]=i\hbar{C^{ab}}_c\,\hat{\gamma}^c,\qquad
\left(\hat{\gamma}^a\right)^\dagger=\hat{\gamma}^a,
\end{equation}
if the symmetry is to be realised in the quantised system as well (operator ordering issues are
at play here and one has to identify a quantisation which preserves these commutation relations
while also producing self-adjoint Noether charges $\hat{\gamma}^a$). Except for the factor $\hbar$
which is easily absorbed through a rescaling of the Noether charges, $\hat{\gamma}^a/\hbar$,
one then has identically the algebra of the abstract Lie algebra, realised in terms of the
quantum observables $\hat{\gamma}^a/\hbar$ on the Hilbert space of the quantised system.
In particular, this implies that finite Lie symmetry group transformations are realised on
the Hilbert space of quantum states by the unitary operators
\begin{equation}
e^{\frac{i}{\hbar}\alpha_a\hat{\gamma}^a},
\end{equation}
defined in terms of the properly normalised Noether charge operators $\frac{1}{\hbar}\hat{\gamma}^a$.

These results thus establish that indeed the Noether charges, as conserved quantities the existence
of which follows from the symmetries, are also the generators of these symmetries, either on phase
space through their Poisson brackets with the phase space coordinates and other observables in the
classical context, or on Hilbert space through their action on quantum states as quantum
operators. Conserved charges are the generators of the symmetries of which they are the Noether
charges. In particular at the quantum level, selecting a maximal subset of Noether charges which
all commute with one another, namely a maximal abelian subalgebra, all these operators may be
diagonalised simultaneously, thereby leading to a basis of states of Hilbert space each of
which element carries specific conserved quantum charges under the symmetry,
namely the associated eigenvalues under the symmetries of
one of its maximal abelian subgroups. For instance returning to the example of SO(3), any maximal
abelian subalgebra of $so(3)$ is one dimensional. Taking for instance the subalgebra generated by
$T^3$, namely rotations around the axis $i=3$, quantum states are then classified in terms of their
charge under $T^3$, namely their angular-momentum component along the axis $i=3$ in units
of $\hbar$. Because of the structure of representations of $so(3)=su(2)$ labelled by
the integer or half-integer spin value $j$ as already discussed in the context of the two dimensional
spherically symmetric harmonic oscillator, these components of angular-momentum for an arbitrary
system are quantised in the range $-j\le m\le j$ in integer steps for the eigenvalues $m$ of $T^3$.

Since there are examples of symmetries of physical relevance which induce a total time derivative
in the action, let us now describe the general result. Though not necessarily
always the case, it may then happen that the algebra of Poisson brackets of the Noether
charges is no longer of the form (\ref{J1.eq:Noetheralgebra1}) but acquires an extra contribution
through one extra term independent of time and the phase space variables, a so-called 
central extension of the Lie algebra of $G$ since this extra contribution has vanishing Poisson
brackets or quantum commutation relations with all elements of the algebra of observables of the
system, whether classical or quantised. Namely one has in general the expression
\begin{equation}
\left\{\gamma^a,\gamma^b\right\}={C^{ab}}_c\,\gamma^c\,+\,C^{ab},
\end{equation}
where the antisymmetric constant coefficients $C^{ab}$ are obtained as follows. Under the group composition law,
one has a certain composition law for the group parameters $\alpha_a$, namely,
\begin{equation}
g[\alpha]\,g[\beta]=g[\eta(\alpha,\beta)].
\end{equation}
As a matter of fact, the structure coefficients of the Lie algebra of $G$ are given as follows
from the functions $\eta_a(\alpha,\beta)$,
\begin{equation}
\frac{\partial^2}{\partial\alpha_a\partial\beta_b}
\left[\eta_c(\alpha,\beta)\,-\,\eta_c(\beta,\alpha)\right]_{|_{\alpha=0=\beta}}=
{C^{ab}}_c.
\end{equation}
By considering then how the Lagrange function transforms under such a composition of
symmetry transformations it may be shown that one has necessarily,
\begin{equation}
\Lambda\left(q^n,t;\gamma(\alpha,\beta)\right)\,-\,
\Lambda\left({q^n}'(q^n,t;\beta),t'(t;\beta);\alpha\right)\,-\,
\Lambda\left(q^n,t;\beta\right)=C\left(\alpha,\beta\right),
\end{equation}
where $C(\alpha,\beta)$ are thus specific functions of the group parameters, in fact
associated to cocycle properties of the symmetry group encoded through the function
$\Lambda(q^n,t;\alpha)$. The quantities $C^{ab}$ appearing in the above general Poisson
brackets of the Noether charges are then given as
\begin{equation}
C^{ab}=\frac{\partial^2}{\partial\alpha_a\partial\beta_b}
\left[C(\alpha,\beta)\,-\,C(\beta,\alpha)\right]_{|_{\alpha=0=\beta}}.
\end{equation}
Incidentally, it may easily be checked that a redefinition of the Lagrange function
by a total time derivative, thus leaving the equations of motion invariant, does not
modify these central extension coefficients $C^{ab}$.

What is interesting about this result is that central extensions are often believed
to arise only at the quantum level, and to correspond then to an explicit breakdown of
the classical symmetry generated by the Noether charges, a phenomenon called 
{\bf a quantum anomaly}. Here we see that classical extensions are also possible,
provided the action varies under a symmetry with a total time derivative which itself
must possess some nontrivial cocycle property, as a consequence of some nontrivial
topology properties in the configuration space of the system. Examples are described
hereafter. At the quantum level, the symmetry is then realised not through a faithful
representation of the symmetry group (the type of representation encountered so
far in terms simply of the action of ${\rm exp}\,(\frac{i}{\hbar}\alpha\hat{\gamma}^a)$),
but in terms of what is called {\bf a projective representation} of the abstract
Lie symmetry group such that symmetry transformations of quantum states are obtained not
only through the action of the unitary operators $e^{\frac{i}{\hbar}\alpha_a\hat{\gamma}^a}$
but also some further phase factor which is function of the group parameters $\alpha_a$ and
in direct correspondence with the total time derivative contribution $\Lambda(q^n,t;\alpha)$
(as may heuristically be understood from the path integral representation of quantum
physics). Even though we shall not make use later on of such features specific to certain
classes of Lie group symmetries, it is important to know about the existence of such
particular situations and the possibility of classical central extensions of Lie algebras
in the classical Poisson bracket algebra of Noether charges.

\vspace{10pt}

\noindent
\underline{\bf Remark}

\vspace{10pt}

One final remark concerning symmetries and their Noether charges may be made in order to
conclude this discussion. The result (\ref{J1.eq:NoetherCharges}) gives the general expression
for these charges in terms of the functions parametrising the linearised variations of
$t$, $q^n$ and the induced total time derivative term $\Lambda$ under the symmetry group $G$.
However, in practice and in almost all cases (except for field theories of supergravity
to the best of this author's knowledge) one may rather easily identify the Noether charges
through the following little trick. In the case that the symmetry parameters $\alpha_a$ are
constant, the action is invariant up to a total time derivative term,
\begin{equation}
S[{q^n}']=S[q^n]+\int dt\ {\rm t.t.d.},
\end{equation}
where ``t.t.d." stands for some unspecified total time derivative contribution. In the case
of a linearised or infinitesimal symmetry transformation, one thus has
\begin{equation}
\delta S[q^n]=\int dt\ {\rm t.t.d.}\ .
\end{equation}
Imagine now for the sake of the argument that the symmetry transformation is considered
for parameters $\alpha_a(t)$ which are arbitrary functions of time. Except when in fact the
symmetry is a local gauge symmetry, the action may then no longer change just by a total
time derivative, since it is then no longer invariant up to such a term.
Rather in general it is expected then not to be invariant up to such terms, but to vary
by terms involving the first order time derivative of the symmetry parameters $\alpha_a(t)$
since the Lagrange function is function of both $q^n(t)$ and $\dot{q}^n(t)$. Hence we should
expect a linearised variation of the form
\begin{equation}
\delta S[q^n]=\int dt\,\left[\frac{d\alpha_a}{dt}\,Q^a\,+\,{\rm t.t.d.}\right],
\end{equation}
where $Q^a(q^n,\dot{q}^n,t)$ are some quantities which could {\it a priori\/} be functions
of the generalised coordinates and their velocities as well as time. Clearly when the
parameters $\alpha_a(t)$ are constants one recovers the symmetry property of the action.
In the latter form one may now integrate by parts and bring that variation of the
action into the form
\begin{equation}
\delta S[q^n]=\int dt\,\left[-\alpha_a\,\frac{dQ^a}{dt}\,+\,{\rm t.t.d.}\right].
\end{equation}
However, since when the parameters $\alpha_a(t)$ are constants the action may only change
by a total time derivative, one must conclude that the quantities $Q^a$ must be conserved
quantities, $dQ^a/dt=0$. In other words, possibly up to an overal sign, necessarily the quantities
$Q^a(q^n,\dot{q}^n,t)$ must coincide with the Noether charges $\gamma^a(q^n,\dot{q}^n,t)$
associated to the symmetry leaving the action invariant up to total time derivatives, which are
in fact readily defined over the phase space of the system. This method is quite generally sufficient
in order to readily identify the Noether charges
given a symmetry of the action, rather than going through the identification of the
functions $\chi^a(t)$, $\phi^{an}(q^n,t)$ and $\Lambda^a(q^n,t)$ introduced in the above general
discussion of Noether's first theorem.

\subsection{Illustrations}

\subsubsection{Time translation invariance}

As an illustration of the general discussion of Noether's first theorem,
let us begin by considering any dynamical system described by an action of the type we have
been using ever since the beginning of these notes,
\begin{equation}
S[q^n]=\int\,dt\,L(q^n,\dot{q}^n).
\end{equation}
Having excluded from the outset any explicit time dependence of the Lagrange function,
it is quite obvious that this dynamics is invariant under arbitrary constant translations
in the time variable $t$. Any transformation of the form
\begin{equation}
t'=t+t_0,\qquad
{q^n}'(t')=q^n(t),
\end{equation}
defines an invariance of the action, with no total time derivative term being
induced, $\Lambda(q^n,t)=0$, simply because we have all along assumed the Lagrange
function not to possess any explicit time dependence. This group of symmetries is
a one dimensional Lie group with as continuous parameter the constant quantity $t_0$ defining
the translation in time. There thus exists a conserved charged associated to
time translation invariance in the evolution parameter $t$ used to parametrise
the dynamics of the system.

For what concerns then linearised or infinitesimal transformations, the index $a$ of
the variables $\alpha_a$ of the general discussion takes only a single value in
the present case, $a=1$. Furthermore, one readily identifies
\begin{equation}
\chi^a(t)=1,\qquad \phi^{an}(q^n,t)=0,\qquad \Lambda^a(q^n,t)=0.
\end{equation}
By direct substitution into the definition (\ref{J1.eq:NoetherCharges}) of the Noether
charges, one finds that the Noether charge associated to this symmetry is
\begin{equation}
\gamma^a=-\left[\dot{q}^n\frac{\partial L}{\partial\dot{q}^n}\,-\,L\right]=
-\,H_0(q^n,p_n).
\end{equation}
Hence in full generality, whenever a system is invariant under translations in its
time evolution parameter, its canonical Hamiltonian is a conserved quantity,
a constant of motion which coincides with the Noether charge for that symmetry.
Furthermore, this Noether charge indeed does generate infinitesimal translations in the time
evolution parameter, namely it is the generator of the time dependence of
the system through the Poisson brackets. This latter fact is an explicit illustration
of the general result that Noether charges are the generators of the symmetries of which
they are the conserved quantities.

This conclusion applies likewise at the quantum level as may easily be seen. Let us
consider an orthonormalised basis of Hilbert space consisting of the energy eigenstates $|E_m\rangle$
where, in keeping with our previous generic notation, in general $m$ stands for a multi-index some
of which components could even take values in a continuous range. As we know the exponentiated action
of the quantum Hamiltonian on quantum states generates their time evolution,
\begin{equation}
|\psi,t\rangle=e^{-\frac{i}{\hbar}(t-t_0)\hat{H}}\,|\psi,t_0\rangle=
\sum_m\,|E_m\rangle\,e^{-\frac{i}{\hbar}(t-t_0)E_m}\,\langle E_m|\psi,t_0\rangle,
\end{equation}
which shows indeed that the quantum Hamiltonian is the generator of constant translations
in time, in the present case by the value $(t-t_0)$ for the quantum evolution operator
$U(t,t_0)=e^{-\frac{i}{\hbar}(t-t_0)\hat{H}}$.

Note that the time evolution parameter does not, in general, necessarily coincide with the physical
time, when one considers parametrised systems such as the relativistic particle, string
theory and general relativity. However, when the time evolution parameter coincides
with the physical time, the above conservation law shows that in direct correspondence
with the invariance of a dynamics under arbitrary time translations its total energy is a conserved
quantity, namely the Noether charge generating that symmetry.

\subsubsection{Nonrelativistic particles}

In order to display other examples of direct relevance even to nonrelativistic
Newton dynamics, we now consider systems of nonrelativistic particles in interaction
through some conservative forces in different situations, and then identify each of the
conservation laws of mechanical energy, momentum and angular-momentum. Consider a system
of $N$ particles of masses $m_\alpha$ ($\alpha=1,2,\ldots,N$), of position vectors $\vec{r}_\alpha(t)$
with respect to some inertial frame, and subjected to conservative forces of total potential
energy $V(\vec{r}_\alpha-\vec{r}_\beta)$, namely some function of pairwise differences of the
position vectors. As is well known the Lagrange function for such a system is
\begin{equation}
L(\vec{r}_\alpha,\dot{\vec{r}}_\alpha)=\sum_{\alpha=1}^N\frac{1}{2}m_\alpha
\dot{\vec{r}}_\alpha\,^2\,-\,V(\vec{r}_\alpha-\vec{r}_\beta).
\end{equation}
For later purposes, the cartesian coordinates of the particles will be denoted
$x^i_\alpha$ with $i=1,2,3$ in three dimensional Euclidean space. These are the
configuration space degrees of freedom of the system, the index $n$ of the general
discussion standing here for the double index $(i,\alpha)$ taking $3N$ values.

\vspace{10pt}

\noindent
\underline{\sl Time translation invariance}

\vspace{10pt}

{}From the previous example in the general case, at once we know that the canonical
Hamiltonian of this system is the Noether charge for time translation invariance
of this system, since the Lagrange function does not possess any explicit time
dependence (this would thus no longer be the case had the particles been coupled
to a time dependent background electromagnetic field $\vec{E}(t,\vec{r}\,)$
and $\vec{B}(t,\vec{r}\,)$, as was already established explicitly earlier in
these notes). Since the time evolution parameter $t$ of the system is also the
physical time, we conclude that the total mechanical energy of the system, namely
its canonical Hamiltonian,
\begin{equation}
H_0=\sum_{\alpha=1}^N\frac{1}{2m_\alpha}\vec{p}_\alpha\,^2\,+\,
V(\vec{r}_\alpha-\vec{r}_\beta),
\end{equation}
is the conserved Noether charge generating infinitesimal time translations.

\vspace{10pt}

\noindent
\underline{\sl Space translation invariance}

\vspace{10pt}

Given that the total potential energy is taken to be function only of the pairwise
differences in the position vectors of the particles, the system is also invariant
under arbitrary constant translations in space, namely
\begin{equation}
t'=t,\qquad
\vec{r}\,'_\alpha(t')=\vec{r}_\alpha(t)+\vec{r}_0.
\end{equation}
These transformations do not induce any total time derivative term in the action,
$\Lambda(\vec{r}_\alpha,t)=0$. The continuous parameters of this Lie symmetry group
are the three cartesian components of the translation vector, $\vec{r}_0=\left(r_{0,a}\right)$,
distinguished by the label $a=1,2,3$ in order to use the notations of the general
discussion and avoid any confusion with the cartesian index $i$ of the coordinates $x^i_\alpha$.
When linearised in the parameters $r_{0,a}$, the above transformations imply
\begin{equation}
\chi^a(t)=0,\qquad
\phi^{ai}_\alpha(\vec{r}_\alpha,t)=\delta^{ai},\qquad
\Lambda^a(\vec{r}_\alpha,t)=0.
\end{equation}
Consequently, using the general definition (\ref{J1.eq:NoetherCharges}), the Noether charges
generating the space translation invariance of the system are
\begin{equation}
\gamma^a=\sum_{\alpha=1}^N\sum_{i=1}^3\phi^{ai}_\alpha\,p_{\alpha,i}=
\sum_{\alpha=1}^N\,p^a_\alpha,
\end{equation}
where $p^a_\alpha=m \dot{x}^a_\alpha$ are the conjugate momenta of the coordinates $x^a_\alpha$, namely
the cartesian components of the momentum of particle $\alpha$.

In conclusion, the Noether
charges related to invariance under translations in space are the components of the total
momentum vector of the system of particles,
\begin{equation}
\vec{\gamma}=\sum_{\alpha=1}^N\,\vec{p}_\alpha=\vec{P}.
\end{equation}
Again this is a most general result valid for any system invariant under space translations.

This conclusion remains also valid at the quantum level. For the sake of pointing this out
explicitly, let us restrict to a single degree of freedom system of cartesian coordinate $x(t)$
with $p(t)$ as its conjugate momentum. At the quantum level the operators $\hat{x}$ and
$\hat{p}$ obey the Heisenberg algebra, $\left[\hat{x},\hat{p}\right]=i\hbar$. Considering for instance
the configuration space representation in terms of configuration space wave functions
$\psi(x)=\langle x|\psi\rangle$, $|x\rangle$ being the normalised position eigenstates, we know that
the momentum operator is represented as,
\begin{equation}
\hat{p}:\qquad -i\hbar\frac{d}{dx}\psi(x).
\end{equation}
In other words, we have,
\begin{equation}
\frac{i}{\hbar}\hat{p}:\qquad \frac{d}{dx}\psi(x).
\end{equation}
That this differential operator is indeed the generator for infinitesimal translations in space
follows from the Taylor series expansion
\begin{equation}
\psi(x+a)=\psi(x)+a\frac{d\psi(x)}{dx}+\frac{1}{2!}a^2\frac{d^2\psi(x)}{dx^2}+\cdots=
\sum_{n=0}^\infty\frac{1}{n!}a^n\frac{d^n\psi(x)}{dx^n}=e^{a\frac{d}{dx}}\,\psi(x)=
e^{\frac{i}{\hbar}a\hat{p}}\,\psi(x).
\end{equation}
An alternative way of seeing this result without relying on the wave function representation
is by considering the following action of the exponentiated operator $\hat{p}$,
\begin{equation}
e^{-\frac{i}{\hbar}a\,\hat{p}}\,|x\rangle.
\end{equation}
In order to identify which quantum state is obtained, let us act on it with the position operator $\hat{x}$ and
use one of the Baker--Campbell--Hausdorff formulae,
\begin{eqnarray}
\hat{x}\,e^{-\frac{i}{\hbar}a\hat{p}}\,|x\rangle &=&
e^{-\frac{i}{\hbar}a\hat{p}}\,e^{\frac{i}{\hbar}a\hat{p}}\,\hat{x}\,e^{-\frac{i}{\hbar}a\hat{p}}|x\rangle \nonumber \\
&=& e^{-\frac{i}{\hbar}a\hat{p}}\left(\hat{x}+\left[\frac{i}{\hbar}a\hat{p},\hat{x}\right]\right)\,|x\rangle \nonumber \\
&=& e^{-\frac{i}{\hbar}a\hat{p}}\left(x+a\right)\,|x\rangle \nonumber \\
&=& \left(x+a\right)\,e^{-\frac{i}{\hbar}a\hat{p}}\,|x\rangle.
\end{eqnarray}
In other words, the state obtained by the action of $e^{-\frac{i}{\hbar}a\hat{p}}$ on $|x\rangle$
is necessarily proportional to the position eigenstate $|x+a\rangle$. In order to identify the corresponding
coefficient or component, we need only project it onto any of the position eigenstate basis vectors $| x'\rangle$,
\begin{eqnarray}
\langle x'|e^{-\frac{i}{\hbar}a\hat{p}}|x\rangle &=&
\int_{-\infty}^{+\infty}dp\,\langle x'|e^{-\frac{i}{\hbar}a\hat{p}}|p\rangle
\langle p|x\rangle \nonumber \\
&=& \int_{-\infty}^{+\infty}\frac{dp}{2\pi\hbar}\,e^{\frac{i}{\hbar}(x'-x-a)p} \nonumber \\
&=& \delta\left(x'-(x+a)\right).
\end{eqnarray}
In other words, we have
\begin{equation}
e^{-\frac{i}{\hbar}a\,\hat{p}}\,|x\rangle=|x+a\rangle .
\end{equation}
From this it also follows that we have for any state $|\psi\rangle$,
\begin{equation}
\langle x|e^{\frac{i}{\hbar}a\,\hat{p}}|\psi\rangle=\langle x+a|\psi\rangle=
\psi(x+a).
\end{equation}
Hence indeed the exponentiation $e^{\frac{i}{\hbar}a\hat{p}}$ of the conjugate momentum
operator $\hat{p}$ generates a finite translation by the value $a$ in the configuration
space coordinate $x$, while $\frac{i}{\hbar}\hat{p}$ is the generator of infinitesimal
translations in $x$. Incidentally, the very last of these relations may also be used as
a starting point to establish the functional representations of the quantum operators $\hat{x}$
and $\hat{p}$ in the configuration space wave function representation of the Hilbert space
realising the Heisenberg algebra.

\vspace{10pt}

\noindent
\underline{\sl Space rotation invariance}

\vspace{10pt}

In order to address the consequences of invariance under space rotations, let us now furthermore
assume that the total potential energy is function only of the pairwise distances
$|\vec{r}_\alpha-\vec{r}_\beta|$ between the particles, $V(|\vec{r}_\alpha-\vec{r}_\beta|)$.
Clearly in such a situation the action is invariant under arbitrary constant rotations
of the position vectors, namely with
\begin{equation}
t'=t,\qquad
\vec{r}\,'_\alpha(t)=R\cdot\vec{r}_\alpha(t),\qquad
\Lambda(\vec{r}_\alpha,t)=0,
\end{equation}
where $R$ stands for the rotation matrix acting on the components of each of the
position vectors $\vec{r}_\alpha$. The parameters of the corresponding Lie symmetry group
are three independent and continuous rotation angles. Using the previous discussion for
such transformations with as generators those that generate rotations of angle $\theta_a$
around the axis $a=1,2,3$, namely the matrices ${\left(T^a\right)^j}_k=-i\epsilon^{ajk}$ constructed
above, the linearised form of these symmetry transformations is
\begin{equation}
\delta t(t)=0,\qquad
\delta x^i_\alpha=i\theta_a{\left(T^a\right)^i}_j\,x^j_\alpha,\qquad
\delta\Lambda(\vec{r}_\alpha,t)=0.
\end{equation}
Consequently we identify
\begin{equation}
\chi^a(t)=0,\qquad
\phi^{ai}_\alpha(\vec{r}_\alpha,t)=\epsilon^{aij}\,x^j_\alpha,\qquad
\Lambda^a(\vec{r}_\alpha,t)=0.
\end{equation}
By substitution into (\ref{J1.eq:NoetherCharges}) the associated Noether charges are
\begin{equation}
\gamma^a=\sum_{\alpha=1}^N\sum_{i=1}^3\,\epsilon^{aij}\,x^i_\alpha\,p_{\alpha,j}.
\end{equation}
In other words, the Noether charges which generate space rotation invariance are the components
of the total angular-momentum vector of the system,
\begin{equation}
\vec{\gamma}=\sum_{\alpha=1}^N\,\vec{r}_\alpha\times\vec{p}_\alpha=\vec{L}.
\end{equation}
Once again this result is most general and applies to any system invariant under 
constant rotations in space.

It thus is quite remarkable that symmetry properties of space and time, in the present case the
symmetries of three and one dimensional Euclidean space and time, when also shared by the
dynamics of a system, imply the existence of conserved quantities whatever the configuration
of the system solving its equations of motion. These conserved quantities are also the generators of these symmetries
on phase space through the Poisson brackets of these conserved quantities with any phase
space observable, beginning with the phase space coordinates. Indeed, within the Hamiltonian
framework it readily follows that the above Noether charges do generate the corresponding
infinitesimal symmetry transformations of the configuration space coordinates, while
their Poisson brackets are isomorphic to the Lie algebra of the Lie symmetry group of which 
they are the Noether charges. In particular the Poisson brackets of the total momentum $\vec{P}$
and angular-momentum $\vec{L}$ components with the generator of time translations are
\begin{equation}
\left\{H,P^i\right\}=0,\qquad
\left\{H,L^i\right\}=0,
\end{equation}
indeed expressing the conservation of these quantities, whereas among themselves they possess
the Poisson brackets,
\begin{eqnarray}
\left\{P^i,P^j\right\} &=& 0,\nonumber \\
\left\{L^i,P^j\right\} &=& \epsilon^{ijk}\,P^k,\nonumber \\
\left\{L^i,L^j\right\} &=& \epsilon^{ijk}\,L^k
\end{eqnarray}
(summation over repeated indices is implicit). This algebra is recognised to be that of
the Euclidean group $E(3)$ in three dimensional Euclidean space. The first set of brackets
represents the fact that any two translations commute with one another, whereas the last
two sets of brackets represent the fact that under space rotations the vector generator of translations
transforms as a vector quantity, and likewise for the vector generator of space rotations.
Note also that the algebra of the angular-momentum coincides with the abstract $so(3)$
algebra constructed previously. This also implies that at the quantum level, this algebra
will be realised in the space of quantum states in terms of their spin values in the
manner already discussed.

\subsubsection{The free nonrelativistic particle}

Now as an illustration of a symmetry leading to a central extension in the algebra
of Poisson brackets of its Noether charges,
consider a single free nonrelativistic particle in Euclidean space,
\begin{equation}
L=\frac{1}{2}m\dot{\vec{q}}\,^2,\qquad
H_0=\frac{1}{2m}\,\vec{p}\,^2.
\end{equation}
Besides the symmetries discussed above, Newton's mechanics is known to be invariant
under Galilei boosts, namely a six dimensional Lie symmetry group, with
\begin{equation}
t'=t,\qquad
\vec{q}\,'(t')=\vec{q}(t)\,+\,\vec{a}\,+\,\vec{V}\,t,
\end{equation}
\begin{equation}
L'=L+m\,\vec{V}\cdot\dot{\vec{q}}\,+\,\frac{1}{2}m\vec{V}\,^2,\qquad
\Lambda\left(\vec{q},t;\vec{a},\vec{V}\right)=
m\,\vec{V}\cdot\vec{q}\,+\,\frac{1}{2}m\,\vec{V}\,^2\,t.
\end{equation}
Hence even though the simplest of systems, a total time derivative is induced
in the action under Galilei boosts, namely space translations which are no
longer constant but linear in time, or of constant velocity, bringing the
system from one inertial frame to another with the relative velocity $\vec{V}$.

Then, with indices $a,b$ in correspondence with the continuous Lie group parameters through
$a,b\leftrightarrow\left(\vec{a},\vec{V}\right)=\left(a^i,V^i\right)$
and $i=1,2,3$ labelling vector components, it readily follows that the coefficients
$C(\alpha,\beta)$ of the general discussion as well as the central
extension constants $C^{ab}$ are given as,
\begin{equation}
C\left(\vec{a}_1,\vec{V}_1;\vec{a}_2,\vec{V}_2\right)=-m\,\vec{V}_1\cdot\vec{a}_2,\qquad
C^{ab}=\left(\begin{array}{cc}
	0 & m\mathbb{I} \\
	- m\mathbb{I} & 0 \end{array}\right).
\end{equation}

Identifying then all the relevant quantities which enter the definition (\ref{J1.eq:NoetherCharges}),
the Noether charges are found to be
\begin{equation}
\gamma^a=\left(p^i,\gamma^i\right),\qquad
p^i=m\dot{q}^i,\qquad
\gamma^i=p^i\,t\,-\,mq^i,
\end{equation}
with the Noether algebra
\begin{equation}
\left\{p^i,p^j\right\}=0,\quad
\left\{p^i,\gamma^j\right\}=m\,\delta^{ij},\quad
\left\{\gamma^i,p^j\right\}=-m\,\delta^{ij},\quad
\left\{\gamma^i,\gamma^j\right\}=0,
\end{equation}
namely
\begin{equation}
\left\{\gamma^a,\gamma^b\right\}=C^{ab}.
\end{equation}
Hence indeed, the generators $\vec{\gamma}=\vec{p}\,t\,-\,m\vec{q}$ of
Galilei boosts have a Poisson algebra with a central extension term which is
the mass of the particle. Such a parameter does not appear in the abstract algebra
of the Galilei group, simply because no such physical parameter is then available.
But in the present system this symmetry is realised even though with a central
extension determined by the unique physical parameter available, the mass $m$ of
the free nonrelativistic particle.

Note than in spite of the explicit time dependence appearing in the definition
of the vector Noether charge generating Galilei boosts, this quantity is
conserved nonetheless. Indeed, the Hamiltonian equations of motion for
the above Noether charges are
\begin{equation}
\frac{dp^i}{dt}=\left\{p^i,H_0\right\}=0,\qquad
\frac{d\gamma^i}{dt}=\frac{\partial\gamma^i}{\partial t}\,
+\,\left\{\gamma^i,H_0\right\}=p^i-p^i=0,
\end{equation}
yet
\begin{equation}
\left\{\gamma^i,H_0\right\}=-p^i\ne 0
\end{equation}
since $\partial\gamma^i/\partial t =p^i \ne 0$. At the quantum level within the
Heisenberg picture of quantum physics, the Schr\"odinger equation for the
quantum operator $\hat{\vec{\gamma}}$ generating Galilei boosts has to be
adapted accordingly, to account for the explicit time dependence of that operator,
\begin{equation}
\hat{\gamma}(t)=\hat{p}(t_0)\,t-m\hat{\vec{q}}(t).
\end{equation}
For an arbitrary quantum operator $\hat{A}$ which may have
in a likewise manner an explicit time dependence, this extension of the Schr\"odinger
equation in the Heisenberg picture reads,
\begin{equation}
i\hbar\frac{d\hat{A}}{dt}=i\hbar\frac{\partial\hat{A}}{\partial t}\,+\,
\left[\hat{A},\hat{H}_0\right].
\end{equation}

\subsubsection{The free extended Landau problem}

As a final illustration of interest, let us consider a charged particle
confined to the two dimensional Euclidean plane and subjected to static and homogeneous
magnetic and electric fields, $\vec{B}$ and $\vec{E}$, the former perpendicular to
the plane and the latter lying within it. In the symmetric gauge this system is described
by the Lagrange function
\begin{equation}
L=\frac{1}{2}m\delta_{ij}\,\dot{x}^i\dot{x}^j\,-\,\frac{1}{2}qB\,\epsilon_{ij}\,
\dot{x}^i x^j\,+\,q\,x^i E_i,\qquad \epsilon_{12}=\epsilon^{12}=+1, \quad i,j=1,2.
\end{equation}

Clearly such a system is invariant under arbitrary translations in the plane, forming
a two parameter abelian Lie group of symmetries. Yet, the
action is not invariant, but transforms by a total time derivative,
\begin{equation}
t'=t,\qquad
{x'}^i(t')=x^i(t)\,+\,a^i,
\end{equation}
\begin{equation}
L'=L-\frac{1}{2}qB\,\epsilon_{ij}\dot{x}^i a^j\,+\,q\,a^i E_i,\qquad
\Lambda\left(\vec{x},t;\vec{a}\right)
=-\frac{1}{2}qB\,\epsilon_{ij}\,x^i a^j\,+\,q\,a^i E_i\,t.
\end{equation}
The associated cocycle function $C(\alpha,\beta)$ of the general discussion
is then found to be
\begin{equation}
C\left(\vec{a},\vec{b}\,\right)=-\frac{1}{2}qB\,\epsilon_{ij}\,a^i b^j,
\end{equation}
leading to the central extension coefficients
\begin{equation}
C^{ij}=-qB\,\epsilon^{ij}.
\end{equation}

The Noether charges that follow read
\begin{equation}
\gamma^i=p^i\,-\,\frac{1}{2}qB\,\epsilon^{ij} x^j\,-\,q E^i\,t,
\end{equation}
with the algebra
\begin{equation}
\left\{\gamma^i,\gamma^j\right\}=-qB\,\epsilon^{ij}=C^{ij},
\end{equation}
and indeed generate translations in the plane as may checked by computing their Poisson
brackets with the phase space coordinates $x^i$ and $p_i$.

Given the canonical Hamiltonian
\begin{equation}
H_0=\frac{1}{2m}\left[p^i+\frac{1}{2}qB\,\epsilon^{ij}\,x^j\right]^2\,-\,q\,x^i E^i,
\end{equation}
the Hamiltonian equations of motion of these generators of space translations are
\begin{equation}
\frac{d\gamma^i}{dt}=\frac{\partial\gamma^i}{\partial t}+\left\{\gamma^i,H_0\right\}=0,
\end{equation}
thus indeed these charges are conserved. Nevertheless, they possess an explicit
time dependence,
\begin{equation}
\left\{\gamma^i,H_0\right\}=-\frac{\partial\gamma^i}{\partial t}=q\,E^i\ne 0
\end{equation}

\vspace{10pt}

\noindent
\underline{\bf Remark} 

\vspace{10pt}

\noindent
It is possible to consider the motion of the magnetic center defined as
\begin{equation}
\vec{C}=\vec{x}\,+\,\frac{m}{q|\vec{B}\,|}\,\dot{\vec{x}}\times\vec{B},
\end{equation}
namely the instantaneous rotation center of the particle. Using the
Lorentz force equation of motion,
\begin{equation}
m\ddot{\vec{x}}=q\,\vec{E}\,+\,q\,\dot{\vec{x}}\times\vec{B},
\end{equation}
one finds
\begin{equation}
\dot{\vec{C}}=\frac{1}{|\vec{B}\,|}\vec{E}\times\vec{B}.
\end{equation}
Hence when the electric field $\vec{E}$ is present the magnetic center follows
a trajectory of constant velocity in a direction perpendicular to the electric field
and given by the vector product $\vec{E}\times\vec{B}$.

\section{Conclusions}
\label{Gov1.Sec6}

The main purpose of the present notes has been two-fold. On the one hand, to introduce to the
basic concepts and formalism of quantum physics in a language which probably emphasizes more the
algebraic and mathematical aspects of that general physics framework than what is usually found in
most textbooks or introductory notes on the subject. On the other hand, at the same time to highlight the
fundamental r\^ole of symmetries in physics, and in particular quantum physics, in relation both to Noether's
(first) theorem and to the representation theory of symmetry groups. All throughout, the general discussion
put within the context of mechanical systems rather than field theories, has been illustrated with rather
familiar examples, and yet, examples which already provide very useful insight into what lies in store
beyond the contents of these notes. For instance, it has been indicated how among symmetries those that are local,
namely the realisation of the principle of gauge invariance, do play a fundamental r\^ole in the modern theories
of the fundamental interactions, be they classical or quantum.

And as a matter of fact the ambition of these notes has been to bring the reader onto the threshold from
where he/she may now embark on his/her own into the study of quantum field theories, their relativistic
quantum particle interpretation and the perturbative description of their interactions. By extending to the
relativistic context the methods and concepts developed in these notes, one quickly comes to the conclusion
that relativistic quantum field theories and relativistic quantum particles are just two dual aspects
of a common underlying physical reality, that of quantum interactions in a relativistic spacetime of
unified matter and radiation phenomena. How this is indeed achieved is discussed in quite many textbooks on
Quantum Field Theory, such as in Ref.\cite{Gov1.Peskin}, leading then to the remarkable achievements of perturbative
renormalisation theory. However, an introductory exposition to that subject in the spirit of the present notes, and which builds
on its contents, may also be found in lectures notes available from other Volumes in the present
Proceedings Series \cite{Gov1.Gov1, Gov1.Gov2}.

Simply, by having interwoven the mathematics and the physics of quantum physics in the present introductory
discussion of the concepts of the quantum world, it is hoped that both our more mathematics or more physics
inclined readers will find sufficient inspiration from these notes to launch their own line of study and
research into this world of physics at the frontiers of today in both mathematics and physics, in search of
a fundamental unification of the conceptual representation of the physical Universe.

\section*{Acknowledgements}

These notes grew out of lectures combining different material, delivered not only at the Third International COPROMAPH Summer School
in Cotonou, but also on other occasions, in particular twice already at the African Institute for
Mathematical Sciences (AIMS, {\tt http://www.aims.ac.za/english/}, Muizenberg, South Africa).
The author is most grateful to all participants to these lectures, whether in Benin or at AIMS, for their keen and genuine
interest and their many inquisitive questions, and wishes to thank Profs.~Fritz Hahne (AIMS) and M. Norbert Hounkonnou (ICMPA-UNESCO)
for their support and invitation to deliver these lectures. The author also acknowledges the Abdus Salam International
Centre for Theoretical Physics (ICTP, Trieste, Italy) Visiting Scholar Programme in support of
a Visiting Professorship at the ICMPA-UNESCO (Republic of Benin).
His work is also supported by the Institut Interuniversitaire des Sciences Nucl\'eaires, and by
the Belgian Federal Office for Scientific, Technical and Cultural Affairs through
the Interuniversity Attraction Poles (IAP) P6/11.


\begin{thebibliography}{99}

\bibitem{Gov1.Gov1}
J. Govaerts, {\it The Quantum Geometer's Universe: Particles, Interactions and Topology},
in the {\sl Proceedings of the Second International Workshop on Contemporary Problems in Mathematical Physics},
eds. J.~Govaerts, M.~N.~Hounkonnou and A.~Z.~Msezane (World Scientific, Singapore, 2002), pp.~79--212
[e-print {\tt arXiv:hep-th/0207276} (July 2002)].

\bibitem{Gov1.Gov2}
J. Govaerts, {\it On the Road Towards the Quantum Geometer's Universe:
An Introduction to Four-Dimensional Supersymmetric Quantum Field  Theories},
in the {\sl Proceedings of the Third International Workshop on Contemporary Problems
in Mathematical Physics}, eds. J. Govaerts, M. N. Hounkonnou and A. Z. Msezane (World Scientific, 
Singapore, 2004), pp.~94--150 [e-print {\tt arXiv:hep-th/0408021} (August 2004)].

\bibitem{Gov1.GovBook}
J. Govaerts, {\sl Hamiltonian Quantisation and Constrained Dynamics}
(Leuven University Press, Leuven, 1991).

\bibitem{Gov1.Peskin}
M. E. Peskin and D. V. Schroeder, {\sl An Introduction to Quantum Field Theory}
(Perseus Books Publishing, Cambridge, Massachusetts, 1995).

\bibitem{Gov1.Ali}
S. T. Ali, {\it Quantization Techniques: A Quick Overview},
in the {\sl Proceedings of the Second International Workshop on Contemporary Problems in Mathematical Physics},
eds. J.~Govaerts, M.~N.~Hounkonnou and A.~Z.~Msezane (World Scientific, Singapore, 2002), pp.~3--78.

\bibitem{Gov1.Villa}
J. Govaerts and V. M. Villanueva, {\it Topology Classes of Flat U(1) Bundles and
Diffeomorphic Covariant Representations of the Heisenberg Algebra},
{\sl Int. J. Mod. Phys. A\/} {\bf 15}, 4903--4931 (2000) [e-print {\tt arXiv:quant-ph/9908014} (August 1999)].

\bibitem{Feyn}
R. P. Feynman and A. R. Hibbs, {\sl Quantum Mechanics and 
Path Integrals\/} (McGraw-Hill Book Company, New York, 1965).


\end{thebibliography}
\end{document}